\documentclass[fleqn,usenatbib]{mnras}

\usepackage{newtxtext,newtxmath}

\usepackage[T1]{fontenc}
\usepackage{ae,aecompl}

\usepackage{graphicx}	
\usepackage{graphics}
\usepackage{amsmath}	
\usepackage{caption}
\usepackage{tikz}
\usetikzlibrary{shapes.geometric,arrows}
\usepackage{lipsum,adjustbox}
\usepackage{varwidth}
\usepackage{titlesec}
\usepackage{rotating}
\usepackage{hyperref}
\usepackage{float}
\usepackage[switch]{lineno}
\usepackage{eso-pic}

\titleclass{\subsubsubsection}{straight}[\subsection]

\newcounter{subsubsubsection}[subsubsection]
\renewcommand\thesubsubsubsection{\thesubsubsection.\arabic{subsubsubsection}}
\renewcommand\theparagraph{\thesubsubsubsection.\arabic{paragraph}} 

\titleformat{\subsubsubsection}
{\normalfont\normalsize\itshape}{\thesubsubsubsection}{1em}{}
\titlespacing*{\subsubsubsection}
{0pt}{3.25ex plus 1ex minus .2ex}{1.5ex plus .2ex}
\definecolor{lime}{HTML}{A6CE39}
\DeclareRobustCommand{\orcidicon}{%
	\begin{tikzpicture}
	\draw[lime, fill=lime] (0,0) 
	circle [radius=0.16] 
	node[white] {{\fontfamily{qag}\selectfont \tiny ID}};
	\draw[white, fill=white] (-0.0625,0.095) 
	circle [radius=0.007];
	\end{tikzpicture}
	\hspace{-2mm}
}

\foreach \x in {A, ..., Z}{%
	\expandafter\xdef\csname orcid\x\endcsname{\noexpand\href{https://orcid.org/\csname orcidauthor\x\endcsname}{\noexpand\orcidicon}}
}

\title[DESY3RM-eRASS clusters in high exposure area]{A census of Dark Energy Survey Year 3 redMaPPer galaxy clusters in a high exposure eROSITA All Sky Survey region}

\author[P. A. Giles et al.]
{P. A. Giles$^{1}$\thanks{p.a.giles@sussex.ac.uk}\orcidA{}, 
A. K. Romer$^{1}$\orcidB{},
J. Pilling$^{1}$\orcidL{},
R. Wilkinson$^{1}$\orcidC{},
D. J. Turner$^{2,3,4}$\orcidD{},
\newauthor
T. Jeltema$^{5}$,
A. Farahi$^{6,7}$\orcidI,
M. E. Ramos-Ceja$^{8}$,
F. Balzer$^{8}$,
M. Kluge$^{8}$,
S. Grandis$^{9}$,
\newauthor
E. Bulbul$^{8}$,
M. Hilton$^{10,11}$\orcidE{},
S. Bhargava$^{12}$\orcidM{},
E. Aljamal$^{13}$,
A. E. Evrard$^{13}$,
\newauthor
E. W. Upsdell$^{1}$\orcidH{},
P. T. P. Viana$^{14,15}$\orcidG{},
A. Abdulghafour$^{1}$,
M.~Aguena$^{16,17}$,
\newauthor
S.~S.~Allam$^{18}$,
O.~Alves$^{19}$,
F.~Andrade-Oliveira$^{20}$,
D.~Bacon$^{21}$,
D.~Brooks$^{22}$,
\newauthor
R.~Camilleri$^{23}$,
A.~Carnero~Rosell$^{24,17,25}$,
J.~Carretero$^{26}$,
M.~Costanzi$^{27,16,28}$,
\newauthor
L.~N.~da Costa$^{17}$,
J.~De~Vicente$^{29}$,
S.~Desai$^{30}$,
P.~Doel$^{22}$,
C.~Doux$^{31,32}$,
B.~Flaugher$^{18}$,
\newauthor
J.~Frieman$^{33,18,34}$,
J.~Garc$^{35}$,
M.~Gatti$^{36,34}$,
D.~Gruen$^{37}$,
G.~Gutierrez$^{34}$,
\newauthor
S.~R.~Hinton$^{23}$,
D.~L.~Hollowood$^{5}$,
K.~Honscheid$^{38,39}$,
K.~Kuehn$^{40,41}$,
\newauthor
O.~Lahav$^{22}$,
S.~Lee$^{42}$,
J.~L.~Marshall$^{43}$,
J. Mena-Fern{\'a}ndez$^{44,45}$,
F.~Menanteau$^{46,47}$,
\newauthor
R.~Miquel$^{48,26}$
R. C. Nichol$^{49}$,
R.~L.~C.~Ogando$^{50}$,
A.~A.~Plazas~Malag\'on$^{51,52}$,
\newauthor
S.~Samuroff$^{53,26}$,
E.~Sanchez$^{29}$,
D.~Sanchez Cid$^{29,20}$,
M.~Smith$^{54}$,
\newauthor
M.~E.~da Silva Pereira$^{55}$,
M.~Soares-Santos$^{20}$,
E.~Suchyta$^{56}$,
M.~E.~C.~Swanson$^{46}$,
\newauthor
C.~To$^{33}$,
D.~L.~Tucker$^{18}$,
V.~Vikram$^{57}$,
A.~R.~Walker$^{57}$,
N.~Weaverdyck$^{58,59}$,
J.~Weller$^{8,60}$
\\
Affiliations are listed at the end of the paper
}

\date{Accepted 2026 August 24. Received 2026 August 24; in original form 2026 February 20}

\pubyear{2026}

\AddToShipoutPictureBG*{%
  \AtPageUpperLeft{%
    \hspace{0.75\paperwidth}%
    \raisebox{-3.5\baselineskip}{%
      \makebox[0pt][l]{\textnormal{DES-2025-0901}}
}}}%

\AddToShipoutPictureBG*{%
  \AtPageUpperLeft{%
    \hspace{0.75\paperwidth}%
    \raisebox{-4.5\baselineskip}{%
      \makebox[0pt][l]{\textnormal{FERMILAB-PUB-25-0884-PPD}}
}}}%

\begin{document}
\label{firstpage}
\pagerange{\pageref{firstpage}--\pageref{lastpage}}
\maketitle

\begin{abstract}
We present the cross-match of galaxy clusters detected by redMaPPer using the first three years of Dark Energy Survey (DESY3RM) operations, and the first release of the {\em eROSITA}-All Sky Survey (eRASS1) taken by {\em eROSITA} on board the Spectrum Roentgen Gamma telescope.  We define a sky region ($60^{\circ}<{\rm R.A.}<90^{\circ}$ and $-58^{\circ}<{\rm Dec}<-48^{\circ}$) whereby the mean eRASS1 exposure time of clusters falling within this footprint is $\sim$2.5 times longer than the mean exposure time of the full eRASS1--DESY3RM overlap.  Limiting to a redshift range $0.2 < z < 0.65$ and redMaPPer estimated richnesses to $>$50, we obtain a sample of 57 clusters.  We match these clusters to the cluster catalogue constructed for the eRASS1, finding 45 clusters with a match and 12 with no match in the catalogue. X-ray properties of the 12 unmatched clusters are estimated via a forced photometry analysis.  We investigate the form of the luminosity-richness relation, finding the relation using the full 57 cluster sample is statistically consistent with using only the 45 matched clusters, and will require the deeper data of future eRASS releases (or dedicated X-ray follow-up) to explore further potential differences.  Using external data from {\em XMM}, ACT and SPT, we find detections of 5 of the 12 unmatched clusters, indicating the presence of virialised gas.  Finally, we found the 12 unmatched clusters occupy a higher redshift range and lower richness range (compared to the matched clusters), and find no evidence for an underlying physical reason for the lack of X-ray emission.    
\end{abstract}




\renewcommand{\thefootnote}{\textsuperscript{\arabic{footnote}}}

\section{Introduction}

Enabled by a new generation of imaging surveys, from across the electromagnetic spectrum, galaxy clusters are expected to play an important role in forthcoming attempts to measure cosmological parameters to percent level accuracy.  Several detection methods will be used to deliver cluster samples of sufficient size, quality, and redshift grasp to meet the requirements of Stage IV (and beyond) Dark Energy Experiments \citep[][]{2016arXiv160407626D}. These methods include detections of spectral distortions to the Cosmic Microwave Background (CMB), of extended X-ray emission, and of projected overdensities (in the optical/near-IR band) of member galaxies. Ongoing, or soon to begin, large survey experiments aiming to construct large galaxy cluster samples include the CMB surveys undertaken by the South Pole Telescope \citep[e.g.,][]{2020ApJS..247...25B} and the Atacama Cosmology Telescope \citep[e.g.,][]{2021ApJS..253....3H}, X-ray surveys such as those undertaken by the {\em eROSITA} telescope \citep[e.g.,][]{2021A&A...647A...1P,2022A&A...661A...2L,erass1clusters}, and optical surveys from the Dark Energy Survey \citep[e.g.,][]{2016ApJS..224....1R}, the Hyper Suprime-Cam Subaru Strategic Program \citep[e.g.,][]{2018PASJ...70S..20O}, the Euclid mission \citep[e.g.,][]{2026A&A...711A..34E} and the upcoming Legacy Survey of Space and Time \citep{2018arXiv180901669T}.  

The full power of these single-waveband-regime surveys are unlocked when they are combined.  For example, the Dark Energy Survey (DES) used external X-ray observations in order to model the miscentering effects of galaxy clusters detected by the red-sequence Matched-filter Probabilistic Percolation \citep[redMaPPer,][]{2014ApJ...785..104R} algorithm.  A model description of the distribution of the distance between the optical center and X-ray center \citep[e.g.,][]{2019MNRAS.487.2578Z,kelly24} can then be used to model this miscentering systematic in subsequent galaxy cluster cosmology analysis \citep[e.g.,][]{PhysRevD.102.023509}.  However, the work of \cite{2019MNRAS.487.2578Z} and \cite{kelly24} relied on X-ray information from the {\em XMM} and {\em Chandra} archives i.e., non-contiguous coverage, which will suffer from observer biases.  To overcome this, ideally analysis would include overlapping contiguous multi-wavelength surveys.  Previous work utilising overlapping contiguous X-ray and optical survey areas are typically limited to $\sim$10's of square degrees \citep[e.g.,][]{giles_gama,upsdelldes}.  A drawback of using small area surveys is the lack of higher mass clusters in the survey field, due to the steeper slope of the cluster mass function at these high masses (compared to low mass clusters).  With the launch, and subsequent release of the {\em eROSITA}-All Sky Survey (eRASS) at X-ray wavelengths \citep{2024A&A...682A..34M}, the overlap of X-ray and optical survey areas has the potential to grow by many orders of magnitude.  

For example, the full overlap of the eRASS with the Dark Energy Survey footprint is $\approx$4,200~deg$^{2}$.  For comparison, the work of \cite{upsdelldes} presented the cross-match of clusters detected in the DES with four contiguous {\em XMM} survey regions, comprising of two fields of the {\em XMM}-XXL survey \citep[][]{2016A&A...592A...1P}, and two fields of the {\em XMM}-SERVS \citep{2021ApJS..256...21N}, which when combined, total an area of $\approx$60~deg$^{2}$.  However, one limitation of the eRASS is the somewhat short exposure time (a necessity due to the fact that eRASS surveyed the entire sky).  Currently, only one full X-ray scan of the sky has been publicly released\footnote{Note that only the Western hemisphere of the sky has been released by the German {\em eROSITA} consortium, see \href{https://erosita.mpe.mpg.de/dr1/index.html}{https://erosita.mpe.mpg.de/dr1/index.html}}, with an exposure time of between $\sim100-1200$~seconds depending on the sky position due to the scanning strategy of the eRASS \cite[see][for full details, and in particular Figure~3 of that paper]{2024A&A...682A..34M}.  For comparison, while smaller in area, the median exposure times of the {\em XMM} survey regions studied in \cite{upsdelldes} is 10,000~seconds, and 30,000~seconds in two of the smaller regions studied.  While the eRASS has been revolutionary in the population studies of X-ray selected galaxy clusters \citep[e.g.,][]{erass1clusters,2024A&A...689A.298G,2024A&A...688A.210K,2025arXiv251114356R}, the short exposure could prohibit the investigation of optically selected clusters.  For example, of particular interest over the last decade has been optically selected clusters that appear X-ray under-luminous \citep[e.g.,][]{andreon16,andreon22}.  In order to understand these systems, whether these optically selected clusters represent the full underlying scatter of the cluster scaling relations \cite[][]{andreon16, giles_gama}, and their physical nature \citep[e.g.,][]{2024MNRAS.527..895P}, the ideal scenario would be complete X-ray detections of optically selected clusters over some wide survey area. 

Therefore, the work presented herein is motivated by the desire to understand what is currently achievable with the highest exposure regions of eRASS1 data in the DES footprint.  The outline of this paper is as follows, in Section~\ref{sec:sampleanalysis} we discuss the DESY3RM cluster sample used in this work and the current eRASS1 data available. The X-ray analysis of these clusters is described in Section~\ref{sec:specanalysis}.  We go on to present scaling relations between X-ray and redMaPPer observables, and their associated scatter, in Section~\ref{sec:relations}.  In Section~\ref{sec:discussion} we explore future prospects with subsequent releases of the eRASS.  Conclusions are summarised in Section~\ref{sec:conc}. Throughout this paper we assume a cosmology of $\Omega_{M}$=0.3, $\Omega_{\Lambda}$=0.7 and $H_{0}$=70 km s$^{-1}$ Mpc$^{-1}$.     

\section{Development of the DESY3-eRASS cluster sample}
\label{sec:sampleanalysis}

In this section, we describe the construction of the cluster samples used throughout this work. The process starts with the parent DESY3 optical cluster catalogue (Section~\ref{sec:desrm}), which is then matched to the public eRASS cluster catalogue to determine the presence of X-ray emission (Section~\ref{sec:desy3-erass}).  

\subsection{The DES redMaPPer cluster catalogue}
\label{sec:desrm}

The redMaPPer cluster finding algorithm \citep{2014ApJ...785..104R} is a powerful tool for finding clusters from optical/near-IR photometric survey data and has already been successfully applied to the Sloan Digital Sky Survey, SDSS \citep{2014ApJ...785..104R} and DES \citep{2016ApJS..224....1R}.  redMaPPer self-trains the red sequence model to any available spectroscopic redshifts, and then calculates, in an iterative fashion, photometric redshifts for each cluster identified.  The richness estimated by redMaPPer (hereafter, $\lambda_{\rm RM}$) of each cluster is calculated as the sum of membership probabilities over all galaxies within a scale radius, R$_{\lambda}$, where R$_{\lambda}=1.0h^{-1}{\rm Mpc}(\lambda / 100)^{0.2}$.  The specific redMaPPer cluster sample used throughout this work is based upon the redMaPPer v6.4.22+2 run on the DES Y3 data (Y3A2 Gold 2.2.1), restricting ourselves to the volume-limited catalogue.  In total, this catalogue catalog contains 21,092 clusters above a richness of $\lambda>20$ and within a photometric redmapper redshift ($z_{\rm RM}$) of $0.1 < z_{\rm RM} < 0.93$.  The redMaPPer catalogue can be downloaded from \href{https://des.ncsa.illinois.edu/releases/y3a2/Y3key-cluster}{https://des.ncsa.illinois.edu/releases/y3a2/Y3key-cluster}.


\subsection{Cross-matching the DESY3RM sample with eRASS extended sources}
\label{sec:desy3-erass}

For this study, we limit ourselves to a sky area enclosed by $60^{\circ}<{\rm R.A.}<90^{\circ}$ and $-58^{\circ}<{\rm Dec}<-48^{\circ}$, which totals a survey area $\sim$180~deg$^{2}$.  This region, near the southern galactic pole, is scanned multiple times due to the eRASS scanning strategy, and therefore represents a higher than average X-ray exposure than the full overlap between DES and the eRASS1 (the average exposure time of the full DES and eRASS overlap is $\sim$190 seconds).  Within this area, the mean eRASS1 X-ray exposure time of the redMaPPer clusters in Section~\ref{sec:desrm} is Exp$_{eRASS}\approx 500$s, a factor 2.5 times higher than the average eRASS1 exposure time.  Next, we impose a redshift range of the sample of $0.2<z<0.65$, the same redshift limit used in the DES cluster cosmological analysis \citep[see][]{PhysRevD.102.023509,2025PhRvD.112h3535A}. Finally, we limit the DESY3 redMaPPer catalogue in Section~\ref{sec:desy3-erass} to richnesses of $\lambda_{\rm RM}>50$.  By imposing this high richness limit, this aids in the interpretation of the results of this study by minimising the effects of projection on the detection of redMaPPer clusters at lower richnesses \citep[e.g.,][]{2021MNRAS.505...33M,2025PhRvD.112d3517C}. It has been found that the fraction of projection remains flat at richnesses $\gtrsim$55 \cite{2025MNRAS.544.2080M}.  Additionally, a higher richness cutoff limits the number of non-detections by eRASS of DES clusters, which could also effect the interpretation of the results.  This results in a final sample of 57 clusters for study in this work (referred to hereafter as the DESY3RM cluster sample).  We note that imposing a hard richness cutoff will introduce a Malmquist bias in richness, as richness is not equivalent to a mass limit.  Nonetheless, this sample represents a complete richness selected sample, with an aim of understanding the X-ray properties of this complete sample (see below). 

Next, we matched the DESY3RM sample to the galaxy clusters detected by {\rm eROSITA} in its first six months of operation \citep{erass1clusters}.  This cluster sample is constructed from the first six months of {\rm eROSITA} survey operations, covering a total of 13,116 deg$^{2}$ in the western Galactic half of the sky.  The catalogue contains 14,818 X-ray detections with an extent likelihood $>3$, detected in the 0.2-2.3~keV band.  Using the {\tt eROMaPPer} algorithm \citep{2020MNRAS.499.4768I}, optical counterparts are found for 12,554 sources, utilising legacy survey DR9-North and DR10 data.  Note that spectroscopic redshift data is used where possible, and preferred over the photometric estimates.  In cases where no spectroscopic or photometric estimate are available, literature values are used.  For full details of the optical confirmation and redshift estimates of the eRASS1 catalogue, see \cite{2024A&A...688A.210K}.  Finally, after removing contaminants, the \cite{erass1clusters} cluster sample contains 12,247 optically confirmed clusters.  Limiting to the R.A. and Dec region outlined above, the sample reduces to 572 clusters, hereafter referred to as the eRASS1-high-exp sample.  

\begin{figure*}
\begin{center}
\begin{tabular}{ccc}
\includegraphics[width=5.5cm]{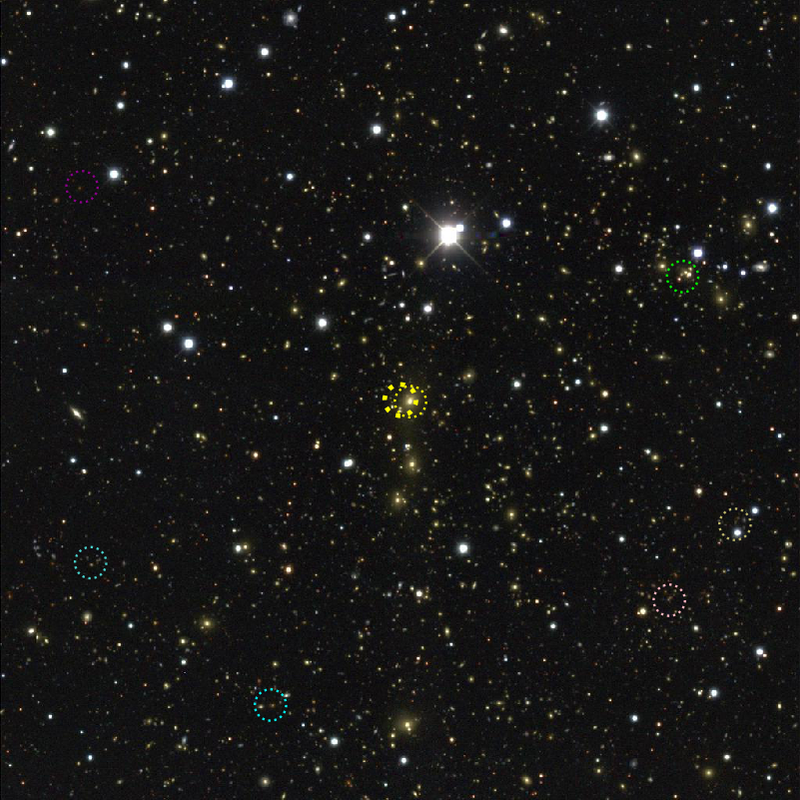} &
\includegraphics[width=5.5cm]{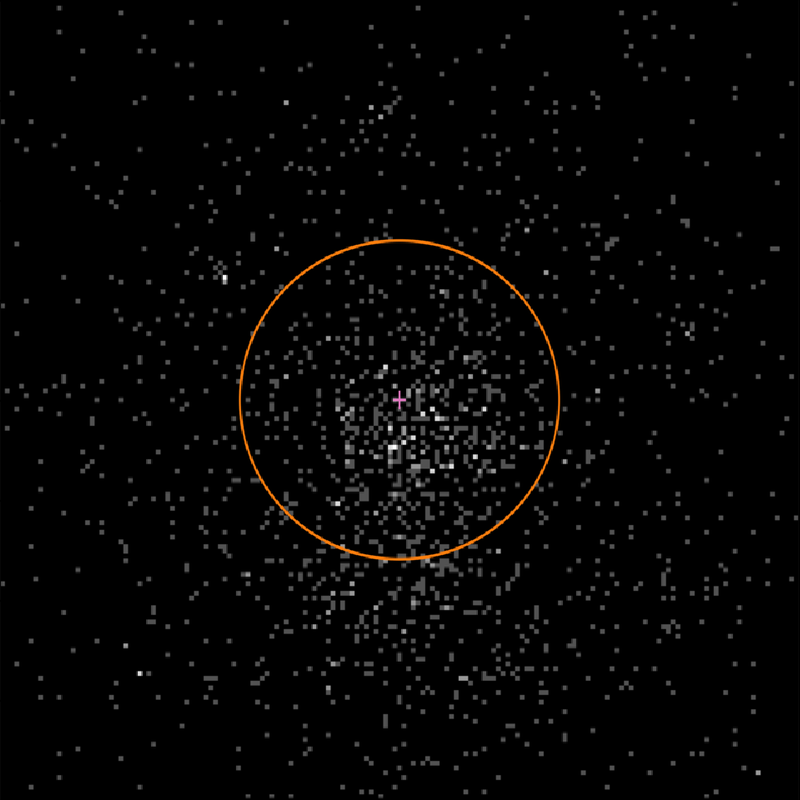} &
\includegraphics[width=5.5cm]{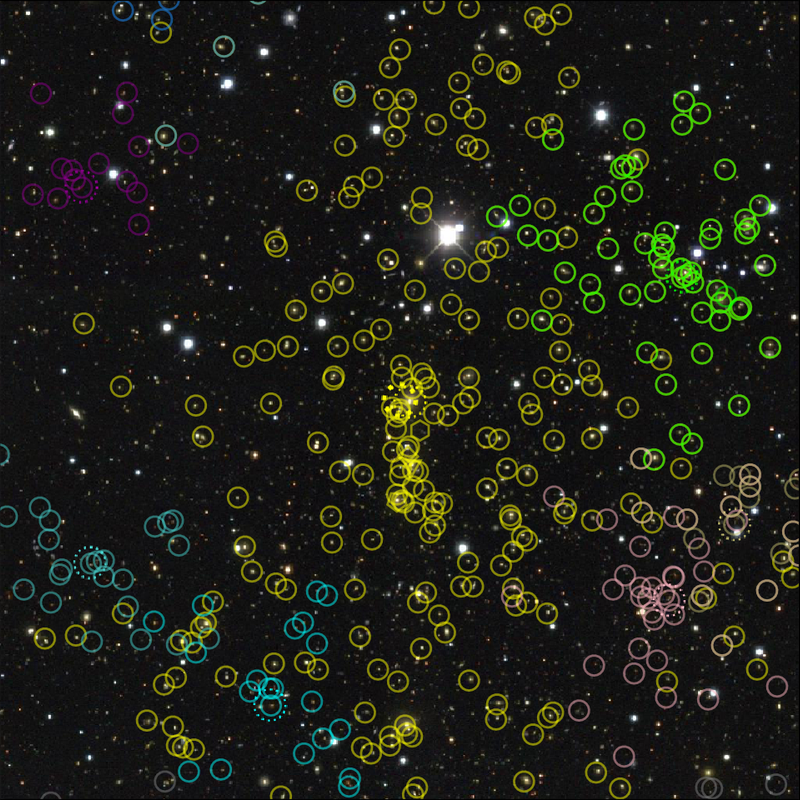} \\
\end{tabular}
\end{center}
\caption[]{An example of a cluster in the DESY3RM sample. Left: DES optical image of the cluster with redMaPPer ID = 4, at a redshift of $z=0.299$ and richness of $\lambda_{\rm RM}=$207.  The dashed yellow circle represents the 1st most probable central galaxy of the cluster as determined by redMaPPer. Middle: Soft-band (0.2 - 2.3 keV) eRASS1 X-ray cutout of the matched eRASS extended source.  The orange circle represents a radius of 600 kpc. Right:  DES optical image, with member galaxies of the cluster overplotted as yellow solid circles.  Galaxy members of other nearby redMaPPer clusters nearby are circled in different colours (e.g. green, blue, pink).}
\label{fig:cutouts}
\end{figure*}

To match the two samples, we take the position of the DESY3RM clusters and find all eRASS1-high-exp clusters within 2~Mpc (calculated from the redshift of the DESY3RM cluster).  Of the 57 DESY3RM clusters, 46 have a eRASS1-high-exp cluster within 2~Mpc, and the remaining 11 have no eRASS1-high-exp cluster within 2~Mpc.  Next, to confirm an association of the 46 DESY3RM clusters with the matched eRASS1-high-exp cluster, we used the Object Classification Tool for Astronomy VIsUaliSation (OCTAVIUS\footnote{\href{https://furciferi.github.io/OCTAVIUS}{https://furciferi.github.io/OCTAVIUS}}), to visually inspect each cluster.  OCTAVIUS is a web-based visual inspection tool that can display images in order for classification.  For this, X-ray cutout images around the DESY3RM clusters were provided by the German {\em eROSITA} consortium, which were 2$\times$2~Mpc on a side and centered on the DESY3RM position.  Corresponding DESY3 images were constructed of the same size.  An example of these images are given in Figure~\ref{fig:cutouts}, for the DESY3RM cluster with redMaPPer ID RMID=4.  These images, along with auxiliary information for each cluster (e.g., redMaPPer redshift, eRASS redshift, redMaPPer R.A., redMaPPer Dec, eRASS R.A., eRASS Dec etc.), we find that the 45 DESY3RM clusters have a corresponding X-ray emission matched in the eRASS1-high-exp, we denoted these clusters as the DESY3RM-eRASS-match sample.  During this visual inspection process, we found that one of the 46 DESY3RM clusters matched to a cluster in projection.  The cluster redMaPPer ID 3834 is matched to a nearby eRASS confirmed cluster (i.e., an extended source detected) in projection.  This eRASS cluster is at a redshift of $z=0.19$ (as given in the eRASS catalogue), compared to the redshift of redMaPPer ID 3834 of $z=0.61$, and is therefore unassociated with 3834.  We therefore class redMaPPer ID 3834 as an unmatched cluster.  This results in 12 clusters having no corresponding eRASS match, denoted as DESY3RM-eRASS-unmatched. Additionally, during our visual inspection, we found that one of the DESY3RM clusters suffered from a mis-percolation effect. A full description of mis-percolation can be found in \citet[][see their Section 3.7]{2019ApJS..244...22H}, but briefly, redMaPPer can incorrectly split a single massive halo into multiple separate clusters, typically assigning large richness to the smaller system and a small richness to the larger system.  We found the cluster redMaPPer ID = 596 (assigned $\lambda_{\rm RM}=58$ and $z_{\rm RM}= 0.28$) had a large offset with the matched eRASS cluster 1eRASS J053155.6-552032.  Through the visual inspection, we found the redMaPPer cluster ID = 591 (assigned $\lambda_{\rm RM}=13$ and $z_{\rm RM}= 0.29$) was coincident with the aforementioned eRASS cluster.  Therefore, as per the procedure in \cite{2019ApJS..244...22H}, we take the redMaPPer ID = 591 as the cluster that should be contained in our sample, and simply swap the richness values.  The properties of the DESY3RM-eRASS-match sample is given in Table~\ref{tab:matched_props}, and the properties of the DESY3RM-eRASS-unmatched sample is given in Table~\ref{tab:unmatched_props}.

\subsubsection{DESY3-{\em eRASS}-unmatched clusters associated with eRASS point sources}
\label{sec:undet_point}

For the 12 DESY3-eRASS-unmatched clusters, we matched them to the full eRASS source catalogue \citep{2024A&A...682A..34M}.  This was done as it includes extended source that were not confirmed as clusters, and it has recently been shown that there are a large number of point source detections in eRASS1 that are in fact clusters \citep[see][]{2025A&A...701A.283B}.  Indeed, the Phoenix cluster, one of the highest X-ray flux clusters in the sky, was detected as a point source in eRASS1.   Additionally, when estimating the properties of these unmatched clusters, if a point source has been detected that is in fact due to cluster emission, then one would want to include that emission in subsequent analysis (see Section~\ref{sec:undet_lx}).  

We follow as similar process as outlined in Section~\ref{sec:desy3-erass}, finding all sources within 2~Mpc of the redMaPPer cluster positions.  Since point sources are numerous in X-rays, it is unsurprising that all 12 DESY3-eRASS-unmatched clusters have a point source within this matching radius.  To determine whether the point source is associated with the cluster in question, we used OCTAVIUS to flag whether the point source was likely associated with the cluster.  In general, we associated the point source with the cluster if it was coincident with a collection of redMaPPer member galaxies.  We find that there are 7 point sources that appear coincident with the redMaPPer position and we flag all of these as cluster emission.  There are two points to note in this process.  The cluster redMaPPer ID = 2096 appears to have visually identified extended emission, with this cluster being detected as a point source.  This cluster also has a clear detection in {\em XMM} data (see Section~\ref{sec:additional_X-ray}).  Second, all visually identified point sources flagged as associated with the cluster also appear in the point source cluster catalogue of \cite{2025A&A...701A.283B}.

\subsubsection{An assessment of chance associations}

In order to determine the rate of chance associations between the DESY3RM and eRASS catalogues, we make use of the DESY3RM random catalogue.  The random catalog is constructed such as to map the detectability of clusters as a function of redshift and richness, taking into account the large-scale structure that is already imprinted on the galaxy catalog.  The details of the construction of the random catalogue follows that described in \citet[][see Section~3.6 of that paper]{2016ApJS..224....1R}.  We restrict the random catalogue to the RA, Dec range considered in this work, and limit to $0.2 < z < 0.65$ and $\lambda_{\rm RM} > 50$, resulting in 10,411 random clusters.  We draw at random from the these random clusters and create samples of equal size to the DESY3RM sample (i.e. 57 clusters), resulting in 178 separate catalogues of random clusters.  For each of the 178 samples, we match to the eRASS catalogue using a 2~Mpc matching radius.  We find that with a sample of 57 clusters and a matching radius of 2~Mpc, we would find $5.77\pm2.14$ chance associations.  We note however, that while we use a 2~Mpc matching radius, in practice, the distance between the redMaPPer coordinates and the X-ray coordinates of the DESY3-eRASS-matched sample are all $<0.5$~Mpc.  Therefore, we perform the same analysis described above, but retain on random matches with 0.5~Mpc.  We find that in ~ 65\% of random samples there would be no chance associations, in ~25\% of cases 1 chance association and in ~5\% of cases 2 chance associations.  This agrees well with the case of 3834, representing one case of a cluster in projection for our sample. 

Next, we focus on the matching of the random clusters with eRASS detected point sources.  This is of particular interest for the DESY3-eRASS-unmatched sample which are matched to point sources in Section~\ref{sec:undet_point}, as we have a population of point sources that we have visually identified as associated with the DESY3-eRASS-unmatched sample.  To assess the rate of chance associations of eRASS point sources, we again make use of the DESY3RM random catalogue.  In this instance we split the random catalogue into samples of 12 clusters (the size of our DESY3-eRASS-unmatched cluster sample), resulting in 801 samples of random clusters.  All of the eRASS point sources identified as being associated with an unmatched cluster through visual inspection (see Section~\ref{sec:undet_point}) are within 30$^{\prime\prime}$ of the RM position.  Therefore, for each random sample, we match to the eRASS point source catalogue within this 30$^{\prime\prime}$ matching radius.  We find that in 74\% of cases there is no match between the clusters and a point source, in 22\% of cases there is one match, 3.5\% of cases there are two matches and in 0.5\% of cases there are three matches.  Therefore, the probability that all 7 point sources deemed associated with the DESY3-eRASS-unmatched clusters would be ruled out as chance associations based upon this analysis.

\section{X-ray Analysis of the DESY3-eRASS sample}
\label{sec:specanalysis}

The eRASS1 cluster catalogue of \cite{erass1clusters} provides numerous measured X-ray properties of their detected clusters.  The X-ray property we are concerned with in this work is the X-ray luminosity.  In particular, we focus on the soft-band (0.5--2.0~keV) luminosity estimated within two radii, a physical 300~kpc radius, and an overdensity radius of r$_{500}$, hereafter denoted L$_{X,0520}^{300kpc}$ and L$_{X,0520}^{500}$ respectively.  These properties are available for all of the DESY3-eRASS-matched clusters \citep[as taken from][]{erass1clusters}, and we make use of them in this work, we detail here the salient points of the analysis.  The value of r$_{500}$ is calculated from the estimated mass of the cluster, which itself is determined from a calibrated count-rate - mass scaling relation with weak lensing shear data \citep[see][for full details]{2024A&A...689A.298G}. 

To estimate the X-ray properties of the eRASS cluster sample, the MultiBand Projector in 2D ({\tt MBProj2D}) tool \citep{2018MNRAS.474.1065S} was used.  Briefly, {\tt MBProj2D} uses a forward modeling approach of background-included X-ray images of galaxy clusters to fit cluster and background emission simultaneously.  For each cluster, images and exposure maps are created in 7 energy bands (in units of keV): [0.3–0.6], [0.6–1.0], [1.0–1.6], [1.6–2.2], [2.2–3.5], [3.5–5.0], [5.0–7.0].  Cluster profiles are converted into 3D emissivity profiles for each energy band, using {\tt Xspec} \citep{1996ASPC..101...17A} with the {\tt APEC} model \citep{2001ApJ...556L..91S} to convert between temperature, density, metallicity to emissivity, assuming a Galactic absorption using {\tt Tbabs} \cite{2000ApJ...542..914W}.  The profiles are projected onto the sky to make a rate image, given a parametrized cluster central position, out to some fixed physical cluster radius.   Given the combined model count image, the mask image, and the observed count image, a Poisson likelihood is computed. The likelihoods for each input band and image are combined to give a total Poisson likelihood.  An Markov chain Monte Carlo (MCMC) chain produces posterior distributions of free parameters, and X-ray properties such as flux and luminosity as functions of radius and energy bands are then derived from the chains.

\begin{table*}
    \centering
    \begin{tabular}{ccccccl}
    \hline
    \hline
  redMaPPer ID & R.A. & Dec. & $z_{RM}$ & $\lambda_{RM}$ & eRASS NAME & L$_{X,0520}^{r500}$ \\
   & (deg) & (deg) & & & & (10$^{42}$~erg~s$^{-1}$) \\
  (a) & (b) & (c) & (d) & (e) & (f) & (g) \\
  \hline
  4    &    79.16 & -54.50 & 0.299  &   207.2$\pm$7.2 & 1eRASS J051636.6-543031  &    433.2$^{+18.5}_{-15.2}$  \\     
  6    &    62.80 & -48.33 & 0.413  &   178.0$\pm$5.0 & 1eRASS J041116.2-481850  &    427.8$^{+35.7}_{-28.2}$  \\     
  38   &    61.49 & -49.27 & 0.301  &   136.5$\pm$4.8 & 1eRASS J040558.4-491553  &    265.1$^{+19.9}_{-17.0}$  \\     
  48   &    69.57 & -54.32 & 0.423  &   132.0$\pm$4.7 & 1eRASS J043817.8-541917  &    643.0$^{+35.5}_{-32.8}$  \\     
  61   &    89.93 & -52.82 & 0.608  &   142.5$\pm$5.1 & 1eRASS J055942.9-524851  &    242.1$^{+74.8}_{-55.3}$  \\     
  231  &    69.57 & -55.89 & 0.281  &   82.9$\pm$3.8  & 1eRASS J043815.6-555338  &    28.4$^{+4.6}_{-4.0}$      \\  
  353  &    85.05 & -56.28 & 0.447  &   95.0$\pm$4.5  & 1eRASS J054014.3-561721  &    58.9$^{+16.9}_{-13.8}$   \\    
  382  &    69.67 & -49.11 & 0.232  &   74.9$\pm$3.4  & 1eRASS J043839.3-490630  &    36.5$^{+4.9}_{-4.2}$     \\   
  392  &    87.96 & -56.75 & 0.207  &   74.9$\pm$4.6  & 1eRASS J055150.7-564459  &    39.3$^{+4.7}_{-4.5}$     \\   
  591  &    82.99 & -55.34 & 0.297  &   58.1$\pm$3.8  & 1eRASS J053155.6-552032  &    96.5$^{+8.0}_{-8.7}$     \\   
  629  &    80.56 & -48.30 & 0.302  &   84.7$\pm$4.7  & 1eRASS J052215.2-481813  &    147.6$^{+14.3}_{-15.0}$  \\     
  633  &    65.94 & -54.53 & 0.262  &   70.2$\pm$3.8  & 1eRASS J042342.9-543120  &    46.6$^{+5.3}_{-5.5}$     \\   
  674  &    77.34 & -53.70 & 0.455  &   89.4$\pm$5.8  & 1eRASS J050921.2-534213  &    298.8$^{+24.0}_{-22.8}$  \\    
  694  &    71.85 & -50.92 & 0.387  &   63.1$\pm$3.1  & 1eRASS J044724.0-505505  &    98.0$^{+10.7}_{-10.4}$   \\    
  813  &    87.54 & -54.70 & 0.582  &   63.8$\pm$3.6  & 1eRASS J055007.6-544137  &    70.7$^{+19.3}_{-16.9}$   \\    
  814  &    89.94 & -55.75 & 0.397  &   75.7$\pm$4.4  & 1eRASS J055943.5-554514  &    65.8$^{+12.8}_{-12.1}$   \\    
  829  &    70.01 & -48.72 & 0.229  &   69.1$\pm$4.6  & 1eRASS J044002.1-484244  &    42.7$^{+4.5}_{-4.8}$     \\   
  913  &    86.16 & -48.15 & 0.456  &   63.1$\pm$3.2  & 1eRASS J054438.5-480842  &    35.7$^{+14.0}_{-13.3}$   \\    
  1014 &    62.80 & -49.17 & 0.324  &   56.6$\pm$3.0  & 1eRASS J041108.0-491015  &    44.5$^{+10.3}_{-9.2}$    \\   
  1181 &    60.68 & -53.51 & 0.294  &   55.7$\pm$3.2  & 1eRASS J040244.1-533034  &    48.9$^{+7.4}_{-6.4}$     \\   
  1265 &    60.97 & -57.32 & 0.464  &   61.9$\pm$3.7  & 1eRASS J040352.5-571940  &    235.5$^{+30.4}_{-24.3}$  \\     
  1358 &    75.18 & -49.04 & 0.216  &   51.7$\pm$3.1  & 1eRASS J050043.3-490251  &    28.7$^{+4.6}_{-4.5}$     \\   
  1397 &    83.36 & -54.34 & 0.297  &   62.9$\pm$3.7  & 1eRASS J053322.1-542008  &    67.0$^{+8.0}_{-7.6}$     \\   
  1403 &    67.70 & -57.62 & 0.248  &   54.9$\pm$3.1  & 1eRASS J043048.1-573715  &    69.3$^{+5.0}_{-5.6}$     \\   
  1423 &    69.93 & -53.51 & 0.426  &   55.8$\pm$3.1  & 1eRASS J043942.1-533024  &    114.0$^{+14.2}_{-12.1}$  \\     
  1432 &    80.50 & -50.45 & 0.554  &   80.8$\pm$5.1  & 1eRASS J052202.8-502705  &    105.5$^{+24.5}_{-20.5}$  \\     
  1437 &    74.12 & -51.28 & 0.565  &   79.9$\pm$5.9  & 1eRASS J045628.0-511644  &    250.1$^{+29.3}_{-28.4}$  \\     
  1464 &    86.03 & -53.16 & 0.305  &   69.2$\pm$4.6  & 1eRASS J054417.8-531032  &    83.0$^{+10.9}_{-10.8}$   \\    
  1481 &    68.99 & -53.90 & 0.278  &   56.0$\pm$3.3  & 1eRASS J043556.3-535352  &    32.7$^{+5.3}_{-4.9}$     \\   
  1545 &    67.30 & -56.08 & 0.264  &   52.1$\pm$3.4  & 1eRASS J042912.4-560327  &    45.9$^{+5.8}_{-5.3}$     \\   
  1638 &    66.52 & -54.93 & 0.634  &   62.9$\pm$4.0  & 1eRASS J042604.2-545516  &    254.0$^{+37.7}_{-35.0}$  \\     
  1643 &    71.13 & -52.69 & 0.549  &   57.5$\pm$3.9  & 1eRASS J044431.5-524134  &    48.5$^{+14.4}_{-11.4}$   \\    
  1648 &    77.13 & -51.19 & 0.468  &   66.9$\pm$5.6  & 1eRASS J050832.2-511133  &    42.6$^{+12.5}_{-11.8}$   \\    
  1682 &    79.61 & -56.01 & 0.320  &   51.6$\pm$3.1  & 1eRASS J051826.5-560055  &    58.7$^{+7.7}_{-7.7}$     \\   
  1977 &    77.92 & -56.40 & 0.590  &   58.0$\pm$4.1  & 1eRASS J051137.4-562332  &    81.4$^{+19.3}_{-16.1}$   \\    
  2043 &    78.18 & -51.66 & 0.643  &   77.0$\pm$5.0  & 1eRASS J051240.8-514007  &    264.6$^{+46.1}_{-36.5}$  \\     
  2216 &    75.53 & -49.54 & 0.213  &   59.1$\pm$5.2  & 1eRASS J050211.9-493257  &    35.3$^{+5.4}_{-5.1}$     \\   
  2364 &    73.01 & -48.10 & 0.413  &   52.4$\pm$3.3  & 1eRASS J045158.1-480627  &    93.7$^{+19.0}_{-16.6}$   \\    
  2377 &    62.54 & -51.19 & 0.346  &   52.2$\pm$3.3  & 1eRASS J041010.5-511134  &    22.4$^{+6.9}_{-5.8}$     \\   
  2691 &    60.14 & -52.25 & 0.341  &   62.4$\pm$5.8  & 1eRASS J040034.0-521445  &    59.4$^{+10.0}_{-10.3}$   \\    
  2696 &    73.05 & -54.66 & 0.484  &   52.5$\pm$3.5  & 1eRASS J045216.3-543939  &    75.1$^{+16.0}_{-12.3}$   \\    
  2904 &    72.79 & -52.17 & 0.636  &   60.2$\pm$4.3  & 1eRASS J045111.4-521001  &    64.4$^{+22.3}_{-16.4}$   \\    
  3307 &    88.58 & -56.62 & 0.209  &   52.0$\pm$5.5  & 1eRASS J055421.5-563718  &    33.9$^{+5.2}_{-4.6}$     \\   
  3419 &    60.40 & -50.74 & 0.299  &   51.2$\pm$3.6  & 1eRASS J040134.8-504419  &    21.2$^{+6.8}_{-6.9}$     \\   
  3619 &    75.29 & -49.28 & 0.601  &   54.5$\pm$4.1  & 1eRASS J050105.3-491704  &    90.4$^{+35.9}_{-25.5}$   \\
  \hline
    \end{tabular}
    \caption[]{Properties of the DESY3-eRASS-matched sample (see Section~\ref{sec:desy3-erass} \&~\ref{sec:specanalysis}) used in this work.  The properties provided are (a) redMaPPer ID; (b) redmaPPer right ascension; (c) redMaPPer declination; (d) redMaPPer redshift; (e) redMaPPer richness; (f) matched eRASS catalogue Name; (g) eROSITA soft band X-ray luminosity (0.5--2.0 keV) within $r_{500}$ as taken from \cite{erass1clusters}.}
    \label{tab:matched_props}
\end{table*}

\begin{table*}
    \centering
    \begin{tabular}{cccccl}
    \hline
    \hline
  redMaPPer ID & R.A. & Dec. & $z_{RM}$ & $\lambda_{RM}$ & L$_{X,0520}^{500}$ \\
  & (deg) & (dec) & & & (10$^{42}$~erg~s$^{-1}$) \\
  (a) & (b) & (c) & (d) & (e) & (f) \\
  \hline
  \hline
  655  &    61.01 & -48.22 & 0.433  &   76.4$\pm$3.6  &    30.3$^{+10.8}_{-8.3}$    \\
  1812 &    71.89 & -51.74 & 0.402  &   58.6$\pm$3.5  &    17.8$^{+5.5}_{-4.3}$     \\   
  2096 &    65.61 & -51.68 & 0.573  &   60.3$\pm$4.4  &    72.6$^{+17.4}_{-14.4}$   \\    
  2326 &    78.19 & -50.34 & 0.577  &   68.6$\pm$4.6  &    19.3$^{+13.1}_{-8.4}$    \\    
  2442 &    69.84 & -50.98 & 0.550  &   52.9$\pm$3.3  &    21.6$^{+10.5}_{-7.4}$    \\    
  2451 &    71.62 & -51.66 & 0.476  &   51.2$\pm$3.4  &    2.6$^{+5.4}_{-1.8}$     \\   
  2609 &    63.60 & -50.68 & 0.572  &   52.1$\pm$3.7  &    40.6$^{+14.5}_{-10.6}$   \\    
  3363 &    73.23 & -51.65 & 0.395  &   51.1$\pm$3.1  &    4.0$^{+6.6}_{-2.8}$      \\ 
  3470 &    85.88 & -50.95 & 0.426  &   57.0$\pm$6.1  &    17.2$^{+8.6}_{-6.2}$      \\  
  3834 &    72.46 & -54.10 & 0.615  &   54.6$\pm$4.0  &    26.4$^{+12.7}_{-8.8}$   \\    
  4156 &    82.73 & -54.80 & 0.408  &   51.8$\pm$4.0  &    4.7$^{+5.3}_{-2.9}$      \\  
  4917 &    68.51 & -51.01 & 0.649  &   51.3$\pm$3.8  &    15.0$^{+15.0}_{-8.7}$   \\
  \hline
    \end{tabular}
    \caption[]{Properties of the DESY3-eRASS-unmatched sample (see Section~\ref{sec:desy3-erass} \& \ref{sec:undet_lx}) used in this work. The properties provided are (a) redMaPPer ID; (b) redmaPPer right ascension; (c) redMaPPer declination; (d) redMaPPer redshift; (e) redMaPPer richness; (f) soft band X-ray luminosity (0.5--2.0 keV) within $r_{500}$ estimated via the forced photometry method in Section~\ref{sec:undet_lx}.}
    \label{tab:unmatched_props}
\end{table*}

\subsection{Luminosities associated with DESY3RM-eRASS-undetected clusters}
\label{sec:undet_lx}

Given the DESY3RM-eRASS-unmatched clusters do not have an entry in the \cite{erass1clusters} catalogue, they will naturally not have an X-ray property measurement. Therefore, to estimate their X-ray properties, we follow a method outlined in \cite{2018MNRAS.477.5517W}.  This method implements a Bayesian approach for calculating X-ray aperture photometry.  

First, we make images and exposure maps centered on the redMaPPer coordinates of the DESY3RM-eRASS-unmatched clusters.  These are constructed by using the {\em eROSITA} Science Analysis Software System Data Release 1 \citep[hereafter eSASS;][]{2022A&A...661A...1B}\footnote{\label{foot:esass}\href{https://erosita.mpe.mpg.de/dr1/eSASS4DR1/}{https://erosita.mpe.mpg.de/dr1/eSASS4DR1/}}. For image generation we use {\tt EVTOOL}, generating images in the 0.5 - 2 keV band. Exposure maps are produced using the {\tt EXPMAP} command with the parameters {\tt withweights=yes}, to weight each of the seven telescope modules by their relative contributions to the effective are, and {\tt withdetmaps=yes}, to exclude exposure contributions from the edges of the detector where the effective area is challenging to characterise reliably.

To estimate the number of source counts, we define an aperture of 300~kpc within which to perform the analysis, centered on the redMaPPer coordinates of the cluster and assuming the redMaPPer redshift.  This radius is chosen as luminosities are reported in \cite{erass1clusters} within 300~kpc, therefore providing a point of comparison to validate the aperture photometry method.  Next, to estimate the background counts, we assumed an annulus of 1200-1500~kpc (i.e., 4$\times$-5$\times$~300~kpc).  For this analysis, point sources in the source and background region were masked based upon the eRASS source catalogue \citep{2024A&A...682A..34M}.  However, point sources that were determined to be coincident with the cluster, and therefore assumed to be cluster emission (see Section~\ref{sec:undet_point}), were retained in the analysis.  This is a drawback of the analysis presented herein, and could overestimate the cluster flux if point source emission is in fact associated with e.g., AGN emission.   

The counts ($C$) within the area of the source aperture ($A_{s}$), is a sum of the counts from the source ($S$) and the background ($B_{s}$).  The background counts ($B$) are estimated within an area of $A_{b}$.  Therefore, the observed counts are given by Poisson distributions where:
\begin{align}
    C \sim {\rm Poisson}\left(\frac{S + B_{s}}{T_{s}}\right),
\end{align}
and
\begin{align}
    B_{s} \sim {\rm Poisson}\left(\frac{A_{b}}{A_{s}}\frac{B}{T_{b}}\right).
\end{align}
The factors $T_{s}$ and $T_{b}$ convert the source aperture and background counts into count rates from the average exposure times within $A_{s}$ and $A_{b}$.  Finally, an area correction factor is used to scale between the source and background areas i.e., $A_{b}/A_{s}$.  $S$ is determined from its posterior probability density marginalized over the background,
\begin{align}
    p(S|CB_{s}) = \int dB ~ p(SB|CB_{s}).
\end{align}
Finally, the joint posterior probability of $S$ and $B$ can be rewritten as
\begin{align}
    p(SB|CB_{s}) = \frac{p(S)p(B)p(C|SB)p(B_{s}|B)}{\iint dB~dS~p(S)p(B)p(C|SB)p(B_{s}|B)},
\end{align}
where $p(C|SB)$ and $p(B_{s}|B)$ are Poisson distributions, and $p(S)$ and
$p(B)$ are generalized non-informative $\gamma$-priors \citep[see][for full details]{2018MNRAS.477.5517W}.

The count-rate posterior probability density distributions are then converted into a flux distribution via the use of an Energy Conversion Factor (ECF).  We produced an ECF for each individual cluster using the following method.  Using {\tt Xspec}, we produced a fake spectrum under the assumption of an absorbed {\sc apec} emission model, along with a standard {\em eROSITA} on axis response\footnote{The eROSITA response files were taken from \href{https://erosita.mpe.mpg.de/dr1/eSASS4DR1/eSASS4DR1_arfrmf/}{https://erosita.mpe.mpg.de/dr1/eSASS4DR1/eSASS4DR1\_arfrmf/}}.  The redshift is taken from the redMaPPer catalogue and the value of the absorption ($n_{H}$) taken from \cite{2016A&A...594A.116H} (estimated at the redMaPPer coordinates). The temperature was set to 3~keV and the metal abundance was set to 0.3~Z$_{\odot}$, adopting solar abundances from \citep{2009ARA&A..47..481A}.  The fake spectrum assumes an exposure time of 50~ks.  The ECF is estimated from this spectrum as the ratio of the count rate to the unabsorbed flux.  The count-rate posteriors for each cluster in the DESY3-eRASS-unmatched sample are converted to a flux from their ECF, and converted in to a k-corrected luminosity.

\begin{figure}
\begin{center}
\includegraphics[width=8.3cm]{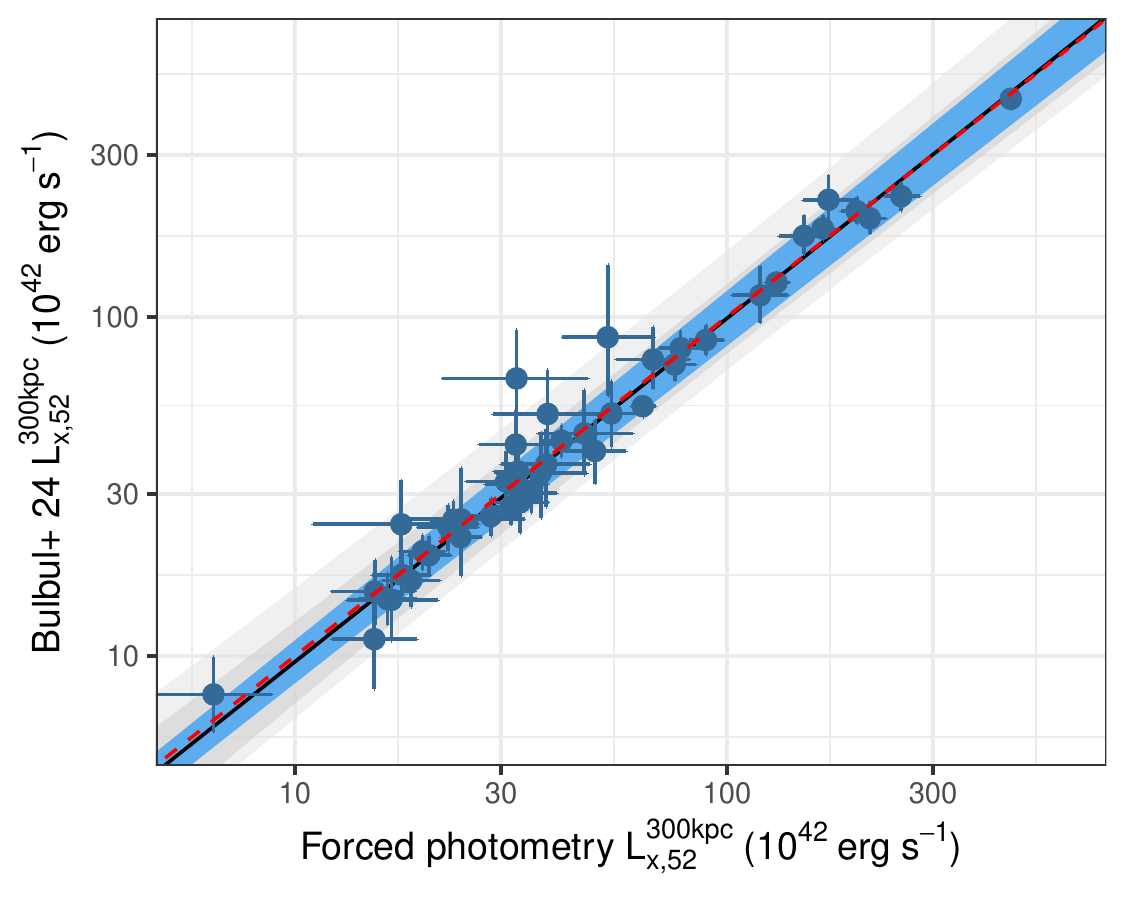}
\end{center}
\caption[]{Comparison of the luminosities for the DESY3RM-eRASS-match sample derived from the forced photometry method outlined in Sect~\ref{sec:undet_lx} to those given in \cite{erass1clusters}.  In each case the luminosities are estimated within a 300kpc aperture.  The best-fit to the data is given by the black solid line (with the blue shaded region the 68\% uncertainty), the grey bands represent the 1, 2, and 3$\sigma$ intrinsic scatter. The red dashed line highlights the 1:1 relation.} 
\label{fig:lx-comp}
\end{figure}

To determine the reliability of the luminosities estimated via this method for the DESY3-eRASS-unmatched clusters, we estimate luminosities of the DESY3RM-eRASS-match sample (i.e., detected clusters) using this forced photometry method, and compare to their luminosities estimated from \cite{erass1clusters} (see Section~\ref{sec:specanalysis}).  The comparison of the two luminosity estimates for the DESY3RM-eRASS-match sample is given in Figure~\ref{fig:lx-comp}, blue points.  For this comparison, the luminosities are compared within the 300~kpc aperture \citep[as available in][]{erass1clusters}.  A fit to the data is performed using {\sc linmix}\footnote{\href{https://github.com/jmeyers314/linmix}{https://github.com/jmeyers314/linmix}}, an implementation of the Bayesian regression fitting model as detailed in \cite{2007ApJ...665.1489K}.  We fit to the data a power-law of the form:
\begin{align}
{\rm log}(L_{B}) &=  {\rm log}(A_{L_{B}L_{G}}) + B_{L_{B}L_{G}}~{\rm log}~(L_{G}) \pm \sigma_{L_{B}|L_{G}},
\label{equ:lcomp}
\end{align}
where $L_{B}$ is the soft-band (0.5-2.0~keV) luminosity within 300~kpc from \cite{erass1clusters}, and $L_{G}$ is the luminosity as estimated from the forced photometry method outlined above (0.5-2.0~keV and within 300~kpc).  The factors $A_{L_{B}L_{G}}$, $B_{L_{B}L_{G}}$ and $\sigma_{L_{B}|L_{G}}$ represents the normalisation, slope and lognormal intrinsic scatter respectively.  From the power-law fit to the data, we find $A_{L_{B}L_{G}} = 0.94\pm0.12$, $B_{L_{B}L_{G}} = 1.01 \pm 0.03$ and $\sigma_{L_{B}|L_{G}} = 0.05 \pm 0.03$.  The fit to the data is given by the black solid line in Figure~\ref{fig:lx-comp}.  For comparison, the 1:1 relation is given by the red dashed line.  While the fit is consistent with a 1:1 relation, we scale the forced photometry luminosities of the DESY3RM-eRASS-unmatched clusters by this power-law relation to be consistent with the \cite{erass1clusters} luminosities.

\begin{figure}
\begin{center}
\includegraphics[width=8.3cm]{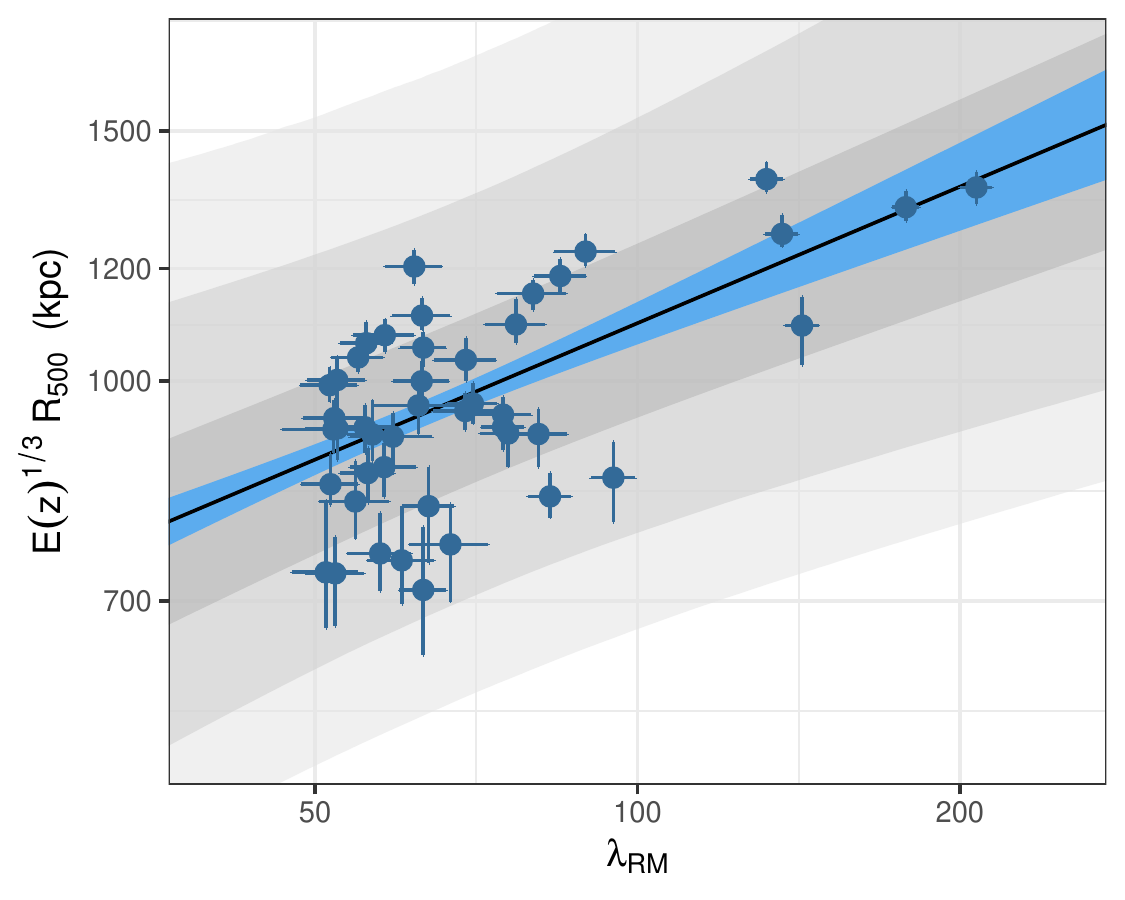}
\end{center}
\caption[]{r$_{500}$-$\lambda_{\rm RM}$ relation for the DESY3RM-eRASS-matched clusters (blue points).  The best-fit to the data is given by the black solid line (with the blue shaded region the 68\% uncertainty), the grey bands represent the 1, 2, and 3$\sigma$ intrinsic scatter.} 
\label{fig:r500-lambda}
\end{figure}

Finally, the luminosities estimated via the forced photometry method are within a fixed annular size of 300~kpc.  Therefore, for their use in further analysis (e.g., scaling relations) we scale these luminosities out to an estimate of r$_{500}$.  To preform the extrapolation, an estimate of the r$_{500}$ values for the DESY3RM-eRASS-unmatched clusters is required.  We construct an r$_{500}$-$\lambda_{\rm RM}$ relation using clusters from the DESY3RM-eRASS-matched sample, and use this to estimate r$_{500}$ values based upon the measured $\lambda_{\rm RM}$ for the DESY3RM-eRASS-unmatched clusters.  The r$_{500}$-$\lambda_{\rm RM}$ relation for the DESY3RM-eRASS-matched sample is given in Figure~\ref{fig:r500-lambda}.  The fit of the relation is performed using {\sc linmix}, with the black line representing the best-fit to the data (with the blue shaded region the 68\% confidence interval).  Based upon this relation, an r$_{500}$ can be estimated from $\lambda_{\rm RM}$ via:
\begin{align}
    r_{500} = E(z)^{-1/3}~949\pm19~\left(\frac{\lambda_{\rm RM}}{63}\right)^{0.314\pm0.06}.
\label{equ:r500_lambda}
\end{align}
Based upon the estimates of r$_{500}$ from $\lambda_{\rm RM}$ and Equation~\ref{equ:r500_lambda}, the forced photometry luminosity estimates for the DESY3RM-eRASS-unmatched clusters within 300~kpc are extrapolated out to r$_{500}$ by integrating under a beta profile assuming a slope $\beta=0.667$, taking the analytical form:
\begin{align}
    L_{\rm frac} = \frac{1 - (1+(r_{500}/r_{c})^{2})^{-0.5}}{1 - (1+(300~{\rm kpc}/r_{c})^{2})^{-0.5}}.
\label{equ:beta}
\end{align}
To estimate the core radius for each cluster, we make the assumption r$_{c}$=0.15$\times$r$_{500}$, following \cite{2016A&A...592A...3G}.  To estimate the luminosities within $r_{500}$, the 300~kpc luminosities are multiplied by $L_{\rm frac}$.

\section{Scaling relations derived from the DESY3RM-eRASS samples}
\label{sec:relations}

Fits to the scaling relations were performed in log space using the {\sc linmix} python.  Formally, scaling relations are fitted with a power-law of the form
\begin{align}
{\rm log}(Y) = {\rm log}(A) + B~{\rm log}(X) \pm \sigma^{2}
\end{align}
where $A$, $B$ and $\sigma$ represent the normalisation, slope and lognormal intrinsic scatter respectively.  For simplicity, throughout, the scaling relations are denoted by the cluster properties in question (for example, see Eqution~\ref{equ:l-lambda}).  Note that due to a limitation of {\sc linmix}, inputted errors are treated as symmetric by averaging the upper and lower errors constraints.

\begin{figure*}
\begin{center}
\begin{tabular}{cc}
\includegraphics[width=8.3cm]{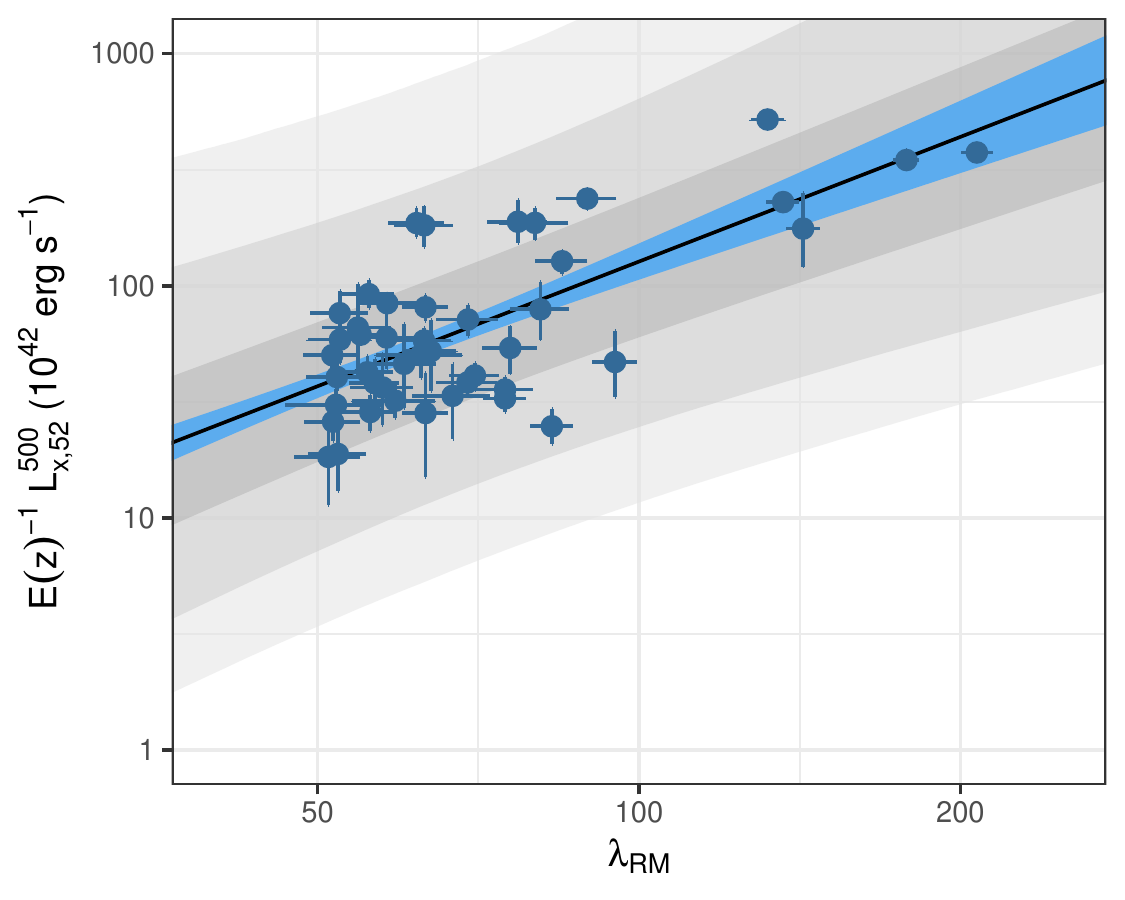} &
\includegraphics[width=8.3cm]{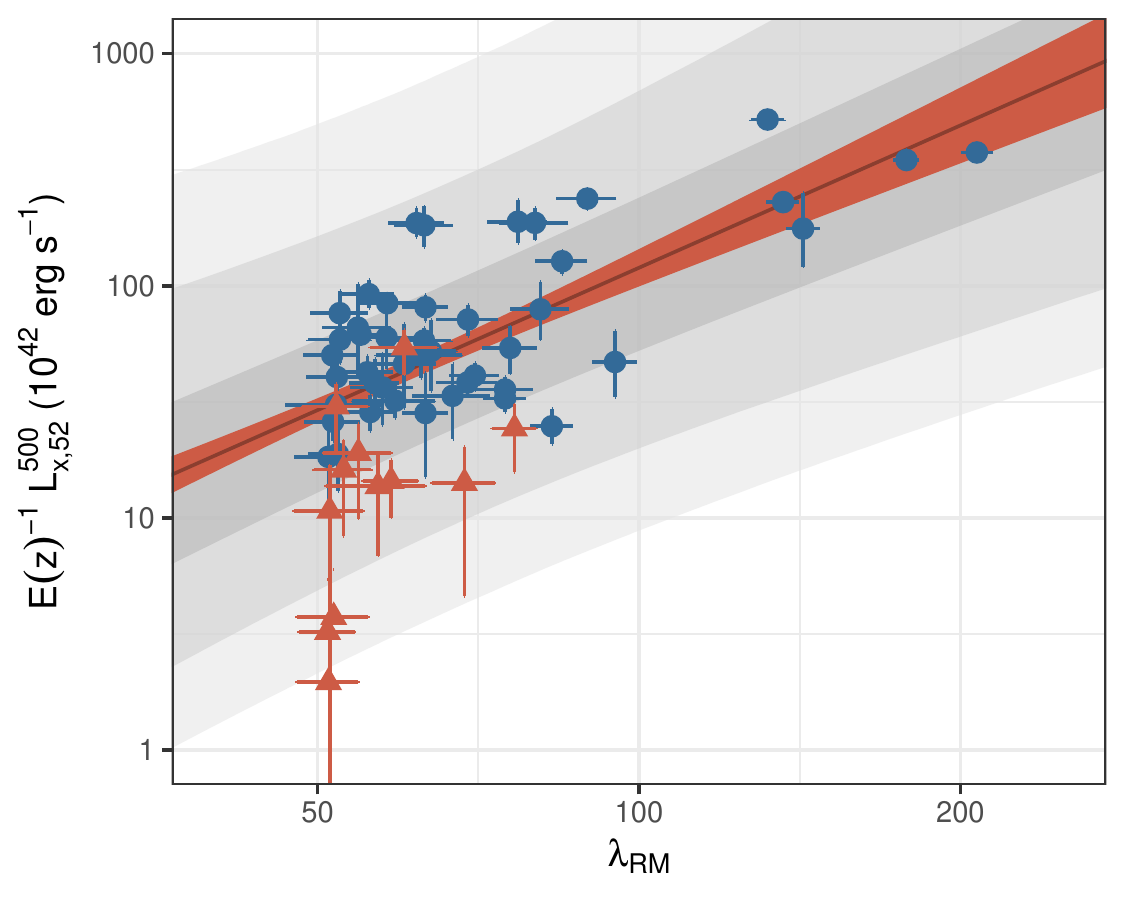} \\
(a) & (b) \\
\end{tabular}
\end{center}
\caption[]{(a) Luminosity-richness relation of the DESY3RM-eRASS-matched cluster sample.  Clusters are given by the blue points.  The black solid line represents a fit to the data with the lightblue shaded region highlighting the 68\% uncertainty.  The grey bands represent the 1, 2, and 3$\sigma$ intrinsic scatter. (b) Luminosity-richness relation of the combination of the DESY3RM-eRASS-matched (blue points) and DESY3RM-eRASS-unmatched (coral triangles) cluster samples.  The fit to the data, given by the coral solid line and coral shaded region, includes both the blue and coral points.  The description of the fit and shaded regions are the same as Figure (a).} 
\label{fig:lx-lambda}
\end{figure*}

The $L_{X,0520}^{r500}-\lambda_{\rm RM}$ relation for the DESY3RM-eRASS-matched sample is shown in Figure~\ref{fig:lx-lambda} (a).  We fit for the $L_{X,0520}^{r500}-\lambda_{\rm RM}$ relation, which we express as:
\begin{align}
{\rm log}\left(\frac{L_{X,0520}^{r500}}{E(z)^{\gamma_{L\lambda}}}\right) &=  {\rm log}(A_{L\lambda}) + B_{L\lambda}~{\rm log}\left(\frac{\lambda_{\rm RM}}{\lambda_{0}}\right) \pm \sigma_{L|\lambda},
\label{equ:l-lambda}
\end{align}
where $A_{L\lambda}$ and $B_{L\lambda}$ represents the normalisation and the slope of the relation respectively, and the intrinsic scatter given by $\sigma_{L\lambda}$.  In this relation we assume $\lambda_{0}=60$ (roughly the median value of the sample), and assume a self-similar evolution where $\gamma_{L\lambda}=1$.  The fit to the data is given by the black line and the blue shaded region represents the 68\% confidence interval.  The best-fit parameters of the relation are given in Table~\ref{tab:bestfit}.  Next, we combine the DESY3-eRASS-matched sample with the DESY3-eRASS-unmatched sample, to refit for the relation.  The combined sample we define as the DESY3-eRASS-combined sample.  This relation therefore includes all clusters within our survey area detected by redMaPPer (above $\lambda_{\rm RM}>50$ and within $0.2<z<0.65$), with complete X-ray coverage.  The relation is shown in Figure~\ref{fig:lx-lambda}(b), where the DESY3-eRASS-matched and DESY3-eRASS-unmatched clusters are given by the blue and black points respectively.  Note that the relation is fit to all data combined.  The fit to the data is given by Equation~\ref{equ:l-lambda}, and the best-fit parameters given in Table~\ref{tab:bestfit}.

One immediate conclusion of the scaling relation analysis is that there is no significant difference between the relation when the DESY3-eRASS-unmatched clusters are included/excluded.  While we find that the slope of the relation steepens when including these clusters, the steepening is not significant.  However, a steepening of the relation when including the DESY3-eRASS-unmatched clusters would be expected.  If including just X-ray detected clusters in the relation, then this will suffer from the usual X-ray selection effects (e.g., Malmquist bias), whereby we would select clusters that are more luminous at a given mass leading to a flattening of the relation.  The same argument is made for the difference in the normalisation.  Therefore, while we find the two relations are consistent, the DESY3-eRASS-Combined relation is representative of the full cluster population.  We find very little difference in the scatter of the relation, but we note there are a small population of the DESY3-eRASS-unmatched clusters that appear qualitatively offset from the full population, hinting at a lack of X-ray gas or a different evolutionary formation path.  However, the somewhat larger errors on the estimated luminosities of the DESY3-eRASS-unmatched clusters makes interpretation of the results difficult, and will require additional X-ray data, or the deeper releases of the eRASS (see Section~\ref{sec:erass4}).  

\begin{table*}           
\begin{center}
\caption[]{{\small Best-fit parameters of the luminosity-richness ($L_{X,0520}^{r500}-\lambda_{\rm RM}$) relations studied in this work.  Parameters are given for the DESY3RM-eRASS-matched sample and the combination of the DESY3RM-eRASS-matched and DESY3RM-eRASS-unmatched samples (defined as the DESY3RM-eRASS-combined sample), see Section~\ref{sec:relations}}.  The column of the sample DESY3-eRASS-prediction contains the best-fit parameters of randomised luminosities, drawn from the current luminosities and within the error constraints, with a precision expected from eRASS:5 (see Section~\ref{sec:erass4}).  Finally, the column of the sample DESY3-eRASS-$\lambda_{\rm RM}>20$ contains the best-fit parameters for all DESY3RM clusters in our survey volume with a $\lambda_{\rm RM}>20$ with a corresponding match in the eRASS1 cluster catalogue (see Section~\ref{sec:low_richness}). \label{tab:bestfit}}
\vspace{1mm}
\begin{tabular}{cccccc}
\hline
\hline
 Sample & Survey &  Normalisation & Slope & Scatter & Relevant \\
 & & $A_{L\lambda}$ & $B_{L\lambda}$ & $\sigma_{L\lambda}$ & section \\
\hline
DESY3RM-eRASS-matched & eRASS1 & 51.11$\pm$5.40 & 1.78$\pm$0.28 & 0.26$\pm$0.03 & Section~\ref{sec:relations} \\
DESY3RM-eRASS-combined & eRASS1 & 42.24$\pm$4.40 & 2.04$\pm$0.30 & 0.28$\pm$0.03 & Section~\ref{sec:relations} \\
DESY3RM-eRASS-predicted & eRASS:5 (prediction) & 37.43$\pm$4.52 & 2.13$\pm$0.37 & 0.35$\pm$0.04 & Section~\ref{sec:erass4} \\
DESY3-eRASS-$\lambda_{\rm RM}>20$ & eRASS1 & 59.36$\pm$4.01 & 1.18$\pm$0.12 & 0.26$\pm$0.02 & Section~\ref{sec:low_richness} \\
\hline
\end{tabular}
\end{center}
\end{table*}

\section{Discussion}
\label{sec:discussion}

\subsection{The use of external data}
\label{sec:externaldata}

\subsubsection{Additional X-ray data}
\label{sec:additional_X-ray}

Here we utilise data from other X-ray and Sunyeav-Zel'Dovich (SZ) effect data, with a particular emphasis on whether the DESY3-eRASS-unmatched clusters are detected from other multi-wavelength data.  The first external data we make use of is the {\em XMM} Cluster Survey \citep[XCS,][]{2001ApJ...547..594R}.  The XCS takes all publicly available {\em XMM}, processes, and runs X-ray source detection in order to find extended sources that are galaxy cluster candidates.  An outline of the process can be found in \cite{2011MNRAS.418...14L}, with the XCS being successfully used to provide X-ray information and analysis to SDSS redMaPPer clusters \citep{gilessdss} and DES redMaPPer clusters \citep[e.g.,][]{2019MNRAS.487.2578Z,2019MNRAS.490.3341F,kelly24}.  Therefore, we make use of XCS here to match to our cluster samples to determine which clusters may be detected by {\em XMM}.  Note however that due to the non-contiguous nature of {\em XMM} coverage in this field, we only use the XCS data as a form of validation.  We follow the same process as outlined in Section~\ref{sec:desy3-erass} to match the DESY3RM clusters with XCS detected extended sources, a 2~Mpc matching radius and a visual inspection process.  Of the DESY3RM-eRASS-matched clusters, 6 fall on an {\em XMM} observation and all have high-SNR detections.  Of the DESY3RM-eRASS-unmatched clusters, only one is covered by an {\em XMM} observation, redMaPPer ID = 2096, and has a clear detection (an {\em XMM} image of the cluster is given in Figure~\ref{fig:2096XMM}).  These 7 clusters detected by {\em XMM} are given in Table~\ref{tab:xmmprops}.  Many of these XCS clusters are contained in the catalogue of \citet[][hereafter K24]{kelly24}, however, that catalogue was based upon an XCS source catalogue from 2020.  The matching above has been performed with an XCS source catalogue produced in Feb. 2024.  An additional source of X-ray information comes from {\em Chandra}.  The work of K24 also included information on DESY3 clusters that fell on {\em Chandra} observations.  Therefore, we searched K24 catalogue for clusters in common with the sample in this work.  We first identified redMaPPer ID's that were included in both our sample and that of K24, resulting in 8 clusters in common.  All of these clusters are contained in the DESY3-eRASS-matched sample, and none of the DESY3-eRASS-unmatched had matches.  Next, we searched the {\em Chandra} archive for more recent observations of our sample, but none were found.  The {\em Chandra} data for these 8 clusters are given in Table~\ref{tab:chandradata}.  Note that 5 of these clusters are in common with the XCS data above.

\begin{figure}
\begin{center}
\includegraphics[width=8.3cm]{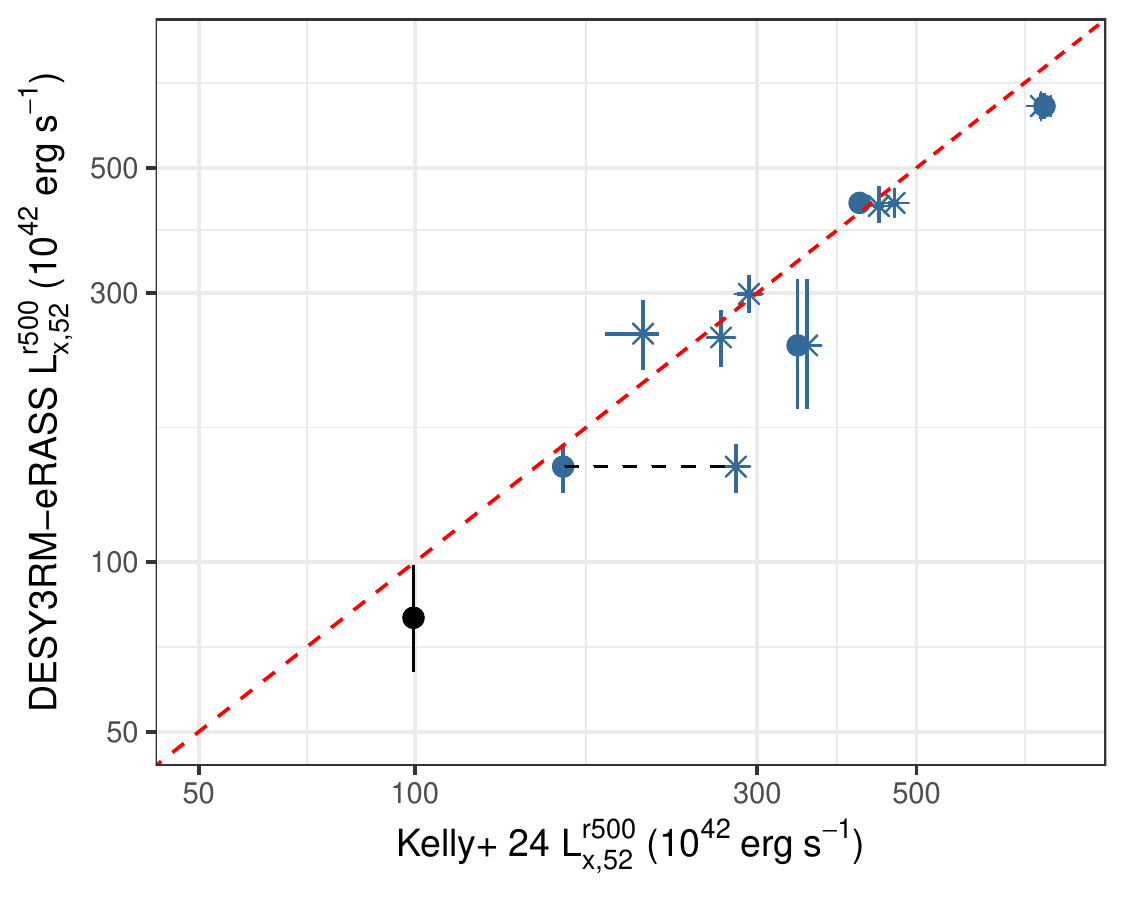}
\end{center}
\caption[]{Comparison of the soft-band $r_{500}$ luminosities used in this work compared to those estimated from available {\em XMM} or {\em Chandra} data as given in \cite{kelly24}.  Circles indicate clusters where the comparison luminosity was estimated from {\em XMM} and crosses estimated from {\em Chandra}.  Blue points belong to the DESY3-eRASS-matched sample, and the single black point represents redMaPPer ID = 2096, belonging to the DESY3-eRASS-unmatched sample.  This cluster had the luminosity estimated from the forced photometry method (see Section~\ref{sec:undet_lx}).  A 1:1 relation is highlighted by the red dashed line.} 
\label{fig:lx_comp}
\end{figure}

As a source of validation, we compare the luminosities from eRASS for our clusters to those estimated from the clusters with available {\em XMM} and {\em Chandra} above.  For this, we use the analysis outlined in K24 as the basis for our luminosity estimates.  Note that r$_{500}$ luminosities were not provided in K24, only r$_{500}$ temperatures.  Therefore, the luminosity values were provided from K24 via private communication, and can be found in Tables~\ref{tab:xmmprops} and~\ref{tab:chandradata} for {\em XMM} and {\em Chandra} respectively.  For the {\em Chandra} clusters, the analysis follows that as given in K24 via the MATCHa pipeline \citep{2019ApJS..244...22H}.  For the {\em XMM} clusters, the analysis follows K24 via the XCS processing pipeline as outlined fully in \cite{gilessdss}. The comparison of the luminosities is given in Figure~\ref{fig:lx_comp}.  Given the somewhat small number of data available, we forego a quantitative assessment of the luminosity comparison, noting that the luminosities are broadly consistent.  We do note however that \cite{erass1clusters} found that {\em eROSITA} luminosities were on average 15\% lower than those measured by {\em Chandra}.  One notable exception to the luminosity comparison is the cluster redMaPPer ID = 629, where the {\em Chandra} luminosity is $\sim$2$\times$ larger than the eRASS estimate. However, the {\em XMM} estimate is consistent with the eRASS measurement (these are connected by the black dashed line in Figure~\ref{fig:lx_comp}). Investigating further, we find that the $r_{2500}$ luminosity measured by {\em Chandra} ($L_{X,52}^{r2500}=1.06\times10^{44}$~erg~s$^{-1}$) and {\em XMM} ($L_{X,52}^{r2500}=1.17\times10^{44}$~erg~s$^{-1}$) are consistent, indicating a potential issue with the {\em Chandra} measured $r_{500}$ value.  Additional analysis of $r_{500}$ luminosity discrepancy is outside the scope of this paper.  Furthermore, while only 1 data point, we note that redMaPPer ID 2096 (belonging to the DESY3-eRASS-unmatched sample, highlighted by the black point in Fig.~\ref{fig:lx_comp}), is very consistent with the {\em XMM} estimate.  

\subsubsection{The use of Sunyaev-Zel'Dovich effect data}
\label{sec:additional_sz}

Next, we make use of two large area Sunyaev-Zel'Dovich (SZ) effect surveys that overlap with our sky region, namely the Atacama Cosmology Telescope’s (ACT) most recent cluster catalogue \citep[DR6;][]{2026OJAp....955863A}, containing 10,040 optically confirmed clusters from a sky survey area of 16,293 deg$^{2}$, and the South Pole Telescope's \citep[SPT;][]{2011PASP..123..568C} 2500~deg$^{2}$ SPT-SZ survey cluster catalogue containing 677 galaxy clusters from \cite{2015ApJS..216...27B}, with updated redshifts as given in \cite{2019ApJ...878...55B}.  Within the area considered in the work, the ACT and SPT samples contain 73 and 62 clusters respectively. Note we do not initially add a redshift cutoff, to allow for some redshift scatter between redMaPPer and the SZ catalogue redshifts.  We then match between the DESY3RM sample and these ACT and SPT samples, using a 5 arc-minute radius match and ensuring a redshift match of 0.05/($1+z_{\rm RM}$) between redMaPPer and the SZ catalogues.  The statistics of the matches are given below. 

{\em ACT matching}: Of the 57 DESY3RM clusters, 34 have a corresponding match in the ACT catalogue.  Of the 45 DESY3RM-eRASS-matched clusters, 29 are matched to an ACT source, with only 5 of the 12 DESY3RM-eRASS-unmatched clusters detected in ACT (redMaPPer IDs = 655, 2096, 2326, 2609, 3834).  The cluster redMaPPer ID = 3834 (Figure~\ref{fig:undet_images}, j) was found to have a nearby extended source associated with another redMaPPer cluster in the field.  While this may complicate the match, the SZ coordinates are only 15$^{\prime\prime}$ away from the redMaPPer coordinates.  

{\em SPT matching}: Of the 57 DESY3RM clusters, 13 have a corresponding match in the SPT catalogue.  Of the 45 DESY3RM-eRASS-matched clusters, 12 area matched, with only 1 DESY3RM-eRASS-unmatched cluster having a match to the SPT catalogue.  This cluster (redMaPPer ID = 2096), is matched to the cluster SPT-CL J0422-5140 with an SPT SNR = 5.86.  As previously found in Section~\ref{sec:additional_X-ray}, this cluster has clear visually identified X-ray emission from an {\em XMM} observation (Fig.~\ref{fig:2096XMM}), likely from the cluster.

Given the overlap of redMaPPer ID = 2096 in both ACT and SPT data, we have found 5 individual DESY3-eRASS-unmatched clusters have been detected in the SZ regime.  We also note that of these 5 clusters, 3 of them were detected as point sources in the eRASS catalogue (see Section~\ref{sec:undet_point}). By utilising the additional data, we have provided validation of the properties measured in this work through the use of external X-ray data, and evidence of the presence virialsed gas associated with some of the DESY3-eRASS-unmatched clusters via X-ray and SZ data.  While these detections represent only 40\% of the DESY3-eRASS-unmatched sample, it provides confidence of the presence of gas that could be detected with deeper X-ray data, either through future releases of the eRASS or dedicated follow-up using e.g., {\em XMM}.

\subsection{Understanding the physical nature of the DESY3RM-eRASS-unmatched clusters}
\label{sec:mag_gap}

\begin{figure*}
\begin{center}
\begin{tabular}{cc}
\includegraphics[width=8.5cm]{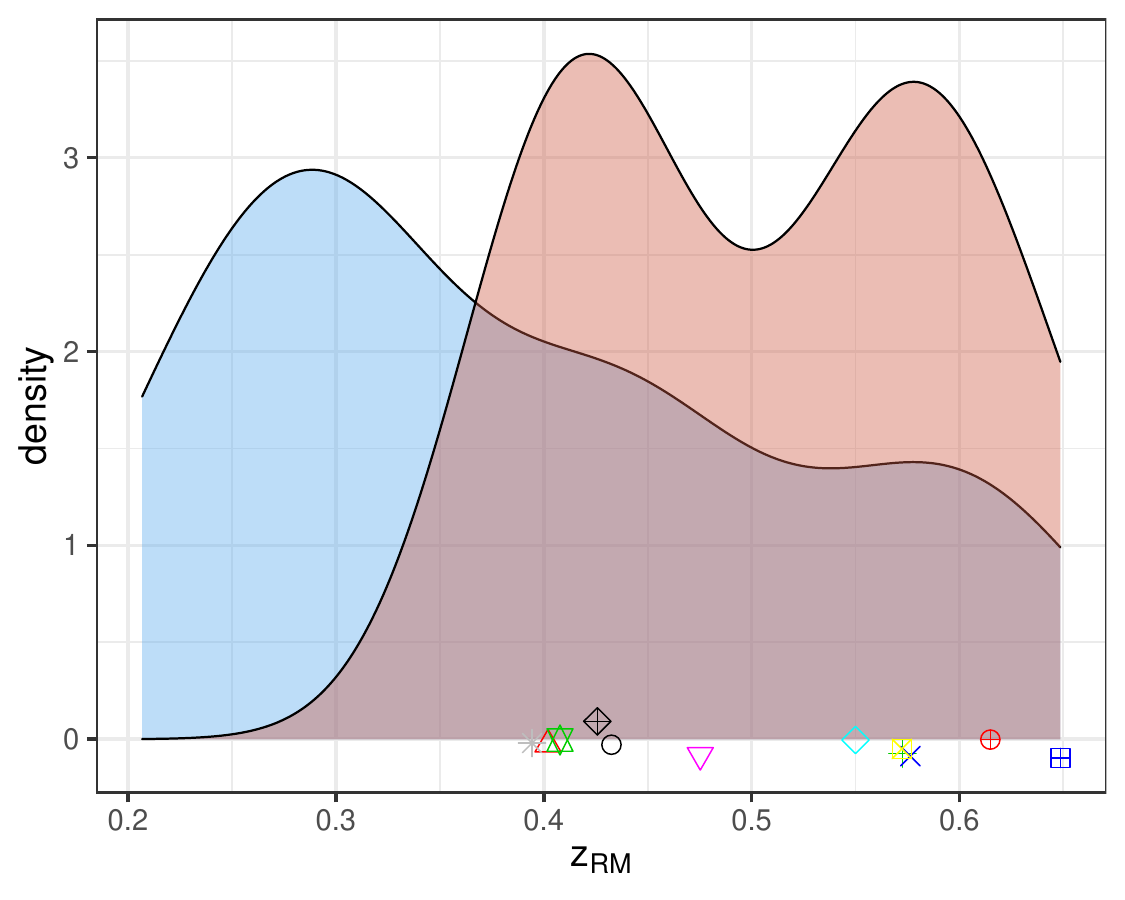} &
\includegraphics[width=8.5cm]{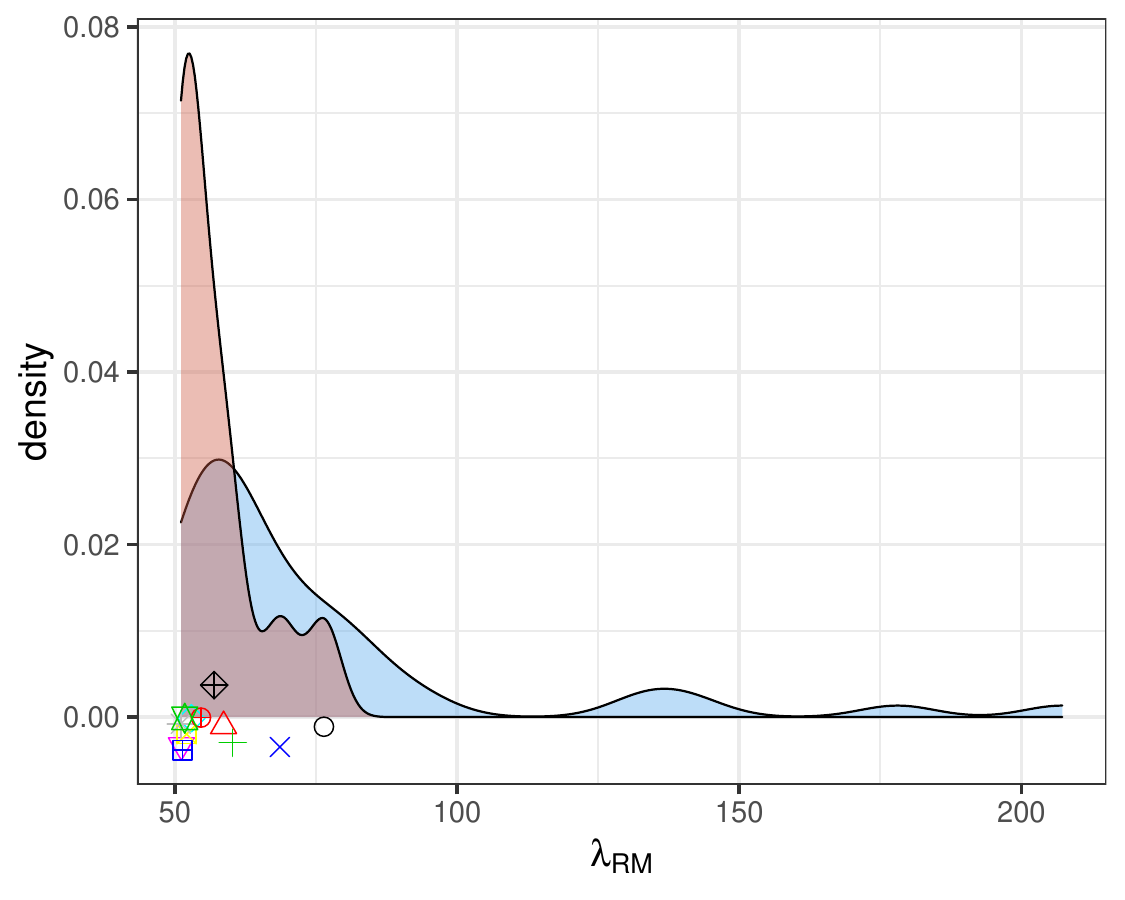} \\
(a) & (b) \\
\includegraphics[width=8.5cm]{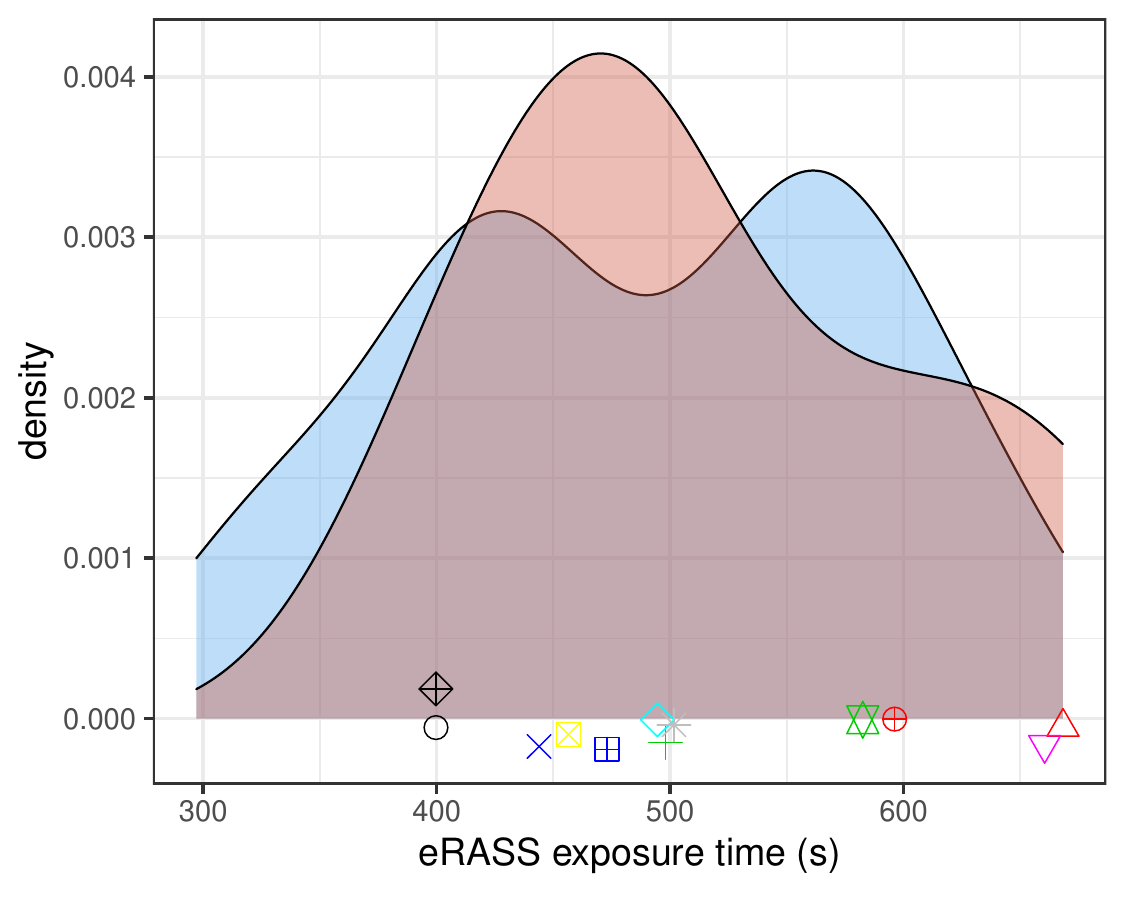} &
\includegraphics[width=8.5cm]{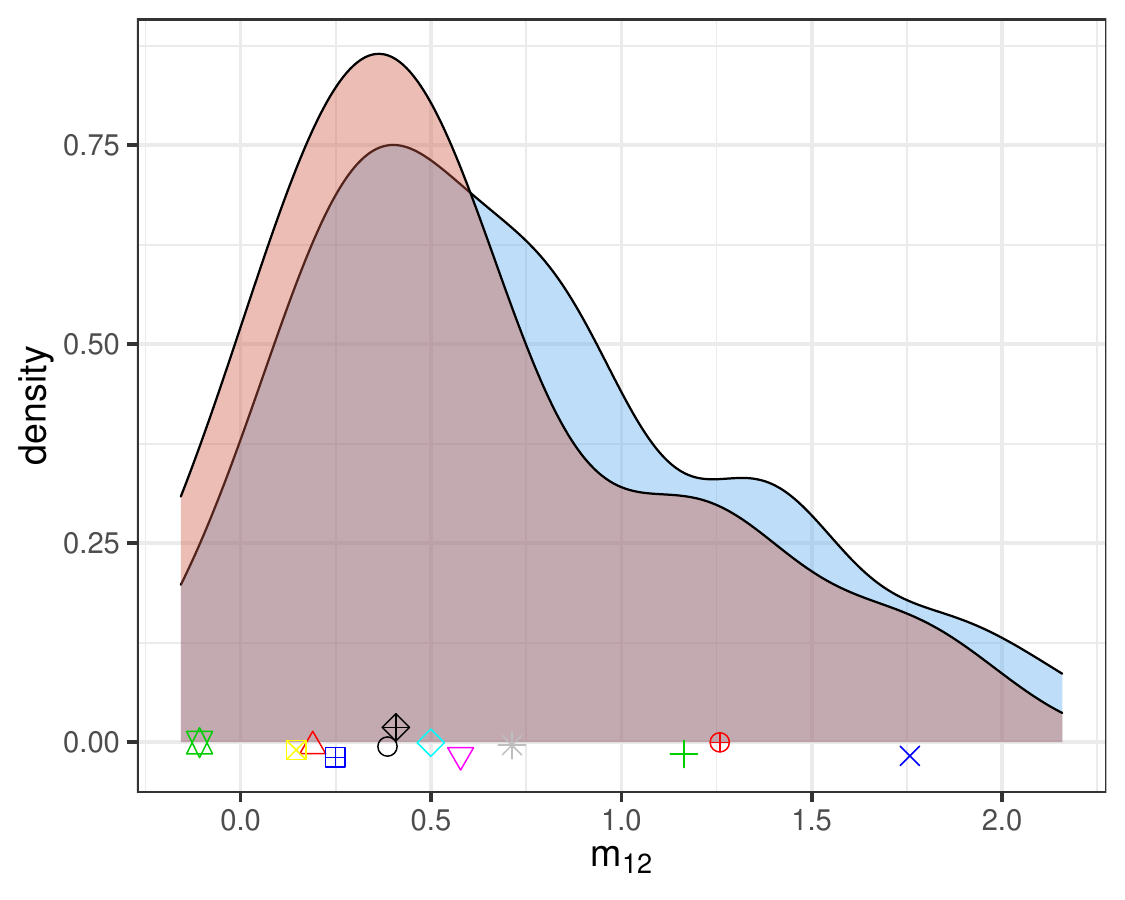} \\
(c) & (d) \\
\end{tabular}
\end{center}
\caption[]{Density distributions of (a) redMaPPer redshift; (b) richness; (c) eRASS exposure time and (d) the magnitude gap between the first and second brightest galaxies, $m_{12}$.  In each case, the distributions are given for the DESY3RM-eRASS-matched (blue) and DESY3RM-eRASS-unmatched (red).  Each data point represents a cluster from the DESY3RM-eRASS-unmatched sample (the offsets of the points from zero on the y-axis are to limit confusion of overlapping points).} 
\label{fig:density}
\end{figure*}

\begin{figure*}
\begin{center}
\begin{tabular}{cc}
\includegraphics[width=6.5cm]{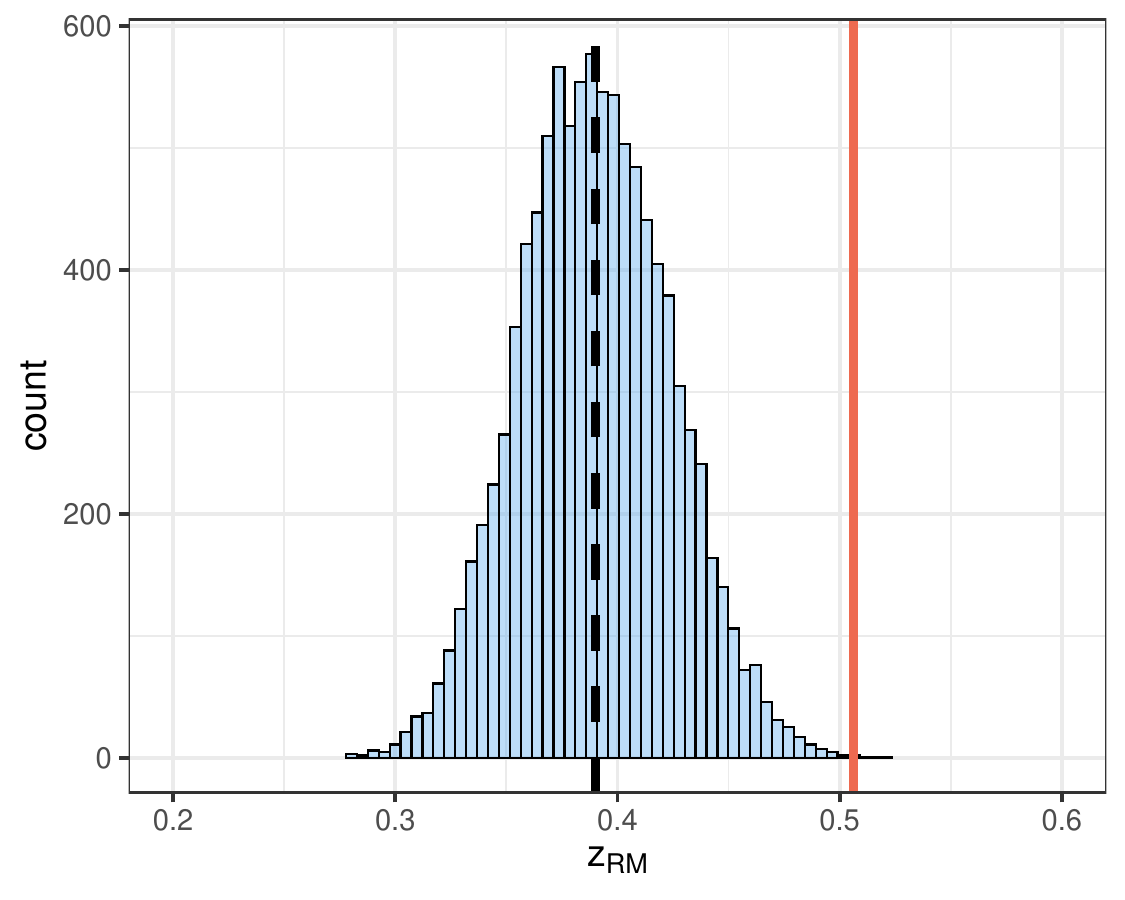} &
\includegraphics[width=6.5cm]{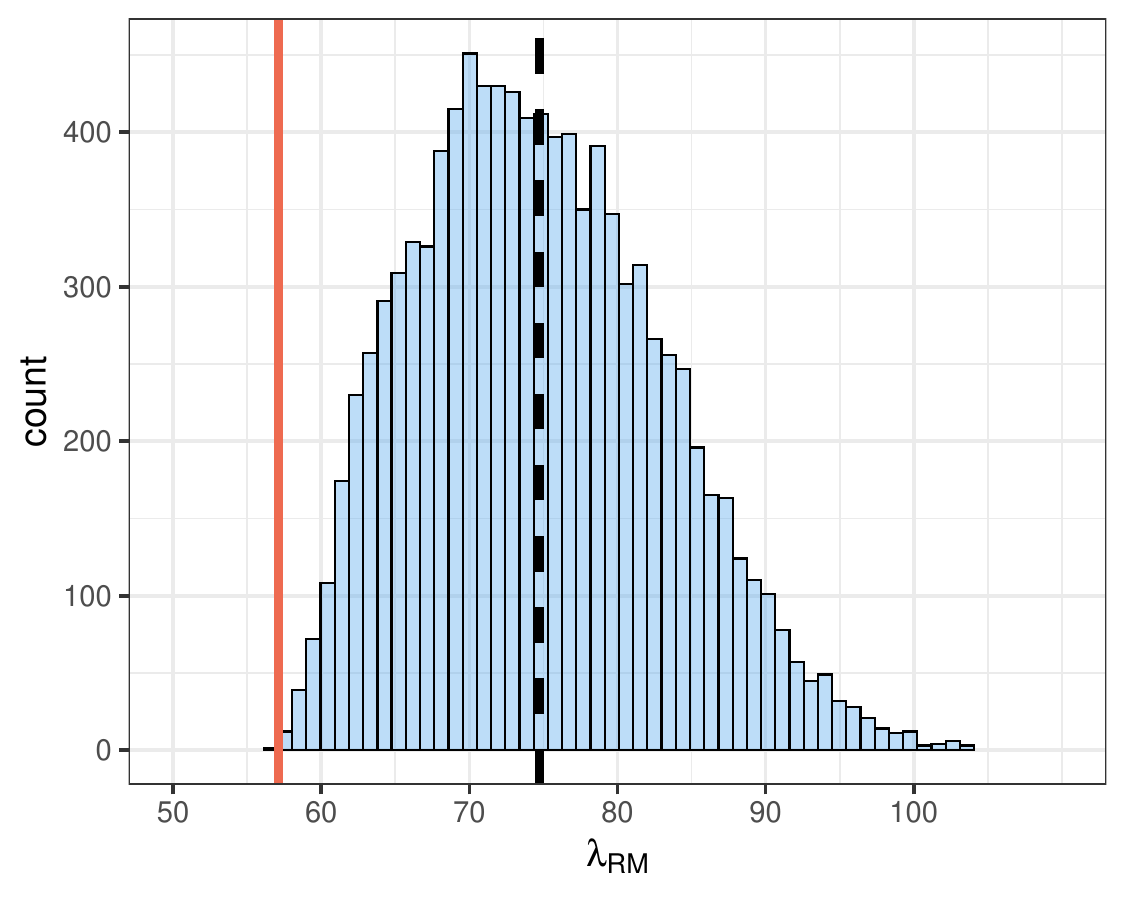} \\
(a) & (b) \\
\includegraphics[width=6.5cm]{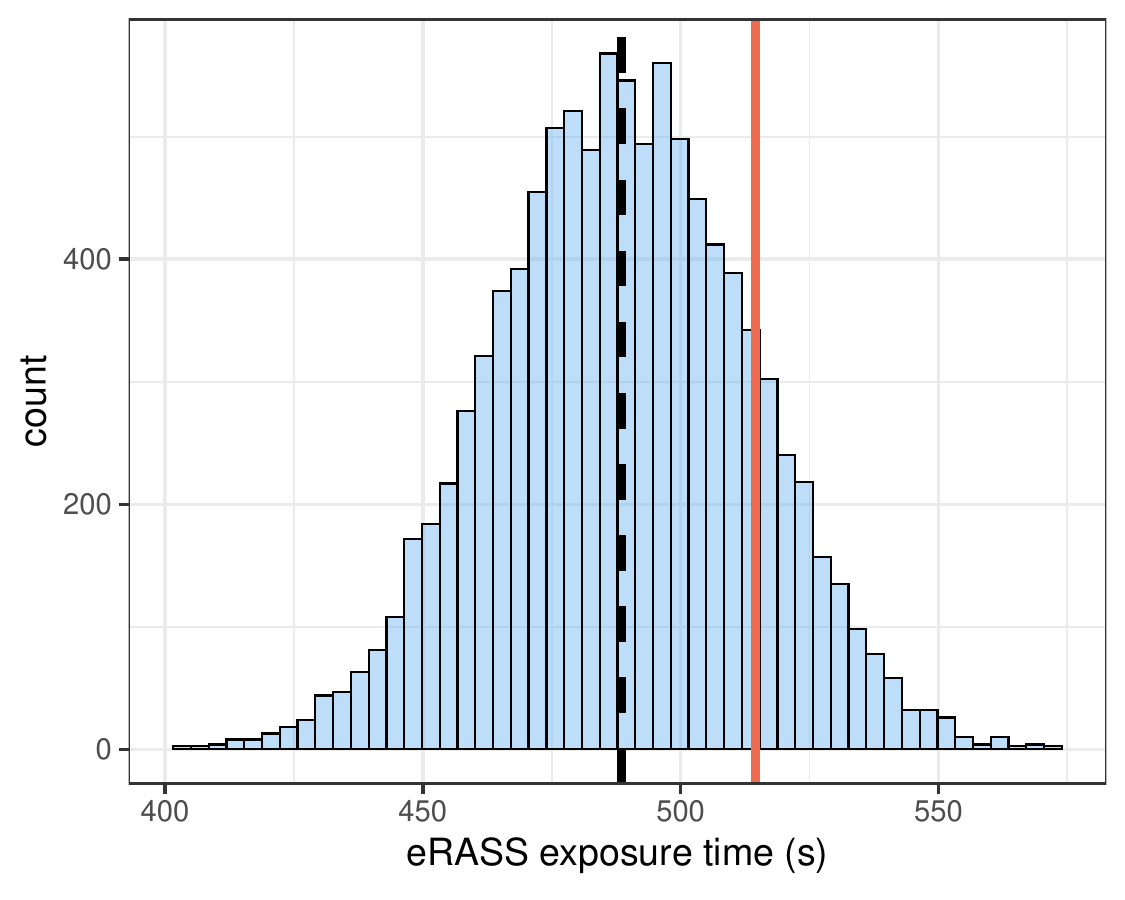} &
\includegraphics[width=6.5cm]{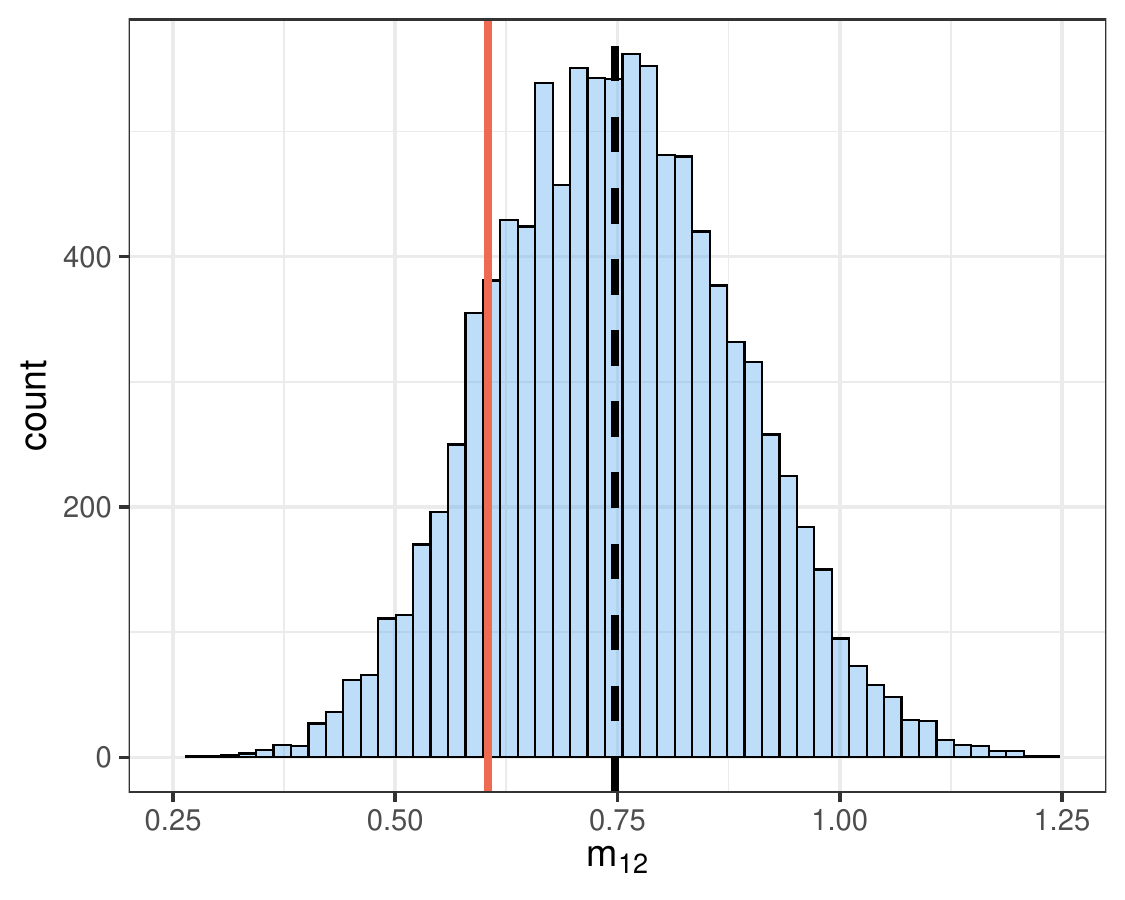} \\
(c) & (d) \\
\end{tabular}
\end{center}
\caption[]{Distributions of the means of 10000 randomly drawn DESY3RM-eRASS-matched clusters.  Each mean is estimated from a random sample of 12 DESY3RM-eRASS-matched clusters (see Sect.~\ref{sec:mag_gap}).  The black dashed line represents the mean of these 10000 drawn samples.  The red solid line represents the mean of the 12 DESY3RM-eRASS-unmatched clusters for the respective property.  The distributions are drawn for the (a) redMaPPer redshift; (b) redMaPPer richness; (c) eRASS exposure time and (d) the maginitude gap.} 
\label{fig:random_dist}
\end{figure*}

In order to understand the lack of apparent X-ray emission of the DESY3RM-eRASS-unmatched clusters, we examine a range of properties of these clusters relating to both their intrinsic properties, and observational effects.  For each of these properties, we construct density plots of both the DESY3RM-eRASS-unmatched and DESY3RM-eRASS-matched cluster samples for comparison.  These density plots are shown in Figure~\ref{fig:density}(a)-(d), with the DESY3RM-eRASS-unmatched and DESY3RM-eRASS-matched clusters shown in red and blue respectively.  From visual inspection of the redshift and richness distributions, it appears that the DESY3RM-eRASS-unmatched clusters fall at higher redshifts and lower richnesses than the DESY3RM-eRASS-matched clusters.  There appears to be no difference in the exposures times of the two samples.

The final property we investigate is the magnitude gap.  The recent work by \citet[][hereafter P24]{2024MNRAS.527..895P}, found that the magnitude gap of the detected clusters in their cluster sample were $>$3$\sigma$ larger than the undetected clusters.  The magnitude gap gives an indication of the age of the system, and therefore, P24 argue that the detected clusters are older than the undetected clusters.  We estimate the magnitude gap of the DESY3-eRASS clusters in our sample, matching the method in P24, by calculating the difference in $r$-band magnitude of the first and second brightest galaxies in each cluster ($m_{12}$).  The density distribution of $m_{12}$ for our two samples is shown in Figure~\ref{fig:density}(d).  It appears that the DESY3RM-eRASS-matched and DESY3RM-eRASS-unmatched clusters have the same distribution.

To determine if the DESY3RM-eRASS-unmatched clusters have properties that are significantly different to the DESY3RM-eRASS-matched clusters, we follow a process outlined in P24. We randomly draw samples from the DESY3RM-eRASS-matched clusters of the same size as the DESY3RM-eRASS-unmatched sample (12), repeating this 10000 times.  For each randomly drawn sample we calculate the mean of the sample and plot these distributions in Figure~\ref{fig:random_dist}(a)-(d).  The black dashed line in each histogram represents the mean of the distributions, and the coral solid line represents the mean of the DESY3RM-eRASS-unmatched for each property.  We find that the mean of the DESY3RM-eRASS-unmatched clusters is $>$3$\sigma$ larger and smaller for the redshift and richness respectively.  We find that the exposure times and $m_{12}$ are consistent between the matched and unmatched clusters.  For $m_{12}$, we repeat the process above, but limiting the richnesses of the DESY3RM-eRASS-matched clusters to $<$70.  We find no change of the distribution.  Therefore, our results are in tension with that of P24.  We note however, that it is possible the difference in our results is due to the different population of clusters probed by P24.  For example, all of our matched (and 9 out of 12 unmatched) clusters have an X-ray luminosity of $>$10$^{43}$~erg~s$^{-1}$, whereas a significant proportion of the clusters in P24 have $<$10$^{43}$~erg~s$^{-1}$.  This indicates that we are probing a different halo mass range, where the physical nature of the unmatched clusters play a less crucial role in their lack of X-ray emission.  We therefore conclude this lack of emission is due to a challenging observational parameter space i.e., high redshift and lower richness, rather than not having virialised X-ray emitting gas if they were young systems still in formation.  Given this, it provides hope that with future, deeper releases of the eRASS\footnote{Future releases of eRASS can be found at \href{https://erosita.mpe.mpg.de/erass/}{https://erosita.mpe.mpg.de/erass/}}, these clusters may have detectable X-ray emission with measured X-ray properties (see Section~\ref{sec:erass4}).

\subsection{Prospects for eRASS:5}
\label{sec:erass4}

Here we investigate further releases of the eRASS, and how the increase in data quality will; i) aid in the detectability of the DESY3-eRASS-unmatched sample; and ii) enhance our understanding of the cluster scaling relations.  

\begin{figure}
\begin{center}
\includegraphics[width=8.3cm]{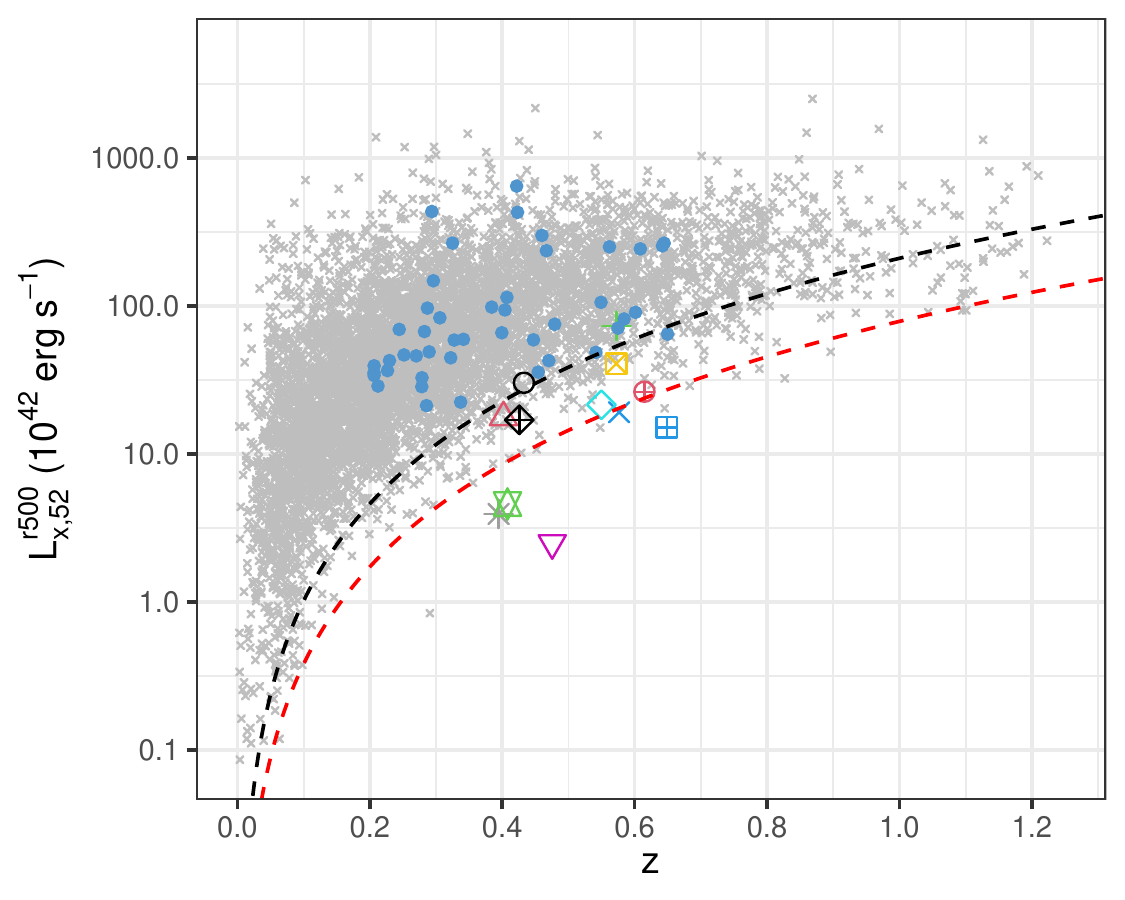} \\
\end{center}
\caption[]{The soft-band (0.5--2.0 keV) luminosity within $r_{500}$ as a function of redshift.  The clusters from the eRASS1 catalogue of \cite{erass1clusters} are given by the grey points, for clusters with an extent likelihood of $\mathcal{L}_{ext}>$6.  The lightblue points are clusters from the DESY3-eRASS-matched clusters.  Each individual DESY3-eRASS-unmatched cluster is given by a discrete shape and colour data point.  The black and red dashed lines represent an flux limit of 4$\times$10$^{-14}$~erg~s$^{-1}$~cm$^{-2}$ and 1.5$\times$10$^{-14}$~erg~s$^{-1}$~cm$^{-2}$ respectively.} 
\label{fig:flux-lx-dist}
\end{figure}

It has not been confirmed what will be included in DR2 of the eRASS (e.g., the number of passes), due to the current pause of eRASS operations.  Therefore, we make the assumption that any future releases will only include up to 4.5 passes of the sky i.e., offering 4.5 times the exposure of the current eRASS used in this work.  To investigate potential detections, we replicate the plot from \cite[][see their Fig~14]{erass1clusters}, which provides some qualitative information on the detection as a function of survey depth.  We examine the luminosity (soft-band, within $r_{500}$) as a function of redshift, displayed in Figure~\ref{fig:flux-lx-dist}.  For reference, the black dashed line represents a flux of 4$\times$10$^{-14}$~erg~s$^{-1}$~cm$^{-2}$, a rough estimate of the eRASS1 survey flux limit \cite[see][]{erass1clusters}.  We stress however that the eRASS is not a flux limited survey and this line is only for reference.  We observe that only 2 of the DESY3-eRASS-unmatched clusters fall above the black dashed flux limit.  Next, we compare to a flux limit of 1.5$\times$10$^{-14}$~erg~s$^{-1}$~cm$^{-2}$ (red dashed line in Figure~\ref{fig:flux-lx-dist}, a).  This flux was chosen as it represents the approximate flux limit of the eFEDS survey region.  The eFEDS region has a somewhat similar exposure to 5$\times$ the survey region defined in this work, and therefore represents a point of comparison.

It is found that now seven of the DESY3-eRASS-unmatched clusters fall above the flux limit in the luminosity-redshift plane.  However, given this flux limit represents an eFEDS completeness of 40\% \citep[see][]{2022A&A...661A...2L}, it is not unreasonable to assume that not all of the DESY3-eRASS-unmatched sample will be detected at the depth of eRASS:5.  Therefore, future work cross-matching DES clusters with the eRASS will want to consider carefully the selection criteria (e.g., increasing the richness limit, or decreasing the redshift ranges) if the requirement is the detection of X-ray emission.  This also has implications for ongoing (e.g., Euclid) and upcoming (e.g., LSST) optical cluster surveys that may require, or enhance their analysis, the use of X-ray information.

In addition to exploring whether the DESY3-eRASS-unmatched clusters will be detected, with the increase in exposure time, it will naturally lead to a decrease in the uncertainties of the measured luminosities of these clusters (and indeed all clusters in this work).  The increase in precision of the cluster properties for these clusters will lead to an increase in the statistical power of these clusters in the scaling relation analysis.  Therefore, we investigate this potential effect on the scaling relation analysis considered in this work.  First, for the DESY3-eRASS-unmatched clusters, in the forced photometry method employed (see Section~\ref{sec:undet_lx}) we artificially increase the exposure time by a factor of 4.5 in the analysis, leading to an increase in the number of source/background counts by a factor 4.5.  Using this artificially increased exposure, the errors on the resulting count rates decrease by $\approx$2.  Based upon this, we take all clusters in the DESY3-eRASS-combined sample (that used to fit for the luminosity-richness relation in Section~\ref{sec:relations}), and draw a new randomised value of the luminosity within the reported uncertainties.  To assign an uncertainty to each randomised luminosity we take the current fractional uncertainty on the luminosity, divide by 2, and apply this reduced fractional uncertainty to the randomised value.  We repeat this process 1000 times, and for each, we refit for the luminosity-richness relation following Section~\ref{sec:relations}.  The final parameters of the scaling relation are taken from the means of the relations of the 1000 samples.  The parameters of these are reported in Table~\ref{tab:bestfit}.  We find that the normalisation has decreased and the slope and scatter has increased, consistent with the increased power of the DESY3-eRASS-unmatched clusters with reduced errors.  We note that this conclusion would only be valid if the measured luminosities from future eRASS data were consistent within our current luminosity estimates.  However, this process highlights how the scaling relation parameters can change with improved data quality, either with future releases of the eRASS, or with dedicated follow-up with another X-ray telescope e.g., {\em XMM}.

\subsection{Probing the $\lambda<50$ population}
\label{sec:low_richness}

So far we have focused on the $\lambda_{\rm RM}>50$ population of redMaPPer clusters.  However, there are an order of magnitude more clusters in this defined area if we consider lower richness clusters.  Within the sky region and redshift range used in this work, there are 597 clusters above a richness of $\lambda_{\rm RM}>20$ (this includes the 57 studied throughout this work).  It is prudent to explore what the eRASS may be able to probe at these lower richnesses (i.e., masses).  We expect a large amount of incompleteness in this test, and the effects of projection on the redMaPPer catalogue at these lower richnesses are increased, hindering the interpretation of the results in the absence of an X-ray detection.  Therefore we limit this exploration to DESY3RM clusters that have a detection in the eRASS, and forgo a forced photometry analysis for all undetected clusters (the result of which would likely be an increased number of upper limit measurements).  To generate the sample of $20 < \lambda_{\rm RM} < 50$ DESY3RM clusters with corresponding eRASS catalogue matches (to supplement the 57 $\lambda_{\rm RM}>50$ studied in this work), we performed a simple matching process to find all eRASS clusters within a 5$^{\prime}$ radius of the 597 DESY3RM clusters.  Next, we cut the matches to ensure the matched clusters have a redshift difference of $z_{\rm diff}$ = 0.05/($1+z_{\rm RM}$).  Finally, if a DESY3RM cluster is matched to multiple eRASS clusters, then we take the closest match.  The result of this matching procedure is outlined in Table~\ref{tab:all_richness_matches}.  The table outlines a series of richness limits, along with the number of DESY3RM clusters above the limit and the corresponding number of DESY3RM clusters with an eRASS catalogue match.  Finally, the fraction of DESY3RM clusters with an eRASS match is listed.  Taking the richness limit of the DESY3RM catalogue (i.e., $\lambda_{\rm RM}>20$), we find that of the 597 clusters, only 179 have a match to the eRASS catalogue, representing a match fraction of 0.30.  This is significantly lower than the fraction of 0.79 of the 57 DESY3RM clusters studied in this work.  As we increase the richness limit of the DESY3RM sample, the fraction of matches steadily increases.

\begin{figure}
\begin{center}
\includegraphics[width=8.3cm]{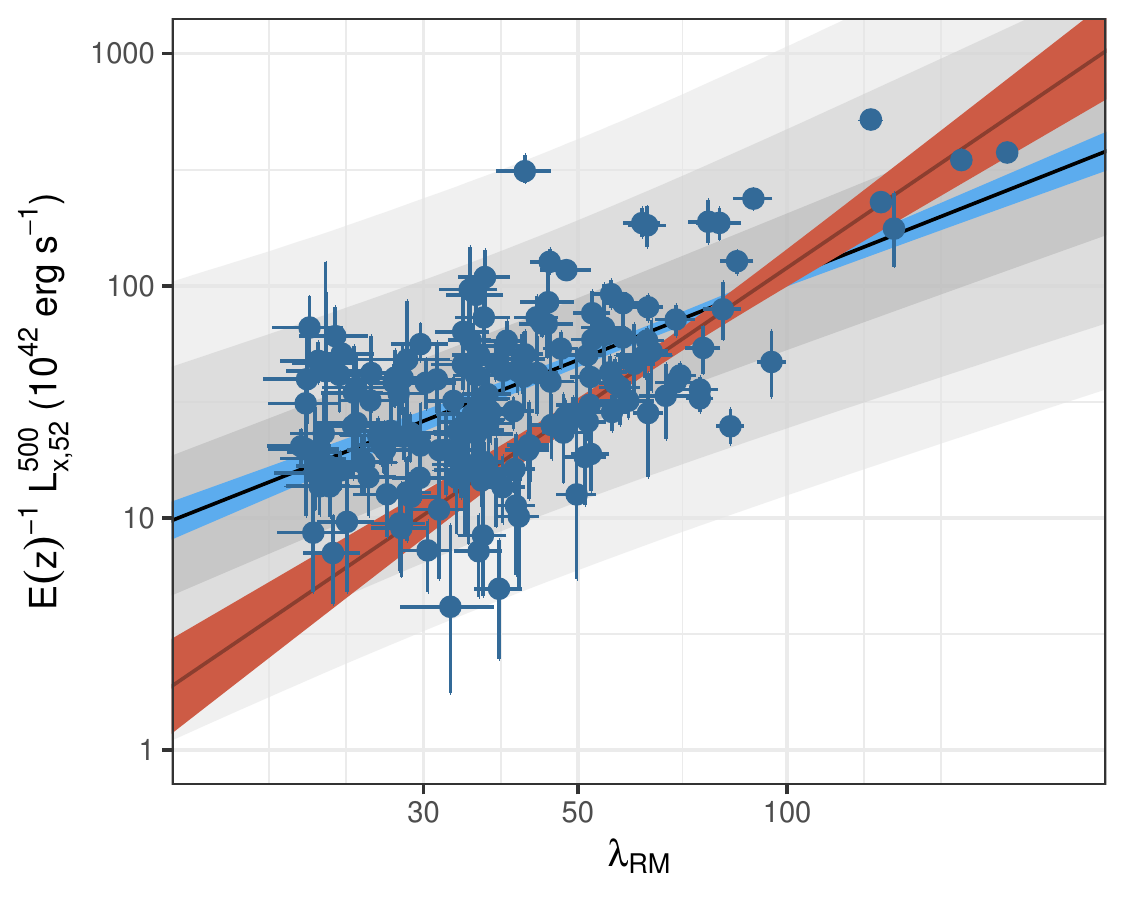}
\end{center}
\caption[]{Luminosity-richness relation of clusters with $\lambda_{\rm RM}>20$ and a corresponding match in the eRASS cluster catalogue (see Section~\ref{sec:low_richness}).  Clusters are given by the blue points.  The black solid line represents a fit to the data (including upper limits) with the lightblue shaded region highlighting the 68\% uncertainty.  The grey bands represent the 1, 2, and 3$\sigma$ intrinsic scatter.  The fit to the DESY3-eRASS-combined sample (which has a limit of $\lambda_{\rm RM}>50$, see Section~\ref{sec:relations}) is given by the coral line and shaded region.} 
\label{fig:low_richness_relation}
\end{figure}

Next, we investigate the form of the luminosity-richness relation using the full richness range of the DESY3 sample with matched eRASS clusters i.e., the 179 clusters defined above, which we label as the DESY3-eRASS-$\lambda_{\rm RM}>20$ sample.  The relation is shown in Figure~\ref{fig:low_richness_relation}, with the data given by the blue circles.  The fit to the data follows the same procedure as in Section~\ref{sec:relations}, and is given by the black solid line (and the 68\% confidence interval by the blue shaded region), with the best-fit parameters given in Table~\ref{tab:bestfit}.  For comparison, the best-fit relation of the DESY3-eRASS-combined sample is given by the coral solid line (with the uncertainty highlighted by the coral shaded region).  We note that while the slope of the DESY3-eRASS-$\lambda_{\rm RM}>20$ relation ($B_{L\lambda}=1.18 \pm 0.12$) appears somewhat shallower than the DESY3-eRASS-combined relation, the difference is only significant at the 2.4$\sigma$ level.  Nonetheless, from this comparison, it is clear that the DESY3-eRASS-$\lambda_{\rm RM}>20$ sample suffers from the effects of X-ray selection biases, based upon the criteria that the DESY3RM cluster was required to have a match to the eRASS cluster catalogue.  From the matched fractions given in Table~\ref{tab:all_richness_matches}, for the lowest richness bin, only 14\% of DESY3RM have a match in the eRASS catalogue, highlighting that only the brightest clusters at a given richness are being included.

\begin{table*}           
\begin{center}
\caption[]{{\small Table outlining the number of matches between the DESY3RM cluster catalogue and the eRASS cluster catalogue of \cite{erass1clusters}.  The clusters are taken within the survey limits outline in Section~\ref{sec:desy3-erass} and the matches are given within two $\lambda_{\rm RM}$ regimes, above increasing $\lambda_{\rm RM}$ limits, and within a $\lambda_{\rm RM}$ range. The bottom row in bold represents the sample studied in this work.}\label{tab:all_richness_matches}}
\vspace{1mm}
\begin{tabular}{cccc|cccc}
\hline
\hline
 $\lambda_{\rm RM}$ limit & No. of DESY3RM &  No. of eRASS & fraction of & $\lambda_{\rm RM}$ range & No. of DESY3RM & No. of eRASS & fraction of \\
 & clusters & matches & matches & & clusters & matches & matches \\
\hline
$>$20 & 597 & 179 & 0.30 & 20$<\lambda_{\rm RM}<$25 & 222 & 30 & 0.14 \\
$>$25 & 375 & 150 & 0.40 & 25$<\lambda_{\rm RM}<$30 & 140 & 25 & 0.18 \\
$>$30 & 235 & 125 & 0.53 & 30$<\lambda_{\rm RM}<$35 & 66 & 19 & 0.29 \\
$>$35 & 169 & 105 & 0.62 & 35$<\lambda_{\rm RM}<$40 & 64 & 30 & 0.47 \\
$>$40 & 105 & 75 & 0.71 & 40$<\lambda_{\rm RM}<$45 & 30 & 18 & 0.60 \\
$>$45 & 75 & 57 & 0.76 & 45$<\lambda_{\rm RM}<$50 & 18 & 12 & 0.66 \\
{\bf $>$50} & {\bf 57} & {\bf 45} & {\bf 0.79} & -- & -- & -- & -- \\
\hline
\end{tabular}
\end{center}
\end{table*}

\section{Summary}
\label{sec:conc}

In this paper, we detail the X-ray analysis of Dark Energy Survey Year 3 redMaPPer clusters using data products from the {\em eROSITA} All Sky Survey. In summary:

\begin{itemize}

\item We define a sky region ($60^{\circ}<{\rm R.A.}<90^{\circ}$ and $-58^{\circ}<{\rm Dec}<-48^{\circ}$) within which the exposure time of the eRASS1 is a factor 2.5 times higher than the average all sky area.
\item Within this region, there are 57 DESY3 redMaPPer clusters with a richness of $\lambda_{\rm RM}>50$ and a redshift range $0.2<{\rm z}<0.65$, the redshift range used for DES Year 1 cluster cosmology.
\item Matching to the public eRASS1 galaxy cluster catalogue of \cite{erass1clusters}, we find that 45 of these DESY3 redMaPPer galaxy clusters are matched to clusters in this catalogue, with the remaining 12 having no match.
\item Of the 12 without a match, 7 of these have an eRASS1 detected point source coincident with the redMaPPer coordinates of the cluster.
\item We investigate the form of the luminosity-richness relation of these clusters, using luminosities from \cite{erass1clusters} for the matched clusters, and employ an X-ray forced photometry method to measure the fluxes (and hence luminosities) of the unmatched clusters.
\item Comparing the luminosity-richness relation of the matched cluster sample, to the relation when we include both the matched and unmatched clusters, we find a weak steepening of the slope of the relation.  However, the two relations are statistically consistent.
\item By examining additional X-ray data, along with SZ data from ACT and SPT, we find evidence for the presence of virialised gas in 5 of the 12 unmatched clusters.  Of these 5 clusters, 3 were detected as eRASS1 point sources.
\item We find that the lack of X-ray emission for the unmatched clusters is due to their low richness and high redshift, compared to the matched clusters, rather than an underlying physical reason.
\item We investigate the possibility of detection of the unmatched clusters with future releases of the eRASS.  We find that with eRASS5 (which would be roughly the depth of the eFEDS survey in our high exposure region), a number of our clusters fall below the approximate flux limit of the eFEDS, resulting in some form of incompleteness of X-ray detection for future eRASS releases.

The release of the eROSITA-All Sky Survey has allowed us for the first time to study a sample of Dark Energy Survey redMaPPer selected clusters with complete X-ray coverage (i.e., X-ray measured properties).  While the depth of the eRASS1 has limited the interpretation of the derived scaling relations, future (deeper) releases of the eRASS will allow for the study of optically selected clusters over a wider area and lower richness range considered in this work.  We highlight however, that with the hard richness limit imposed in this work we have introduced optical selection effects that will impact the derived scaling relations, even though we are free from X-ray selection effects given our complete X-ray coverage.  Future work taking into account the redMaPPer selection function, would allow one to probe richness scaling relations of the full underlying cluster population, and could be folded into optical cluster cosmology efforts \citep[e.g.,][]{2019arXiv191013548K}

\end{itemize}

\section*{Data availability}

The Dark Energy Survey Year 3 redMaPPer catalogue from which the clusters were initially selected can be found at data underlying this work can be found at: \href{https://des.ncsa.illinois.edu/releases/y3a2/Y3key-cluster}{https://des.ncsa.illinois.edu/releases/y3a2/Y3key-cluster}.  The {\em eROSITA}-All Sky Survey data used for X-ray information can be found at \href{https://erosita.mpe.mpg.de/dr1/AllSkySurveyData_dr1/Catalogues_dr1/}{https://erosita.mpe.mpg.de/dr1/AllSkySurveyData\_dr1/Catalogues\_dr1/}.  The data used in the scientific analysis presented in this study can be found throughout the article.

\section*{Acknowledgements}

PG, KR, RW, JP, and EU recognise support from the UK Science and Technology Facilities Council via grants ST/T000473/1 and ST/X001040/1 (PG, KR), ST/X508822/1 (JP), ST/P006760/1 (RW), ST/T506461/1 (EU).  DT is supported by NASA under award number 80GSFC24M0006.  MH acknowledges financial support from the National Research Foundation.  
TJ acknowledges support by the U.S. Department of Energy, Office of Science, Office of
High Energy Physics, under Award Numbers DE-SC0010107 and A00-1465-001.  AF acknowledges support from the National Science Foundation under Cooperative Agreement 2421782 and the Simons Foundation grant MPS-AI-00010515 awarded to NSF-Simons AI Institute for Cosmic Origins (CosmicAI, \href{https://www.cosmicai.org/}{https://www.cosmicai.org/}).

Funding for the DES Projects has been provided by the U.S. Department of Energy, the U.S. National Science Foundation, the Ministry of Science and Education of Spain, the Science and Technology Facilities Council of the United Kingdom, the Higher Education Funding Council for England, the National Center for Supercomputing Applications at the University of Illinois at Urbana-Champaign, the Kavli Institute of Cosmological Physics at the University of Chicago, the Center for Cosmology and Astro-Particle Physics at the Ohio State University, the Mitchell Institute for Fundamental Physics and Astronomy at Texas A\&M University, Financiadora de Estudos e Projetos, Funda{\c c}{\~a}o Carlos Chagas Filho de Amparo {\`a} Pesquisa do Estado do Rio de Janeiro, Conselho Nacional de Desenvolvimento Cient{\'i}fico e Tecnol{\'o}gico and the Minist{\'e}rio da Ci{\^e}ncia, Tecnologia e Inova{\c c}{\~a}o, the Deutsche Forschungsgemeinschaft and the Collaborating Institutions in the Dark Energy Survey. 

The Collaborating Institutions are Argonne National Laboratory, the University of California at Santa Cruz, the University of Cambridge, Centro de Investigaciones Energ{\'e}ticas, Medioambientales y Tecnol{\'o}gicas-Madrid, the University of Chicago, University College London, the DES-Brazil Consortium, the University of Edinburgh, the Eidgen{\"o}ssische Technische Hochschule (ETH) Z{\"u}rich, Fermi National Accelerator Laboratory, the University of Illinois at Urbana-Champaign, the Institut de Ci{\`e}ncies de l'Espai (IEEC/CSIC), the Institut de F{\'i}sica d'Altes Energies, Lawrence Berkeley National Laboratory, the Ludwig-Maximilians Universit{\"a}t M{\"u}nchen and the associated Excellence Cluster Universe, the University of Michigan, NSF NOIRLab, the University of Nottingham, The Ohio State University, the University of Pennsylvania, the University of Portsmouth, SLAC National Accelerator Laboratory, Stanford University, the University of Sussex, Texas A\&M University, and the OzDES Membership Consortium.

Based in part on observations at NSF Cerro Tololo Inter-American Observatory at NSF NOIRLab (NOIRLab Prop. ID 2012B-0001; PI: J. Frieman), which is managed by the Association of Universities for Research in Astronomy (AURA) under a cooperative agreement with the National Science Foundation.

The DES data management system is supported by the National Science Foundation under Grant Numbers AST-1138766 and AST-1536171.

Data access is enabled by Jetstream2 and OSN at Indiana University through allocation PHY240006: Dark Energy Survey from the Advanced Cyberinfrastructure Coordination Ecosystem: Services and Support (ACCESS) program, which is supported by U.S. National Science Foundation grants 2138259, 2138286, 2138307, 2137603, and 2138296.

The DES participants from Spanish institutions are partially supported by MICINN under grants PID2021-123012, PID2021-128989 PID2022-141079, SEV-2016-0588, CEX2020-001058-M and CEX2020-001007-S, some of which include ERDF funds from the European Union. IFAE is partially funded by the CERCA program of the Generalitat de Catalunya.

We  acknowledge support from the Brazilian Instituto Nacional de Ci\^encia e Tecnologia (INCT) do e-Universo (CNPq grant 465376/2014-2).

This document was prepared by the DES Collaboration using the resources of the Fermi National Accelerator Laboratory (Fermilab), a U.S. Department of Energy, Office of Science, Office of High Energy Physics HEP User Facility. Fermilab is managed by Fermi Forward Discovery Group, LLC, acting under Contract No. 89243024CSC000002.

This work is based on data from eROSITA, the soft X-ray instrument aboard SRG, a joint Russian-German science mission supported by the Russian Space Agency (Roskosmos), in the interests of the Russian Academy of Sciences represented by its Space Research Institute (IKI), and the Deutsches Zentrum für Luft- und Raumfahrt (DLR). The SRG spacecraft was built by Lavochkin Association (NPOL) and its subcontractors, and is operated by NPOL with support from the Max Planck Institute for Extraterrestrial Physics (MPE). The development and construction of the eROSITA X-ray instrument was led by MPE, with contributions from the Dr. Karl Remeis Observatory Bamberg \& ECAP (FAU Erlangen-Nuernberg), the University of Hamburg Observatory, the Leibniz Institute for Astrophysics Potsdam (AIP), and the Institute for Astronomy and Astrophysics of the University of Tübingen, with the support of DLR and the Max Planck Society. The Argelander Institute for Astronomy of the University of Bonn and the Ludwig Maximilians Universität Munich also participated in the science preparation for eROSITA. 

The eROSITA data were processed using the eSASS software system developed by the German eROSITA consortium.




\bibliographystyle{mnras}

\bibliography{desy3-erass} 

@ARTICLE{2019MNRAS.487.2578Z,
       author = {{Zhang}, Y. and {Jeltema}, T. and {Hollowood}, D.~L. and {Everett}, S. and {Rozo}, E. and {Farahi}, A. and {Bermeo}, A. and {Bhargava}, S. and {Giles}, P. and {Romer}, A.~K. and {Wilkinson}, R. and {Rykoff}, E.~S. and {Mantz}, A. and {Diehl}, H.~T. and {Evrard}, A.~E. and {Stern}, C. and {Gruen}, D. and {von der Linden}, A. and {Splettstoesser}, M. and {Chen}, X. and {Costanzi}, M. and {Allen}, S. and {Collins}, C. and {Hilton}, M. and {Klein}, M. and {Mann}, R.~G. and {Manolopoulou}, M. and {Morris}, G. and {Mayers}, J. and {Sahlen}, M. and {Stott}, J. and {Vergara Cervantes}, C. and {Viana}, P.~T.~P. and {Wechsler}, R.~H. and {Allam}, S. and {Avila}, S. and {Bechtol}, K. and {Bertin}, E. and {Brooks}, D. and {Burke}, D.~L. and {Carnero Rosell}, A. and {Carrasco Kind}, M. and {Carretero}, J. and {Castander}, F.~J. and {da Costa}, L.~N. and {De Vicente}, J. and {Desai}, S. and {Dietrich}, J.~P. and {Doel}, P. and {Flaugher}, B. and {Fosalba}, P. and {Frieman}, J. and {Garc{\'\i}a-Bellido}, J. and {Gaztanaga}, E. and {Gruendl}, R.~A. and {Gschwend}, J. and {Gutierrez}, G. and {Hartley}, W.~G. and {Honscheid}, K. and {Hoyle}, B. and {Krause}, E. and {Kuehn}, K. and {Kuropatkin}, N. and {Lima}, M. and {Maia}, M.~A.~G. and {Marshall}, J.~L. and {Melchior}, P. and {Menanteau}, F. and {Miller}, C.~J. and {Miquel}, R. and {Ogando}, R.~L.~C. and {Plazas}, A.~A. and {Sanchez}, E. and {Scarpine}, V. and {Schindler}, R. and {Serrano}, S. and {Sevilla-Noarbe}, I. and {Smith}, M. and {Soares-Santos}, M. and {Suchyta}, E. and {Swanson}, M.~E.~C. and {Tarle}, G. and {Thomas}, D. and {Tucker}, D.~L. and {Vikram}, V. and {Wester}, W. and {DES Collaboration}},
        title = "{Dark Energy Surveyed Year 1 results: calibration of cluster mis-centring in the redMaPPer catalogues}",
      journal = {\mnras},
         year = 2019,
        month = aug,
       volume = {487},
       number = {2},
        pages = {2578-2593},
          doi = {10.1093/mnras/stz1361},
archivePrefix = {arXiv},
       eprint = {1901.07119},
 primaryClass = {astro-ph.CO},
       adsurl = {https://ui.adsabs.harvard.edu/abs/2019MNRAS.487.2578Z}
}

@ARTICLE{2001ApJ...547..594R,
       author = {{Romer}, A. Kathy and {Viana}, Pedro T.~P. and {Liddle}, Andrew R. and {Mann}, Robert G.},
        title = "{A Serendipitous Galaxy Cluster Survey with XMM: Expected Catalog Properties and Scientific Applications}",
      journal = {\apj},
         year = 2001,
        month = feb,
       volume = {547},
       number = {2},
        pages = {594-608},
          doi = {10.1086/318382},
       adsurl = {https://ui.adsabs.harvard.edu/abs/2001ApJ...547..594R}
}

@ARTICLE{2001ApJ...556L..91S,
   author = {{Smith}, R.~K. and {Brickhouse}, N.~S. and {Liedahl}, D.~A. and 
	{Raymond}, J.~C.},
    title = "{Collisional Plasma Models with APEC/APED: Emission-Line Diagnostics of Hydrogen-like and Helium-like Ions}",
  journal = {\apjl},
   eprint = {arXiv:astro-ph/0106478},
     year = 2001,
    month = aug,
   volume = 556,
    pages = {L91-L95},
      doi = {10.1086/322992},
   adsurl = {http://adsabs.harvard.edu/abs/2001ApJ...556L..91S}
}

@ARTICLE{2007ApJ...665.1489K,
   author = {{Kelly}, B.~C.},
    title = "{Some Aspects of Measurement Error in Linear Regression of Astronomical Data}",
  journal = {\apj},
archivePrefix = "arXiv",
   eprint = {0705.2774},
     year = 2007,
    month = aug,
   volume = 665,
    pages = {1489-1506},
      doi = {10.1086/519947},
   adsurl = {http://adsabs.harvard.edu/abs/2007ApJ...665.1489K}
}

@ARTICLE{2016A&A...592A...3G,
   author = {{Giles}, P.~A. and {Maughan}, B.~J. and {Pacaud}, F. and {Lieu}, M. and 
	{Clerc}, N. and {Pierre}, M. and {Adami}, C. and {Chiappetti}, L. and 
	{D{\'e}mocl{\'e}s}, J. and {Ettori}, S. and {Le F{\'e}vre}, J.~P. and 
	{Ponman}, T. and {Sadibekova}, T. and {Smith}, G.~P. and {Willis}, J.~P. and 
	{Ziparo}, F.},
    title = "{The XXL Survey. III. Luminosity-temperature relation of the bright cluster sample}",
  journal = {\aap},
archivePrefix = "arXiv",
   eprint = {1512.03833},
     year = 2016,
    month = jun,
   volume = 592,
      eid = {A3},
    pages = {A3},
      doi = {10.1051/0004-6361/201526886},
   adsurl = {http://adsabs.harvard.edu/abs/2016A%26A...592A...3G}
}

@ARTICLE{2016A&A...592A...1P,
   author = {{Pierre}, M. and {Pacaud}, F. and {Adami}, C. and {Alis}, S. and 
	{Altieri}, B. and {Baran}, N. and {Benoist}, C. and {Birkinshaw}, M. and 
	{Bongiorno}, A. and {Bremer}, M.~N. and {Brusa}, M. and {Butler}, A. and 
	{Ciliegi}, P. and {Chiappetti}, L. and {Clerc}, N. and {Corasaniti}, P.~S. and 
	{Coupon}, J. and {De Breuck}, C. and {Democles}, J. and {Desai}, S. and 
	{Delhaize}, J. and {Devriendt}, J. and {Dubois}, Y. and {Eckert}, D. and 
	{Elyiv}, A. and {Ettori}, S. and {Evrard}, A. and {Faccioli}, L. and 
	{Farahi}, A. and {Ferrari}, C. and {Finet}, F. and {Fotopoulou}, S. and 
	{Fourmanoit}, N. and {Gandhi}, P. and {Gastaldello}, F. and 
	{Gastaud}, R. and {Georgantopoulos}, I. and {Giles}, P. and 
	{Guennou}, L. and {Guglielmo}, V. and {Horellou}, C. and {Husband}, K. and 
	{Huynh}, M. and {Iovino}, A. and {Kilbinger}, M. and {Koulouridis}, E. and 
	{Lavoie}, S. and {Le Brun}, A.~M.~C. and {Le Fevre}, J.~P. and 
	{Lidman}, C. and {Lieu}, M. and {Lin}, C.~A. and {Mantz}, A. and 
	{Maughan}, B.~J. and {Maurogordato}, S. and {McCarthy}, I.~G. and 
	{McGee}, S. and {Melin}, J.~B. and {Melnyk}, O. and {Menanteau}, F. and 
	{Novak}, M. and {Paltani}, S. and {Plionis}, M. and {Poggianti}, B.~M. and 
	{Pomarede}, D. and {Pompei}, E. and {Ponman}, T.~J. and {Ramos-Ceja}, M.~E. and 
	{Ranalli}, P. and {Rapetti}, D. and {Raychaudury}, S. and {Reiprich}, T.~H. and 
	{Rottgering}, H. and {Rozo}, E. and {Rykoff}, E. and {Sadibekova}, T. and 
	{Santos}, J. and {Sauvageot}, J.~L. and {Schimd}, C. and {Sereno}, M. and 
	{Smith}, G.~P. and {Smol{\v c}i{\'c}}, V. and {Snowden}, S. and 
	{Spergel}, D. and {Stanford}, S. and {Surdej}, J. and {Valageas}, P. and 
	{Valotti}, A. and {Valtchanov}, I. and {Vignali}, C. and {Willis}, J. and 
	{Ziparo}, F.},
    title = "{The XXL Survey. I. Scientific motivations - XMM-Newton observing plan - Follow-up observations and simulation programme}",
  journal = {\aap},
archivePrefix = "arXiv",
   eprint = {1512.04317},
     year = 2016,
    month = jun,
   volume = 592,
      eid = {A1},
    pages = {A1},
      doi = {10.1051/0004-6361/201526766},
   adsurl = {http://adsabs.harvard.edu/abs/2016A%26A...592A...1P}
}

@ARTICLE{2014ApJ...785..104R,
   author = {{Rykoff}, E.~S. and {Rozo}, E. and {Busha}, M.~T. and {Cunha}, C.~E. and 
	{Finoguenov}, A. and {Evrard}, A. and {Hao}, J. and {Koester}, B.~P. and 
	{Leauthaud}, A. and {Nord}, B. and {Pierre}, M. and {Reddick}, R. and 
	{Sadibekova}, T. and {Sheldon}, E.~S. and {Wechsler}, R.~H.},
    title = "{redMaPPer. I. Algorithm and SDSS DR8 Catalog}",
  journal = {\apj},
archivePrefix = "arXiv",
   eprint = {1303.3562},
     year = 2014,
    month = apr,
   volume = 785,
      eid = {104},
    pages = {104},
      doi = {10.1088/0004-637X/785/2/104},
   adsurl = {http://adsabs.harvard.edu/abs/2014ApJ...785..104R}
}

@ARTICLE{2018PASJ...70S..20O,
   author = {{Oguri}, M. and {Lin}, Y.-T. and {Lin}, S.-C. and {Nishizawa}, A.~J. and 
	{More}, A. and {More}, S. and {Hsieh}, B.-C. and {Medezinski}, E. and 
	{Miyatake}, H. and {Jian}, H.-Y. and {Lin}, L. and {Takada}, M. and 
	{Okabe}, N. and {Speagle}, J.~S. and {Coupon}, J. and {Leauthaud}, A. and 
	{Lupton}, R.~H. and {Miyazaki}, S. and {Price}, P.~A. and {Tanaka}, M. and 
	{Chiu}, I.-N. and {Komiyama}, Y. and {Okura}, Y. and {Tanaka}, M.~M. and 
	{Usuda}, T.},
    title = "{An optically-selected cluster catalog at redshift 0.1 $\lt$ z $\lt$ 1.1 from the Hyper Suprime-Cam Subaru Strategic Program S16A data}",
  journal = {\pasj},
archivePrefix = "arXiv",
   eprint = {1701.00818},
     year = 2018,
    month = jan,
   volume = 70,
      eid = {S20},
    pages = {S20},
      doi = {10.1093/pasj/psx042},
   adsurl = {http://adsabs.harvard.edu/abs/2018PASJ...70S..20O}
}

@ARTICLE{2016ApJS..224....1R,
   author = {{Rykoff}, E.~S. and {Rozo}, E. and {Hollowood}, D. and {Bermeo-Hernandez}, A. and 
	{Jeltema}, T. and {Mayers}, J. and {Romer}, A.~K. and {Rooney}, P. and 
	{Saro}, A. and {Vergara Cervantes}, C. and {Wechsler}, R.~H. and 
	{Wilcox}, H. and {Abbott}, T.~M.~C. and {Abdalla}, F.~B. and 
	{Allam}, S. and {Annis}, J. and {Benoit-L{\'e}vy}, A. and {Bernstein}, G.~M. and 
	{Bertin}, E. and {Brooks}, D. and {Burke}, D.~L. and {Capozzi}, D. and 
	{Carnero Rosell}, A. and {Carrasco Kind}, M. and {Castander}, F.~J. and 
	{Childress}, M. and {Collins}, C.~A. and {Cunha}, C.~E. and 
	{D'Andrea}, C.~B. and {da Costa}, L.~N. and {Davis}, T.~M. and 
	{Desai}, S. and {Diehl}, H.~T. and {Dietrich}, J.~P. and {Doel}, P. and 
	{Evrard}, A.~E. and {Finley}, D.~A. and {Flaugher}, B. and {Fosalba}, P. and 
	{Frieman}, J. and {Glazebrook}, K. and {Goldstein}, D.~A. and 
	{Gruen}, D. and {Gruendl}, R.~A. and {Gutierrez}, G. and {Hilton}, M. and 
	{Honscheid}, K. and {Hoyle}, B. and {James}, D.~J. and {Kay}, S.~T. and 
	{Kuehn}, K. and {Kuropatkin}, N. and {Lahav}, O. and {Lewis}, G.~F. and 
	{Lidman}, C. and {Lima}, M. and {Maia}, M.~A.~G. and {Mann}, R.~G. and 
	{Marshall}, J.~L. and {Martini}, P. and {Melchior}, P. and {Miller}, C.~J. and 
	{Miquel}, R. and {Mohr}, J.~J. and {Nichol}, R.~C. and {Nord}, B. and 
	{Ogando}, R. and {Plazas}, A.~A. and {Reil}, K. and {Sahl{\'e}n}, M. and 
	{Sanchez}, E. and {Santiago}, B. and {Scarpine}, V. and {Schubnell}, M. and 
	{Sevilla-Noarbe}, I. and {Smith}, R.~C. and {Soares-Santos}, M. and 
	{Sobreira}, F. and {Stott}, J.~P. and {Suchyta}, E. and {Swanson}, M.~E.~C. and 
	{Tarle}, G. and {Thomas}, D. and {Tucker}, D. and {Uddin}, S. and 
	{Viana}, P.~T.~P. and {Vikram}, V. and {Walker}, A.~R. and {Zhang}, Y. and 
	{DES Collaboration}},
    title = "{The RedMaPPer Galaxy Cluster Catalog From DES Science Verification Data}",
  journal = {\apjs},
archivePrefix = "arXiv",
   eprint = {1601.00621},
     year = 2016,
    month = may,
   volume = 224,
      eid = {1},
    pages = {1},
      doi = {10.3847/0067-0049/224/1/1},
   adsurl = {http://adsabs.harvard.edu/abs/2016ApJS..224....1R}
}

@INPROCEEDINGS{1996ASPC..101...17A,
   author = {{Arnaud}, K.~A.},
    title = "{XSPEC: The First Ten Years}",
booktitle = {Astronomical Data Analysis Software and Systems V},
     year = 1996,
   series = {Astronomical Society of the Pacific Conference Series},
   volume = 101,
   editor = {{Jacoby}, G.~H. and {Barnes}, J.},
    pages = {17},
   adsurl = {http://adsabs.harvard.edu/abs/1996ASPC..101...17A}
}

@ARTICLE{2011MNRAS.418...14L,
   author = {{Lloyd-Davies}, E.~J. and {Romer}, A.~K. and {Mehrtens}, N. and 
	{Hosmer}, M. and {Davidson}, M. and {Sabirli}, K. and {Mann}, R.~G. and 
	{Hilton}, M. and {Liddle}, A.~R. and {Viana}, P.~T.~P. and {Campbell}, H.~C. and 
	{Collins}, C.~A. and {Dubois}, E.~N. and {Freeman}, P. and {Harrison}, C.~D. and 
	{Hoyle}, B. and {Kay}, S.~T. and {Kuwertz}, E. and {Miller}, C.~J. and 
	{Nichol}, R.~C. and {Sahl{\'e}n}, M. and {Stanford}, S.~A. and 
	{Stott}, J.~P.},
    title = "{The XMM Cluster Survey: X-ray analysis methodology}",
  journal = {\mnras},
archivePrefix = "arXiv",
   eprint = {1010.0677},
     year = 2011,
    month = nov,
   volume = 418,
    pages = {14-53},
      doi = {10.1111/j.1365-2966.2011.19117.x},
   adsurl = {http://adsabs.harvard.edu/abs/2011MNRAS.418...14L}
}

@ARTICLE{2019ApJS..244...22H,
       author = {{Hollowood}, Devon L. and {Jeltema}, Tesla and {Chen}, Xinyi and
         {Farahi}, Arya and {Evrard}, August and {Everett}, Spencer and
         {Rozo}, Eduardo and {Rykoff}, Eli and {Bernstein}, Rebecca and
         {Bermeo-Hernandez}, Alberto and {Eiger}, Lena and {Giles}, Paul and
         {Israel}, Holger and {Michel}, Renee and {Noorali}, Raziq and
         {Romer}, A. Kathy and {Rooney}, Philip and {Splettstoesser}, Megan},
        title = "{Chandra Follow-up of the SDSS DR8 Redmapper Catalog Using the MATCha Pipeline}",
      journal = {\apjs},
         year = 2019,
        month = oct,
       volume = {244},
       number = {2},
          eid = {22},
        pages = {22},
          doi = {10.3847/1538-4365/ab3d27},
archivePrefix = {arXiv},
       eprint = {1808.06637},
 primaryClass = {astro-ph.CO},
       adsurl = {https://ui.adsabs.harvard.edu/abs/2019ApJS..244...22H}
}

@ARTICLE{2019MNRAS.490.3341F,
       author = {{Farahi}, A. and {Chen}, X. and {Evrard}, A.~E. and {Hollowood}, D.~L. and
         {Wilkinson}, R. and {Bhargava}, S. and {Giles}, P. and {Romer}, A.~K. and
         {Jeltema}, T. and {Hilton}, M. and {Bermeo}, A. and {Mayers}, J. and
         {Vergara Cervantes}, C. and {Rozo}, E. and {Rykoff}, E.~S. and
         {Collins}, C. and {Costanzi}, M. and {Everett}, S. and {Liddle}, A.~R. and
         {Mann}, R.~G. and {Mantz}, A. and {Rooney}, P. and {Sahlen}, M. and
         {Stott}, J. and {Viana}, P.~T.~P. and {Zhang}, Y. and {Annis}, J. and
         {Avila}, S. and {Brooks}, D. and {Buckley-Geer}, E. and {Burke}, D.~L. and
         {Carnero Rosell}, A. and {Carrasco Kind}, M. and {Carretero}, J. and
         {Castander}, F.~J. and {da Costa}, L.~N. and {De Vicente}, J. and
         {Desai}, S. and {Diehl}, H.~T. and {Dietrich}, J.~P. and {Doel}, P. and
         {Flaugher}, B. and {Fosalba}, P. and {Frieman}, J. and
         {Garc{\'\i}a-Bellido}, J. and {Gaztanaga}, E. and {Gerdes}, D.~W. and
         {Gruen}, D. and {Gruendl}, R.~A. and {Gschwend}, J. and
         {Gutierrez}, G. and {Honscheid}, K. and {James}, D.~J. and
         {Krause}, E. and {Kuehn}, K. and {Kuropatkin}, N. and {Lima}, M. and
         {Maia}, M.~A.~G. and {Marshall}, J.~L. and {Melchior}, P. and
         {Menanteau}, F. and {Miquel}, R. and {Ogando}, R.~L.~C. and
         {Plazas}, A.~A. and {Sanchez}, E. and {Scarpine}, V. and
         {Schubnell}, M. and {Serrano}, S. and {Sevilla-Noarbe}, I. and
         {Smith}, M. and {Sobreira}, F. and {Suchyta}, E. and
         {Swanson}, M.~E.~C. and {Tarle}, G. and {Thomas}, D. and
         {Tucker}, D.~L. and {Vikram}, V. and {Walker}, A.~R. and {Weller}, J. and
         {DES Collaboration}},
        title = "{Mass variance from archival X-ray properties of Dark Energy Survey Year-1 galaxy clusters}",
      journal = {\mnras},
         year = 2019,
        month = dec,
       volume = {490},
       number = {3},
        pages = {3341-3354},
          doi = {10.1093/mnras/stz2689},
archivePrefix = {arXiv},
       eprint = {1903.08042},
 primaryClass = {astro-ph.CO},
       adsurl = {https://ui.adsabs.harvard.edu/abs/2019MNRAS.490.3341F}
}

@article{PhysRevD.102.023509,
  title = {Dark Energy Survey Year 1 Results: Cosmological constraints from cluster abundances and weak lensing},
  author = {Abbott, T. M. C. and Aguena, M. and Alarcon, A. and Allam, S. and Allen, S. and Annis, J. and Avila, S. and Bacon, D. and Bechtol, K. and Bermeo, A. and Bernstein, G. M. and Bertin, E. and Bhargava, S. and Bocquet, S. and Brooks, D. and Brout, D. and Buckley-Geer, E. and Burke, D. L. and Carnero Rosell, A. and Carrasco Kind, M. and Carretero, J. and Castander, F. J. and Cawthon, R. and Chang, C. and Chen, X. and Choi, A. and Costanzi, M. and Crocce, M. and da Costa, L. N. and Davis, T. M. and De Vicente, J. and DeRose, J. and Desai, S. and Diehl, H. T. and Dietrich, J. P. and Dodelson, S. and Doel, P. and Drlica-Wagner, A. and Eckert, K. and Eifler, T. F. and Elvin-Poole, J. and Estrada, J. and Everett, S. and Evrard, A. E. and Farahi, A. and Ferrero, I. and Flaugher, B. and Fosalba, P. and Frieman, J. and Garc\'{\i}a-Bellido, J. and Gatti, M. and Gaztanaga, E. and Gerdes, D. W. and Giannantonio, T. and Giles, P. and Grandis, S. and Gruen, D. and Gruendl, R. A. and Gschwend, J. and Gutierrez, G. and Hartley, W. G. and Hinton, S. R. and Hollowood, D. L. and Honscheid, K. and Hoyle, B. and Huterer, D. and James, D. J. and Jarvis, M. and Jeltema, T. and Johnson, M. W. G. and Johnson, M. D. and Kent, S. and Krause, E. and Kron, R. and Kuehn, K. and Kuropatkin, N. and Lahav, O. and Li, T. S. and Lidman, C. and Lima, M. and Lin, H. and MacCrann, N. and Maia, M. A. G. and Mantz, A. and Marshall, J. L. and Martini, P. and Mayers, J. and Melchior, P. and Mena-Fern\'andez, J. and Menanteau, F. and Miquel, R. and Mohr, J. J. and Nichol, R. C. and Nord, B. and Ogando, R. L. C. and Palmese, A. and Paz-Chinch\'on, F. and Plazas, A. A. and Prat, J. and Rau, M. M. and Romer, A. K. and Roodman, A. and Rooney, P. and Rozo, E. and Rykoff, E. S. and Sako, M. and Samuroff, S. and S\'anchez, C. and Sanchez, E. and Saro, A. and Scarpine, V. and Schubnell, M. and Scolnic, D. and Serrano, S. and Sevilla-Noarbe, I. and Sheldon, E. and Smith, J. Allyn. and Smith, M. and Suchyta, E. and Swanson, M. E. C. and Tarle, G. and Thomas, D. and To, C. and Troxel, M. A. and Tucker, D. L. and Varga, T. N. and von der Linden, A. and Walker, A. R. and Wechsler, R. H. and Weller, J. and Wilkinson, R. D. and Wu, H. and Yanny, B. and Zhang, Y. and Zhang, Z. and Zuntz, J.},
  collaboration = {DES Collaboration},
  journal = {Phys. Rev. D},
  volume = {102},
  issue = {2},
  pages = {023509},
  numpages = {34},
  year = {2020},
  month = {Jul},
  publisher = {American Physical Society},
  doi = {10.1103/PhysRevD.102.023509},
  url = {https://link.aps.org/doi/10.1103/PhysRevD.102.023509}
}

@ARTICLE{2019ApJ...878...55B,
       author = {{Bocquet}, S. and {Dietrich}, J.~P. and {Schrabback}, T. and
         {Bleem}, L.~E. and {Klein}, M. and {Allen}, S.~W. and
         {Applegate}, D.~E. and {Ashby}, M.~L.~N. and {Bautz}, M. and
         {Bayliss}, M. and {Benson}, B.~A. and {Brodwin}, M. and {Bulbul}, E. and
         {Canning}, R.~E.~A. and {Capasso}, R. and {Carlstrom}, J.~E. and
         {Chang}, C.~L. and {Chiu}, I. and {Cho}, H. -M. and {Clocchiatti}, A. and
         {Crawford}, T.~M. and {Crites}, A.~T. and {de Haan}, T. and
         {Desai}, S. and {Dobbs}, M.~A. and {Foley}, R.~J. and {Forman}, W.~R. and
         {Garmire}, G.~P. and {George}, E.~M. and {Gladders}, M.~D. and
         {Gonzalez}, A.~H. and {Grandis}, S. and {Gupta}, N. and
         {Halverson}, N.~W. and {Hlavacek-Larrondo}, J. and {Hoekstra}, H. and
         {Holder}, G.~P. and {Holzapfel}, W.~L. and {Hou}, Z. and
         {Hrubes}, J.~D. and {Huang}, N. and {Jones}, C. and {Khullar}, G. and
         {Knox}, L. and {Kraft}, R. and {Lee}, A.~T. and {von der Linden}, A. and
         {Luong-Van}, D. and {Mantz}, A. and {Marrone}, D.~P. and
         {McDonald}, M. and {McMahon}, J.~J. and {Meyer}, S.~S. and
         {Mocanu}, L.~M. and {Mohr}, J.~J. and {Morris}, R.~G. and {Padin}, S. and
         {Patil}, S. and {Pryke}, C. and {Rapetti}, D. and {Reichardt}, C.~L. and
         {Rest}, A. and {Ruhl}, J.~E. and {Saliwanchik}, B.~R. and {Saro}, A. and
         {Sayre}, J.~T. and {Schaffer}, K.~K. and {Shirokoff}, E. and
         {Stalder}, B. and {Stanford}, S.~A. and {Staniszewski}, Z. and
         {Stark}, A.~A. and {Story}, K.~T. and {Strazzullo}, V. and
         {Stubbs}, C.~W. and {Vanderlinde}, K. and {Vieira}, J.~D. and
         {Vikhlinin}, A. and {Williamson}, R. and {Zenteno}, A.},
        title = "{Cluster Cosmology Constraints from the 2500 deg$^{2}$ SPT-SZ Survey: Inclusion of Weak Gravitational Lensing Data from Magellan and the Hubble Space Telescope}",
      journal = {\apj},
         year = 2019,
        month = jun,
       volume = {878},
       number = {1},
          eid = {55},
        pages = {55},
          doi = {10.3847/1538-4357/ab1f10},
archivePrefix = {arXiv},
       eprint = {1812.01679},
 primaryClass = {astro-ph.CO},
       adsurl = {https://ui.adsabs.harvard.edu/abs/2019ApJ...878...55B}
}

@ARTICLE{2020ApJS..247...25B,
       author = {{Bleem}, L.~E. and {Bocquet}, S. and {Stalder}, B. and
         {Gladders}, M.~D. and {Ade}, P.~A.~R. and {Allen}, S.~W. and
         {Anderson}, A.~J. and {Annis}, J. and {Ashby}, M.~L.~N. and
         {Austermann}, J.~E. and {Avila}, S. and {Avva}, J.~S. and
         {Bayliss}, M. and {Beall}, J.~A. and {Bechtol}, K. and {Bender}, A.~N. and
         {Benson}, B.~A. and {Bertin}, E. and {Bianchini}, F. and {Blake}, C. and
         {Brodwin}, M. and {Brooks}, D. and {Buckley-Geer}, E. and
         {Burke}, D.~L. and {Carlstrom}, J.~E. and {Rosell}, A. Carnero and
         {Carrasco Kind}, M. and {Carretero}, J. and {Chang}, C.~L. and
         {Chiang}, H.~C. and {Citron}, R. and {Moran}, C. Corbett and
         {Costanzi}, M. and {Crawford}, T.~M. and {Crites}, A.~T. and
         {da Costa}, L.~N. and {de Haan}, T. and {De Vicente}, J. and
         {Desai}, S. and {Diehl}, H.~T. and {Dietrich}, J.~P. and
         {Dobbs}, M.~A. and {Eifler}, T.~F. and {Everett}, W. and
         {Flaugher}, B. and {Floyd}, B. and {Frieman}, J. and {Gallicchio}, J. and
         {Garc{\'\i}a-Bellido}, J. and {George}, E.~M. and {Gerdes}, D.~W. and
         {Gilbert}, A. and {Gruen}, D. and {Gruendl}, R.~A. and {Gschwend}, J. and
         {Gupta}, N. and {Gutierrez}, G. and {Halverson}, N.~W. and
         {Harrington}, N. and {Henning}, J.~W. and {Heymans}, C. and
         {Holder}, G.~P. and {Hollowood}, D.~L. and {Holzapfel}, W.~L. and
         {Honscheid}, K. and {Hrubes}, J.~D. and {Huang}, N. and {Hubmayr}, J. and
         {Irwin}, K.~D. and {James}, D.~J. and {Jeltema}, T. and {Joudaki}, S. and
         {Khullar}, G. and {Klein}, M. and {Knox}, L. and {Kuropatkin}, N. and
         {Lee}, A.~T. and {Li}, D. and {Lidman}, C. and {Lowitz}, A. and
         {MacCrann}, N. and {Mahler}, G. and {Maia}, M.~A.~G. and
         {Marshall}, J.~L. and {McDonald}, M. and {McMahon}, J.~J. and
         {Melchior}, P. and {Menanteau}, F. and {Meyer}, S.~S. and {Miquel}, R. and
         {Mocanu}, L.~M. and {Mohr}, J.~J. and {Montgomery}, J. and
         {Nadolski}, A. and {Natoli}, T. and {Nibarger}, J.~P. and {Noble}, G. and
         {Novosad}, V. and {Padin}, S. and {Palmese}, A. and {Parkinson}, D. and
         {Patil}, S. and {Paz-Chinch{\'o}n}, F. and {Plazas}, A.~A. and
         {Pryke}, C. and {Ramachandra}, N.~S. and {Reichardt}, C.~L. and
         {Remolina Gonz{\'a}lez}, J.~D. and {Romer}, A.~K. and {Roodman}, A. and
         {Ruhl}, J.~E. and {Rykoff}, E.~S. and {Saliwanchik}, B.~R. and
         {Sanchez}, E. and {Saro}, A. and {Sayre}, J.~T. and {Schaffer}, K.~K. and
         {Schrabback}, T. and {Serrano}, S. and {Sharon}, K. and {Sievers}, C. and
         {Smecher}, G. and {Smith}, M. and {Soares-Santos}, M. and
         {Stark}, A.~A. and {Story}, K.~T. and {Suchyta}, E. and {Tarle}, G. and
         {Tucker}, C. and {Vanderlinde}, K. and {Veach}, T. and {Vieira}, J.~D. and
         {Wang}, G. and {Weller}, J. and {Whitehorn}, N. and {Wu}, W.~L.~K. and
         {Yefremenko}, V. and {Zhang}, Y.},
        title = "{The SPTpol Extended Cluster Survey}",
      journal = {\apjs},
         year = 2020,
        month = mar,
       volume = {247},
       number = {1},
          eid = {25},
        pages = {25},
          doi = {10.3847/1538-4365/ab6993},
archivePrefix = {arXiv},
       eprint = {1910.04121},
 primaryClass = {astro-ph.CO},
       adsurl = {https://ui.adsabs.harvard.edu/abs/2020ApJS..247...25B}
}

@ARTICLE{2021ApJS..253....3H,
       author = {{Hilton}, M. and {Sif{\'o}n}, C. and {Naess}, S. and {Madhavacheril}, M. and {Oguri}, M. and {Rozo}, E. and {Rykoff}, E. and {Abbott}, T.~M.~C. and {Adhikari}, S. and {Aguena}, M. and {Aiola}, S. and {Allam}, S. and {Amodeo}, S. and {Amon}, A. and {Annis}, J. and {Ansarinejad}, B. and {Aros-Bunster}, C. and {Austermann}, J.~E. and {Avila}, S. and {Bacon}, D. and {Battaglia}, N. and {Beall}, J.~A. and {Becker}, D.~T. and {Bernstein}, G.~M. and {Bertin}, E. and {Bhandarkar}, T. and {Bhargava}, S. and {Bond}, J.~R. and {Brooks}, D. and {Burke}, D.~L. and {Calabrese}, E. and {Carrasco Kind}, M. and {Carretero}, J. and {Choi}, S.~K. and {Choi}, A. and {Conselice}, C. and {da Costa}, L.~N. and {Costanzi}, M. and {Crichton}, D. and {Crowley}, K.~T. and {D{\"u}nner}, R. and {Denison}, E.~V. and {Devlin}, M.~J. and {Dicker}, S.~R. and {Diehl}, H.~T. and {Dietrich}, J.~P. and {Doel}, P. and {Duff}, S.~M. and {Duivenvoorden}, A.~J. and {Dunkley}, J. and {Everett}, S. and {Ferraro}, S. and {Ferrero}, I. and {Fert{\'e}}, A. and {Flaugher}, B. and {Frieman}, J. and {Gallardo}, P.~A. and {Garc{\'\i}a-Bellido}, J. and {Gaztanaga}, E. and {Gerdes}, D.~W. and {Giles}, P. and {Golec}, J.~E. and {Gralla}, M.~B. and {Grandis}, S. and {Gruen}, D. and {Gruendl}, R.~A. and {Gschwend}, J. and {Gutierrez}, G. and {Han}, D. and {Hartley}, W.~G. and {Hasselfield}, M. and {Hill}, J.~C. and {Hilton}, G.~C. and {Hincks}, A.~D. and {Hinton}, S.~R. and {Ho}, S. -P.~P. and {Honscheid}, K. and {Hoyle}, B. and {Hubmayr}, J. and {Huffenberger}, K.~M. and {Hughes}, J.~P. and {Jaelani}, A.~T. and {Jain}, B. and {James}, D.~J. and {Jeltema}, T. and {Kent}, S. and {Knowles}, K. and {Koopman}, B.~J. and {Kuehn}, K. and {Lahav}, O. and {Lima}, M. and {Lin}, Y. -T. and {Lokken}, M. and {Loubser}, S.~I. and {MacCrann}, N. and {Maia}, M.~A.~G. and {Marriage}, T.~A. and {Martin}, J. and {McMahon}, J. and {Melchior}, P. and {Menanteau}, F. and {Miquel}, R. and {Miyatake}, H. and {Moodley}, K. and {Morgan}, R. and {Mroczkowski}, T. and {Nati}, F. and {Newburgh}, L.~B. and {Niemack}, M.~D. and {Nishizawa}, A.~J. and {Ogando}, R.~L.~C. and {Orlowski-Scherer}, J. and {Page}, L.~A. and {Palmese}, A. and {Partridge}, B. and {Paz-Chinch{\'o}n}, F. and {Phakathi}, P. and {Plazas}, A.~A. and {Robertson}, N.~C. and {Romer}, A.~K. and {Carnero Rosell}, A. and {Salatino}, M. and {Sanchez}, E. and {Schaan}, E. and {Schillaci}, A. and {Sehgal}, N. and {Serrano}, S. and {Shin}, T. and {Simon}, S.~M. and {Smith}, M. and {Soares-Santos}, M. and {Spergel}, D.~N. and {Staggs}, S.~T. and {Storer}, E.~R. and {Suchyta}, E. and {Swanson}, M.~E.~C. and {Tarle}, G. and {Thomas}, D. and {To}, C. and {Trac}, H. and {Ullom}, J.~N. and {Vale}, L.~R. and {Van Lanen}, J. and {Vavagiakis}, E.~M. and {De Vicente}, J. and {Wilkinson}, R.~D. and {Wollack}, E.~J. and {Xu}, Z. and {Zhang}, Y.},
        title = "{The Atacama Cosmology Telescope: A Catalog of >4000 Sunyaev-Zel{\textquoteright}dovich Galaxy Clusters}",
      journal = {\apjs},
         year = 2021,
        month = mar,
       volume = {253},
       number = {1},
          eid = {3},
        pages = {3},
          doi = {10.3847/1538-4365/abd023},
archivePrefix = {arXiv},
       eprint = {2009.11043},
 primaryClass = {astro-ph.CO},
       adsurl = {https://ui.adsabs.harvard.edu/abs/2021ApJS..253....3H}
}

@ARTICLE{2018arXiv180901669T,
       author = {{The LSST Dark Energy Science Collaboration} and {Mandelbaum}, Rachel and
         {Eifler}, Tim and {Hlo{\v{z}}ek}, Ren{\'e}e and {Collett}, Thomas and
         {Gawiser}, Eric and {Scolnic}, Daniel and {Alonso}, David and
         {Awan}, Humna and {Biswas}, Rahul and {Blazek}, Jonathan and
         {Burchat}, Patricia and {Chisari}, Nora Elisa and {Dell'Antonio}, Ian and
         {Digel}, Seth and {Frieman}, Josh and {Goldstein}, Daniel A. and
         {Hook}, Isobel and {Ivezi{\'c}}, {\v{Z}}eljko and {Kahn}, Steven M. and
         {Kamath}, Sowmya and {Kirkby}, David and {Kitching}, Thomas and
         {Krause}, Elisabeth and {Leget}, Pierre-Fran{\c{c}}ois and
         {Marshall}, Philip J. and {Meyers}, Joshua and {Miyatake}, Hironao and
         {Newman}, Jeffrey A. and {Nichol}, Robert and {Rykoff}, Eli and
         {Sanchez}, F. Javier and {Slosar}, An{\v{z}}e and {Sullivan}, Mark and
         {Troxel}, M.~A.},
        title = "{The LSST Dark Energy Science Collaboration (DESC) Science Requirements Document}",
      journal = {arXiv e-prints},
         year = 2018,
        month = sep,
          eid = {arXiv:1809.01669},
        pages = {arXiv:1809.01669},
archivePrefix = {arXiv},
       eprint = {1809.01669},
 primaryClass = {astro-ph.CO},
       adsurl = {https://ui.adsabs.harvard.edu/abs/2018arXiv180901669T}
}

@ARTICLE{2000ApJ...542..914W,
       author = {{Wilms}, J. and {Allen}, A. and {McCray}, R.},
        title = "{On the Absorption of X-Rays in the Interstellar Medium}",
      journal = {\apj},
         year = 2000,
        month = oct,
       volume = {542},
       number = {2},
        pages = {914-924},
          doi = {10.1086/317016},
archivePrefix = {arXiv},
       eprint = {astro-ph/0008425},
 primaryClass = {astro-ph},
       adsurl = {https://ui.adsabs.harvard.edu/abs/2000ApJ...542..914W}
}

@ARTICLE{2016A&A...594A.116H,
       author = {{HI4PI Collaboration} and {Ben Bekhti}, N. and {Fl{\"o}er}, L. and {Keller}, R. and {Kerp}, J. and {Lenz}, D. and {Winkel}, B. and {Bailin}, J. and {Calabretta}, M.~R. and {Dedes}, L. and {Ford}, H.~A. and {Gibson}, B.~K. and {Haud}, U. and {Janowiecki}, S. and {Kalberla}, P.~M.~W. and {Lockman}, F.~J. and {McClure-Griffiths}, N.~M. and {Murphy}, T. and {Nakanishi}, H. and {Pisano}, D.~J. and {Staveley-Smith}, L.},
        title = "{HI4PI: A full-sky H I survey based on EBHIS and GASS}",
      journal = {\aap},
         year = 2016,
        month = oct,
       volume = {594},
          eid = {A116},
        pages = {A116},
          doi = {10.1051/0004-6361/201629178},
archivePrefix = {arXiv},
       eprint = {1610.06175},
 primaryClass = {astro-ph.GA},
       adsurl = {https://ui.adsabs.harvard.edu/abs/2016A&A...594A.116H}
}

@ARTICLE{2021A&A...647A...1P,
       author = {{Predehl}, P. and {Andritschke}, R. and {Arefiev}, V. and {Babyshkin}, V. and {Batanov}, O. and {Becker}, W. and {B{\"o}hringer}, H. and {Bogomolov}, A. and {Boller}, T. and {Borm}, K. and {Bornemann}, W. and {Br{\"a}uninger}, H. and {Br{\"u}ggen}, M. and {Brunner}, H. and {Brusa}, M. and {Bulbul}, E. and {Buntov}, M. and {Burwitz}, V. and {Burkert}, W. and {Clerc}, N. and {Churazov}, E. and {Coutinho}, D. and {Dauser}, T. and {Dennerl}, K. and {Doroshenko}, V. and {Eder}, J. and {Emberger}, V. and {Eraerds}, T. and {Finoguenov}, A. and {Freyberg}, M. and {Friedrich}, P. and {Friedrich}, S. and {F{\"u}rmetz}, M. and {Georgakakis}, A. and {Gilfanov}, M. and {Granato}, S. and {Grossberger}, C. and {Gueguen}, A. and {Gureev}, P. and {Haberl}, F. and {H{\"a}lker}, O. and {Hartner}, G. and {Hasinger}, G. and {Huber}, H. and {Ji}, L. and {Kienlin}, A. v. and {Kink}, W. and {Korotkov}, F. and {Kreykenbohm}, I. and {Lamer}, G. and {Lomakin}, I. and {Lapshov}, I. and {Liu}, T. and {Maitra}, C. and {Meidinger}, N. and {Menz}, B. and {Merloni}, A. and {Mernik}, T. and {Mican}, B. and {Mohr}, J. and {M{\"u}ller}, S. and {Nandra}, K. and {Nazarov}, V. and {Pacaud}, F. and {Pavlinsky}, M. and {Perinati}, E. and {Pfeffermann}, E. and {Pietschner}, D. and {Ramos-Ceja}, M.~E. and {Rau}, A. and {Reiffers}, J. and {Reiprich}, T.~H. and {Robrade}, J. and {Salvato}, M. and {Sanders}, J. and {Santangelo}, A. and {Sasaki}, M. and {Scheuerle}, H. and {Schmid}, C. and {Schmitt}, J. and {Schwope}, A. and {Shirshakov}, A. and {Steinmetz}, M. and {Stewart}, I. and {Str{\"u}der}, L. and {Sunyaev}, R. and {Tenzer}, C. and {Tiedemann}, L. and {Tr{\"u}mper}, J. and {Voron}, V. and {Weber}, P. and {Wilms}, J. and {Yaroshenko}, V.},
        title = "{The eROSITA X-ray telescope on SRG}",
      journal = {\aap},
         year = 2021,
        month = mar,
       volume = {647},
          eid = {A1},
        pages = {A1},
          doi = {10.1051/0004-6361/202039313},
archivePrefix = {arXiv},
       eprint = {2010.03477},
 primaryClass = {astro-ph.HE},
       adsurl = {https://ui.adsabs.harvard.edu/abs/2021A&A...647A...1P}
}

@ARTICLE{2016arXiv160407626D,
       author = {{Dodelson}, Scott and {Heitmann}, Katrin and {Hirata}, Chris and {Honscheid}, Klaus and {Roodman}, Aaron and {Seljak}, Uro{\v{s}} and {Slosar}, An{\v{z}}e and {Trodden}, Mark},
        title = "{Cosmic Visions Dark Energy: Science}",
      journal = {arXiv e-prints},
         year = 2016,
        month = apr,
          eid = {arXiv:1604.07626},
        pages = {arXiv:1604.07626},
archivePrefix = {arXiv},
       eprint = {1604.07626},
 primaryClass = {astro-ph.CO},
       adsurl = {https://ui.adsabs.harvard.edu/abs/2016arXiv160407626D}
}

@ARTICLE{2022A&A...661A...2L,
       author = {{Liu}, A. and {Bulbul}, E. and {Ghirardini}, V. and {Liu}, T. and {Klein}, M. and {Clerc}, N. and {{\"O}zsoy}, Y. and {Ramos-Ceja}, M.~E. and {Pacaud}, F. and {Comparat}, J. and {Okabe}, N. and {Bahar}, Y.~E. and {Biffi}, V. and {Brunner}, H. and {Br{\"u}ggen}, M. and {Buchner}, J. and {Ider Chitham}, J. and {Chiu}, I. and {Dolag}, K. and {Gatuzz}, E. and {Gonzalez}, J. and {Hoang}, D.~N. and {Lamer}, G. and {Merloni}, A. and {Nandra}, K. and {Oguri}, M. and {Ota}, N. and {Predehl}, P. and {Reiprich}, T.~H. and {Salvato}, M. and {Schrabback}, T. and {Sanders}, J.~S. and {Seppi}, R. and {Thibaud}, Q.},
        title = "{The eROSITA Final Equatorial-Depth Survey (eFEDS). Catalog of galaxy clusters and groups}",
      journal = {\aap},
         year = 2022,
        month = may,
       volume = {661},
          eid = {A2},
        pages = {A2},
          doi = {10.1051/0004-6361/202141120},
archivePrefix = {arXiv},
       eprint = {2106.14518},
 primaryClass = {astro-ph.CO},
       adsurl = {https://ui.adsabs.harvard.edu/abs/2022A&A...661A...2L}
}

@ARTICLE{erass1clusters,
       author = {{Bulbul}, E. and {Liu}, A. and {Kluge}, M. and {Zhang}, X. and {Sanders}, J.~S. and {Bahar}, Y.~E. and {Ghirardini}, V. and {Artis}, E. and {Seppi}, R. and {Garrel}, C. and {Ramos-Ceja}, M.~E. and {Comparat}, J. and {Balzer}, F. and {B{\"o}ckmann}, K. and {Br{\"u}ggen}, M. and {Clerc}, N. and {Dennerl}, K. and {Dolag}, K. and {Freyberg}, M. and {Grandis}, S. and {Gruen}, D. and {Kleinebreil}, F. and {Krippendorf}, S. and {Lamer}, G. and {Merloni}, A. and {Migkas}, K. and {Nandra}, K. and {Pacaud}, F. and {Predehl}, P. and {Reiprich}, T.~H. and {Schrabback}, T. and {Veronica}, A. and {Weller}, J. and {Zelmer}, S.},
        title = "{The SRG/eROSITA All-Sky Survey. The first catalog of galaxy clusters and groups in the Western Galactic Hemisphere}",
      journal = {\aap},
         year = 2024,
        month = may,
       volume = {685},
          eid = {A106},
        pages = {A106},
          doi = {10.1051/0004-6361/202348264},
archivePrefix = {arXiv},
       eprint = {2402.08452},
 primaryClass = {astro-ph.CO},
       adsurl = {https://ui.adsabs.harvard.edu/abs/2024A&A...685A.106B}
}

@ARTICLE{2011PASP..123..568C,
       author = {{Carlstrom}, J.~E. and {Ade}, P.~A.~R. and {Aird}, K.~A. and {Benson}, B.~A. and {Bleem}, L.~E. and {Busetti}, S. and {Chang}, C.~L. and {Chauvin}, E. and {Cho}, H. -M. and {Crawford}, T.~M. and {Crites}, A.~T. and {Dobbs}, M.~A. and {Halverson}, N.~W. and {Heimsath}, S. and {Holzapfel}, W.~L. and {Hrubes}, J.~D. and {Joy}, M. and {Keisler}, R. and {Lanting}, T.~M. and {Lee}, A.~T. and {Leitch}, E.~M. and {Leong}, J. and {Lu}, W. and {Lueker}, M. and {Luong-Van}, D. and {McMahon}, J.~J. and {Mehl}, J. and {Meyer}, S.~S. and {Mohr}, J.~J. and {Montroy}, T.~E. and {Padin}, S. and {Plagge}, T. and {Pryke}, C. and {Ruhl}, J.~E. and {Schaffer}, K.~K. and {Schwan}, D. and {Shirokoff}, E. and {Spieler}, H.~G. and {Staniszewski}, Z. and {Stark}, A.~A. and {Tucker}, C. and {Vanderlinde}, K. and {Vieira}, J.~D. and {Williamson}, R.},
        title = "{The 10 Meter South Pole Telescope}",
      journal = {\pasp},
         year = 2011,
        month = may,
       volume = {123},
       number = {903},
        pages = {568},
          doi = {10.1086/659879},
archivePrefix = {arXiv},
       eprint = {0907.4445},
 primaryClass = {astro-ph.IM},
       adsurl = {https://ui.adsabs.harvard.edu/abs/2011PASP..123..568C}
}

@ARTICLE{2015ApJS..216...27B,
       author = {{Bleem}, L.~E. and {Stalder}, B. and {de Haan}, T. and {Aird}, K.~A. and {Allen}, S.~W. and {Applegate}, D.~E. and {Ashby}, M.~L.~N. and {Bautz}, M. and {Bayliss}, M. and {Benson}, B.~A. and {Bocquet}, S. and {Brodwin}, M. and {Carlstrom}, J.~E. and {Chang}, C.~L. and {Chiu}, I. and {Cho}, H.~M. and {Clocchiatti}, A. and {Crawford}, T.~M. and {Crites}, A.~T. and {Desai}, S. and {Dietrich}, J.~P. and {Dobbs}, M.~A. and {Foley}, R.~J. and {Forman}, W.~R. and {George}, E.~M. and {Gladders}, M.~D. and {Gonzalez}, A.~H. and {Halverson}, N.~W. and {Hennig}, C. and {Hoekstra}, H. and {Holder}, G.~P. and {Holzapfel}, W.~L. and {Hrubes}, J.~D. and {Jones}, C. and {Keisler}, R. and {Knox}, L. and {Lee}, A.~T. and {Leitch}, E.~M. and {Liu}, J. and {Lueker}, M. and {Luong-Van}, D. and {Mantz}, A. and {Marrone}, D.~P. and {McDonald}, M. and {McMahon}, J.~J. and {Meyer}, S.~S. and {Mocanu}, L. and {Mohr}, J.~J. and {Murray}, S.~S. and {Padin}, S. and {Pryke}, C. and {Reichardt}, C.~L. and {Rest}, A. and {Ruel}, J. and {Ruhl}, J.~E. and {Saliwanchik}, B.~R. and {Saro}, A. and {Sayre}, J.~T. and {Schaffer}, K.~K. and {Schrabback}, T. and {Shirokoff}, E. and {Song}, J. and {Spieler}, H.~G. and {Stanford}, S.~A. and {Staniszewski}, Z. and {Stark}, A.~A. and {Story}, K.~T. and {Stubbs}, C.~W. and {Vanderlinde}, K. and {Vieira}, J.~D. and {Vikhlinin}, A. and {Williamson}, R. and {Zahn}, O. and {Zenteno}, A.},
        title = "{Galaxy Clusters Discovered via the Sunyaev-Zel'dovich Effect in the 2500-Square-Degree SPT-SZ Survey}",
      journal = {\apjs},
         year = 2015,
        month = feb,
       volume = {216},
       number = {2},
          eid = {27},
        pages = {27},
          doi = {10.1088/0067-0049/216/2/27},
archivePrefix = {arXiv},
       eprint = {1409.0850},
 primaryClass = {astro-ph.CO},
       adsurl = {https://ui.adsabs.harvard.edu/abs/2015ApJS..216...27B}
}

@ARTICLE{2021MNRAS.505...33M,
       author = {{Myles}, J. and {Gruen}, D. and {Mantz}, A.~B. and {Allen}, S.~W. and {Morris}, R.~G. and {Rykoff}, E. and {Costanzi}, M. and {To}, C. and {DeRose}, J. and {Wechsler}, R.~H. and {Rozo}, E. and {Jeltema}, T. and {Carrasco}, E.~R. and {Kremin}, A. and {Kron}, R.},
        title = "{Spectroscopic quantification of projection effects in the SDSS redMaPPer galaxy cluster catalogue}",
      journal = {\mnras},
         year = 2021,
        month = jul,
       volume = {505},
       number = {1},
        pages = {33-44},
          doi = {10.1093/mnras/stab1243},
archivePrefix = {arXiv},
       eprint = {2011.07070},
 primaryClass = {astro-ph.CO},
       adsurl = {https://ui.adsabs.harvard.edu/abs/2021MNRAS.505...33M}
}

@ARTICLE{giles_gama,
       author = {{Giles}, P.~A. and {Robotham}, A. and {Ramos-Ceja}, M.~E. and {Maughan}, B.~J. and {Sereno}, M. and {McGee}, S. and {Phillipps}, S. and {Iovino}, A. and {Turner}, D.~J. and {Adami}, C. and {Brough}, S. and {Chiappetti}, L. and {Driver}, S.~P. and {Evrard}, A.~E. and {Farahi}, A. and {Gastaldello}, F. and {Holwerda}, B.~W. and {Hopkins}, A.~M. and {Koulouridis}, E. and {Pacaud}, F. and {Pierre}, M. and {Pimbblet}, K.~A.},
        title = "{The XXL Survey. XLII. The L$_{X}$ - {\ensuremath{\sigma}}$_{v}$ relation of galaxy groups and clusters detected in the XXL and GAMA surveys}",
      journal = {\mnras},
         year = 2022,
        month = mar,
       volume = {511},
       number = {1},
        pages = {1227-1246},
          doi = {10.1093/mnras/stab3626},
       adsurl = {https://ui.adsabs.harvard.edu/abs/2022MNRAS.511.1227G}
}

@ARTICLE{2024MNRAS.527..895P,
       author = {{Popesso}, P. and {Biviano}, A. and {Bulbul}, E. and {Merloni}, A. and {Comparat}, J. and {Clerc}, N. and {Igo}, Z. and {Liu}, A. and {Driver}, S. and {Salvato}, M. and {Brusa}, M. and {Bahar}, Y.~E. and {Malavasi}, N. and {Ghirardini}, V. and {Robotham}, A. and {Liske}, J. and {Grandis}, S.},
        title = "{The X-ray invisible Universe. A look into the haloes undetected by eROSITA}",
      journal = {\mnras},
         year = 2024,
        month = jan,
       volume = {527},
       number = {1},
        pages = {895-910},
          doi = {10.1093/mnras/stad3253},
archivePrefix = {arXiv},
       eprint = {2302.08405},
 primaryClass = {astro-ph.CO},
       adsurl = {https://ui.adsabs.harvard.edu/abs/2024MNRAS.527..895P}
}

@ARTICLE{2024A&A...682A..34M,
       author = {{Merloni}, A. and {Lamer}, G. and {Liu}, T. and {Ramos-Ceja}, M.~E. and {Brunner}, H. and {Bulbul}, E. and {Dennerl}, K. and {Doroshenko}, V. and {Freyberg}, M.~J. and {Friedrich}, S. and {Gatuzz}, E. and {Georgakakis}, A. and {Haberl}, F. and {Igo}, Z. and {Kreykenbohm}, I. and {Liu}, A. and {Maitra}, C. and {Malyali}, A. and {Mayer}, M.~G.~F. and {Nandra}, K. and {Predehl}, P. and {Robrade}, J. and {Salvato}, M. and {Sanders}, J.~S. and {Stewart}, I. and {Tub{\'\i}n-Arenas}, D. and {Weber}, P. and {Wilms}, J. and {Arcodia}, R. and {Artis}, E. and {Aschersleben}, J. and {Avakyan}, A. and {Aydar}, C. and {Bahar}, Y.~E. and {Balzer}, F. and {Becker}, W. and {Berger}, K. and {Boller}, T. and {Bornemann}, W. and {Br{\"u}ggen}, M. and {Brusa}, M. and {Buchner}, J. and {Burwitz}, V. and {Camilloni}, F. and {Clerc}, N. and {Comparat}, J. and {Coutinho}, D. and {Czesla}, S. and {Dannhauer}, S.~M. and {Dauner}, L. and {Dauser}, T. and {Dietl}, J. and {Dolag}, K. and {Dwelly}, T. and {Egg}, K. and {Ehl}, E. and {Freund}, S. and {Friedrich}, P. and {Gaida}, R. and {Garrel}, C. and {Ghirardini}, V. and {Gokus}, A. and {Gr{\"u}nwald}, G. and {Grandis}, S. and {Grotova}, I. and {Gruen}, D. and {Gueguen}, A. and {H{\"a}mmerich}, S. and {Hamaus}, N. and {Hasinger}, G. and {Haubner}, K. and {Homan}, D. and {Ider Chitham}, J. and {Joseph}, W.~M. and {Joyce}, A. and {K{\"o}nig}, O. and {Kaltenbrunner}, D.~M. and {Khokhriakova}, A. and {Kink}, W. and {Kirsch}, C. and {Kluge}, M. and {Knies}, J. and {Krippendorf}, S. and {Krumpe}, M. and {Kurpas}, J. and {Li}, P. and {Liu}, Z. and {Locatelli}, N. and {Lorenz}, M. and {M{\"u}ller}, S. and {Magaudda}, E. and {Mannes}, C. and {McCall}, H. and {Meidinger}, N. and {Michailidis}, M. and {Migkas}, K. and {Mu{\~n}oz-Giraldo}, D. and {Musiimenta}, B. and {Nguyen-Dang}, N.~T. and {Ni}, Q. and {Olechowska}, A. and {Ota}, N. and {Pacaud}, F. and {Pasini}, T. and {Perinati}, E. and {Pires}, A.~M. and {Pommranz}, C. and {Ponti}, G. and {Poppenhaeger}, K. and {P{\"u}hlhofer}, G. and {Rau}, A. and {Reh}, M. and {Reiprich}, T.~H. and {Roster}, W. and {Saeedi}, S. and {Santangelo}, A. and {Sasaki}, M. and {Schmitt}, J. and {Schneider}, P.~C. and {Schrabback}, T. and {Schuster}, N. and {Schwope}, A. and {Seppi}, R. and {Serim}, M.~M. and {Shreeram}, S. and {Sokolova-Lapa}, E. and {Starck}, H. and {Stelzer}, B. and {Stierhof}, J. and {Suleimanov}, V. and {Tenzer}, C. and {Traulsen}, I. and {Tr{\"u}mper}, J. and {Tsuge}, K. and {Urrutia}, T. and {Veronica}, A. and {Waddell}, S.~G.~H. and {Willer}, R. and {Wolf}, J. and {Yeung}, M.~C.~H. and {Zainab}, A. and {Zangrandi}, F. and {Zhang}, X. and {Zhang}, Y. and {Zheng}, X.},
        title = "{The SRG/eROSITA all-sky survey. First X-ray catalogues and data release of the western Galactic hemisphere}",
      journal = {\aap},
         year = 2024,
        month = feb,
       volume = {682},
          eid = {A34},
        pages = {A34},
          doi = {10.1051/0004-6361/202347165},
archivePrefix = {arXiv},
       eprint = {2401.17274},
 primaryClass = {astro-ph.HE},
       adsurl = {https://ui.adsabs.harvard.edu/abs/2024A&A...682A..34M}
}

@ARTICLE{kelly24,
       author = {{Kelly}, P.~M. and {Jobel}, J. and {Eiger}, O. and {Abd}, A. and {Jeltema}, T.~E. and {Giles}, P. and {Hollowood}, D.~L. and {Wilkinson}, R.~D. and {Turner}, D.~J. and {Bhargava}, S. and {Everett}, S. and {Farahi}, A. and {Romer}, A.~K. and {Rykoff}, E.~S. and {Wang}, F. and {Bocquet}, S. and {Cross}, D. and {Faridjoo}, R. and {Franco}, J. and {Gardner}, G. and {Kwiecien}, M. and {Laubner}, D. and {McDaniel}, A. and {O'Donnell}, J.~H. and {Sanchez}, L. and {Schmidt}, E. and {Sripada}, S. and {Swart}, A. and {Upsdell}, E. and {Webber}, A. and {Aguena}, M. and {Allam}, S. and {Alves}, O. and {Bacon}, D. and {Brooks}, D. and {Burke}, D.~L. and {Carnero Rosell}, A. and {Carretero}, J. and {Collins}, C.~A. and {Costanzi}, M. and {da Costa}, L.~N. and {Pereira}, M.~E.~S. and {Davis}, T.~M. and {Doel}, P. and {Ferrero}, I. and {Frieman}, J. and {Garc{\'\i}a-Bellido}, J. and {Giannini}, G. and {Gruen}, D. and {Gruendl}, R.~A. and {Hilton}, M. and {Hinton}, S.~R. and {Honscheid}, K. and {James}, D.~J. and {Kuehn}, K. and {Mann}, R.~G. and {Marshall}, J.~L. and {Mena-Fern{\'a}ndez}, J. and {Miller}, C.~J. and {Miquel}, R. and {Myles}, J. and {Palmese}, A. and {Pieres}, A. and {Plazas Malag{\'o}n}, A.~A. and {Rooney}, P.~J. and {Sahlen}, M. and {Sanchez}, E. and {Sanchez Cid}, D. and {Schubnell}, M. and {Sevilla-Noarbe}, I. and {Smith}, M. and {Stott}, J.~P. and {Suchyta}, E. and {Swanson}, M.~E.~C. and {Tarle}, G. and {To}, C. and {Viana}, P.~T.~P. and {Weaverdyck}, N. and {Wiseman}, P. and {DES Collaboration}},
        title = "{Dark energy survey year 3 results: miscentring calibration and X-ray-richness scaling relations in redMaPPer clusters}",
      journal = {\mnras},
         year = 2024,
        month = sep,
       volume = {533},
       number = {1},
        pages = {572-588},
          doi = {10.1093/mnras/stae1786},
archivePrefix = {arXiv},
       eprint = {2310.13207},
 primaryClass = {astro-ph.CO},
       adsurl = {https://ui.adsabs.harvard.edu/abs/2024MNRAS.533..572K}
}

@ARTICLE{gilessdss,
       author = {{Giles}, P.~A. and {Romer}, A.~K. and {Wilkinson}, R. and {Bermeo}, A. and {Turner}, D.~J. and {Hilton}, M. and {Upsdell}, E.~W. and {Rooney}, P.~J. and {Bhargava}, S. and {Ebrahimpour}, L. and {Farahi}, A. and {Mann}, R.~G. and {Manolopoulou}, M. and {Mayers}, J. and {Vergara}, C. and {Viana}, P.~T.~P. and {Collins}, C.~A. and {Hollowood}, D. and {Jeltema}, T. and {Miller}, C.~J. and {Nichol}, R.~C. and {Noorali}, R. and {Splettstoesser}, M. and {Stott}, J.~P.},
        title = "{The XMM Cluster Survey analysis of the SDSS DR8 redMaPPer catalogue: implications for scatter, selection bias, and isotropy in cluster scaling relations}",
      journal = {\mnras},
         year = 2022,
        month = nov,
       volume = {516},
       number = {3},
        pages = {3878-3899},
          doi = {10.1093/mnras/stac2414},
archivePrefix = {arXiv},
       eprint = {2202.11107},
 primaryClass = {astro-ph.CO},
       adsurl = {https://ui.adsabs.harvard.edu/abs/2022MNRAS.516.3878G}
}

@ARTICLE{2018MNRAS.477.5517W,
       author = {{Willis}, J.~P. and {Ramos-Ceja}, M.~E. and {Muzzin}, A. and {Pacaud}, F. and {Yee}, H.~K.~C. and {Wilson}, G.},
        title = "{X-ray versus infrared selection of distant galaxy clusters: a case study using the XMM-LSS and SpARCS cluster samples}",
      journal = {\mnras},
         year = 2018,
        month = jul,
       volume = {477},
       number = {4},
        pages = {5517-5535},
          doi = {10.1093/mnras/sty975},
archivePrefix = {arXiv},
       eprint = {1804.06475},
 primaryClass = {astro-ph.CO},
       adsurl = {https://ui.adsabs.harvard.edu/abs/2018MNRAS.477.5517W}
}

@ARTICLE{2025A&A...701A.283B,
       author = {{Balzer}, F. and {Bulbul}, E. and {Kluge}, M. and {Liu}, A. and {Salvato}, M. and {Fabricius}, M. and {Seppi}, R. and {Artis}, E. and {Bahar}, Y.~E. and {Bender}, R. and {Clerc}, N. and {Comparat}, J. and {Ghirardini}, V. and {Grandis}, S. and {Krippendorf}, S. and {Lamer}, G. and {Malavasi}, N. and {Merloni}, A. and {Nandra}, K. and {Ramos-Ceja}, M.~E. and {Sanders}, J.~S. and {Zelmer}, S. and {Zhang}, X.},
        title = "{The first SRG/eROSITA All-Sky Survey: Characterization of clusters of galaxies misclassified in the eRASS1 point source catalog}",
      journal = {\aap},
         year = 2025,
        month = sep,
       volume = {701},
          eid = {A283},
        pages = {A283},
          doi = {10.1051/0004-6361/202553942},
archivePrefix = {arXiv},
       eprint = {2501.17238},
 primaryClass = {astro-ph.GA},
       adsurl = {https://ui.adsabs.harvard.edu/abs/2025A&A...701A.283B}
}

@ARTICLE{upsdelldes,
       author = {{Upsdell}, E.~W. and {Giles}, P.~A. and {Romer}, A.~K. and {Wilkinson}, R. and {Turner}, D.~J. and {Hilton}, M. and {Rykoff}, E. and {Farahi}, A. and {Bhargava}, S. and {Jeltema}, T. and {Klein}, M. and {Bermeo}, A. and {Collins}, C.~A. and {Ebrahimpour}, L. and {Hollowood}, D. and {Mann}, R.~G. and {Manolopoulou}, M. and {Miller}, C.~J. and {Rooney}, P.~J. and {Sahl{\'e}n}, Martin and {Stott}, J.~P. and {Viana}, P.~T.~P. and {Allam}, S. and {Alves}, O. and {Bacon}, D. and {Bertin}, E. and {Bocquet}, S. and {Brooks}, D. and {Burke}, D.~L. and {Carrasco Kind}, M. and {Carretero}, J. and {Costanzi}, M. and {da Costa}, L.~N. and {Pereira}, M.~E.~S. and {De Vicente}, J. and {Desai}, S. and {Diehl}, H.~T. and {Dietrich}, J.~P. and {Everett}, S. and {Ferrero}, I. and {Frieman}, J. and {Garc{\'\i}a-Bellido}, J. and {Gerdes}, D.~W. and {Gutierrez}, G. and {Hinton}, S.~R. and {Honscheid}, K. and {James}, D.~J. and {Kuehn}, K. and {Kuropatkin}, N. and {Lima}, M. and {Marshall}, J.~L. and {Mena-Fern{\'a}ndez}, J. and {Menanteau}, F. and {Miquel}, R. and {Mohr}, J.~J. and {Ogando}, R.~L.~C. and {Pieres}, A. and {Raveri}, M. and {Rodriguez-Monroy}, M. and {Sanchez}, E. and {Scarpine}, V. and {Sevilla-Noarbe}, I. and {Smith}, M. and {Suchyta}, E. and {Swanson}, M.~E.~C. and {Tarle}, G. and {To}, C. and {Weaverdyck}, N. and {Weller}, J. and {Wiseman}, P.},
        title = "{The XMM cluster survey: exploring scaling relations and completeness of the dark energy survey year 3 redMaPPer cluster catalogue}",
      journal = {\mnras},
         year = 2023,
        month = jul,
       volume = {522},
       number = {4},
        pages = {5267-5290},
          doi = {10.1093/mnras/stad1220},
archivePrefix = {arXiv},
       eprint = {2304.13570},
 primaryClass = {astro-ph.CO},
       adsurl = {https://ui.adsabs.harvard.edu/abs/2023MNRAS.522.5267U}
}

@ARTICLE{andreon22,
       author = {{Andreon}, S. and {Trinchieri}, G. and {Moretti}, A.},
        title = "{Low X-ray surface brightness clusters: implications on the scatter of the M-T and L-T relations}",
      journal = {\mnras},
         year = 2022,
        month = apr,
       volume = {511},
       number = {4},
        pages = {4991-4998},
          doi = {10.1093/mnras/stac307},
archivePrefix = {arXiv},
       eprint = {2203.11961},
 primaryClass = {astro-ph.CO},
       adsurl = {https://ui.adsabs.harvard.edu/abs/2022MNRAS.511.4991A}
}

@ARTICLE{andreon16,
       author = {{Andreon}, S. and {Serra}, Ana Laura and {Moretti}, A. and {Trinchieri}, G.},
        title = "{The amazing diversity in the hot gas content of an X-ray unbiased massive galaxy clusters sample}",
      journal = {\aap},
         year = 2016,
        month = jan,
       volume = {585},
          eid = {A147},
        pages = {A147},
          doi = {10.1051/0004-6361/201527408},
archivePrefix = {arXiv},
       eprint = {1511.04195},
 primaryClass = {astro-ph.CO},
       adsurl = {https://ui.adsabs.harvard.edu/abs/2016A&A...585A.147A}
}

@ARTICLE{2024A&A...689A.298G,
       author = {{Ghirardini}, V. and {Bulbul}, E. and {Artis}, E. and {Clerc}, N. and {Garrel}, C. and {Grandis}, S. and {Kluge}, M. and {Liu}, A. and {Bahar}, Y.~E. and {Balzer}, F. and {Chiu}, I. and {Comparat}, J. and {Gruen}, D. and {Kleinebreil}, F. and {Krippendorf}, S. and {Merloni}, A. and {Nandra}, K. and {Okabe}, N. and {Pacaud}, F. and {Predehl}, P. and {Ramos-Ceja}, M.~E. and {Reiprich}, T.~H. and {Sanders}, J.~S. and {Schrabback}, T. and {Seppi}, R. and {Zelmer}, S. and {Zhang}, X. and {Bornemann}, W. and {Brunner}, H. and {Burwitz}, V. and {Coutinho}, D. and {Dennerl}, K. and {Freyberg}, M. and {Friedrich}, S. and {Gaida}, R. and {Gueguen}, A. and {Haberl}, F. and {Kink}, W. and {Lamer}, G. and {Li}, X. and {Liu}, T. and {Maitra}, C. and {Meidinger}, N. and {Mueller}, S. and {Miyatake}, H. and {Miyazaki}, S. and {Robrade}, J. and {Schwope}, A. and {Stewart}, I.},
        title = "{The SRG/eROSITA all-sky survey: Cosmology constraints from cluster abundances in the western Galactic hemisphere}",
      journal = {\aap},
         year = 2024,
        month = sep,
       volume = {689},
          eid = {A298},
        pages = {A298},
          doi = {10.1051/0004-6361/202348852},
archivePrefix = {arXiv},
       eprint = {2402.08458},
 primaryClass = {astro-ph.CO},
       adsurl = {https://ui.adsabs.harvard.edu/abs/2024A&A...689A.298G}
}

@ARTICLE{2025PhRvD.112h3535A,
       author = {{Abbott}, T.~M.~C. and {Aguena}, M. and {Alarcon}, A. and {Amon}, A. and {Anbajagane}, D. and {Andrade-Oliveira}, F. and {Avila}, S. and {Allam}, S.~S. and {Bacon}, D. and {Becker}, M.~R. and {Bhargava}, S. and {Blazek}, J. and {Bocquet}, S. and {Brooks}, D. and {Rosell}, A. Carnero and {Carretero}, J. and {Castander}, F.~J. and {Chang}, C. and {Choi}, A. and {Conselice}, C. and {Costanzi}, M. and {Crocce}, M. and {da Costa}, L.~N. and {Pereira}, M.~E.~S. and {Doux}, C. and {Davis}, T.~M. and {Desai}, S. and {Diehl}, H.~T. and {Dodelson}, S. and {Doel}, P. and {Elvin-Poole}, J. and {Esteves}, J. and {Everett}, S. and {Farahi}, A. and {Fert{\'e}}, A. and {Flaugher}, B. and {Garc{\'\i}a-Bellido}, J. and {Gatti}, M. and {Giannini}, G. and {Giles}, P. and {Grandis}, S. and {Gruen}, D. and {Gruendl}, R.~A. and {Gutierrez}, G. and {Harrison}, I. and {Hinton}, S.~R. and {Hollowood}, D.~L. and {Honscheid}, K. and {Jeffrey}, N. and {Jeltema}, T. and {Krause}, E. and {Lahav}, O. and {Lee}, S. and {Lidman}, C. and {Lima}, M. and {Lin}, H. and {Mohr}, J.~J. and {Marshall}, J.~L. and {McCullough}, J. and {Mena-Fern'andez}, J. and {Miquel}, R. and {Muir}, J. and {Myles}, J. and {Ogando}, R.~L.~C. and {Palmese}, A. and {Paterno}, M. and {Malag{\'o}n}, A.~A. Plazas and {Porredon}, A. and {Prat}, J. and {Romer}, A.~K. and {Roodman}, A. and {Rozo}, E. and {Rykoff}, E.~S. and {Rosenfeld}, R. and {Sanchez}, E. and {Cid}, D. Sanchez and {Sevilla-Noarbe}, I. and {Smith}, M. and {Suchyta}, E. and {Shin}, Tae-hyeon and {Tarle}, G. and {Thomas}, D. and {To}, Chun-Hao and {Troxel}, M.~A. and {Vikram}, V. and {Walker}, A.~R. and {Weinberg}, David H. and {Weaverdyck}, N. and {Wechsler}, R.~H. and {Weller}, J. and {Wu}, H.-Y. and {Yamamoto}, M. and {Yanny}, B. and {Zhang}, Y. and {Zhou}, C. and {DES Collaboration}},
        title = "{Dark energy survey year 3 results: Cosmological constraints from cluster abundances, weak lensing, and galaxy clustering}",
      journal = {\prd},
         year = 2025,
        month = oct,
       volume = {112},
       number = {8},
          eid = {083535},
        pages = {083535},
          doi = {10.1103/3dzh-d8f5},
archivePrefix = {arXiv},
       eprint = {2503.13632},
 primaryClass = {astro-ph.CO},
       adsurl = {https://ui.adsabs.harvard.edu/abs/2025PhRvD.112h3535A}
}

@ARTICLE{2009ARA&A..47..481A,
       author = {{Asplund}, Martin and {Grevesse}, Nicolas and {Sauval}, A. Jacques and {Scott}, Pat},
        title = "{The Chemical Composition of the Sun}",
      journal = {\araa},
         year = 2009,
        month = sep,
       volume = {47},
       number = {1},
        pages = {481-522},
          doi = {10.1146/annurev.astro.46.060407.145222},
archivePrefix = {arXiv},
       eprint = {0909.0948},
 primaryClass = {astro-ph.SR},
       adsurl = {https://ui.adsabs.harvard.edu/abs/2009ARA&A..47..481A}
}

@ARTICLE{2026A&A...711A..34E,
       author = {{Euclid Collaboration} and {Bhargava}, S. and {Benoist}, C. and {Gonzalez}, A.~H. and {Maturi}, M. and {Melin}, J.-B. and {Stanford}, S.~A. and {Munari}, E. and {Vannier}, M. and {Murray}, C. and {Maurogordato}, S. and {Biviano}, A. and {Macias-Perez}, J. and {Bartlett}, J.~G. and {Pacaud}, F. and {Widmer}, A. and {Meneghetti}, M. and {Sartoris}, B. and {Aguena}, M. and {Alguero}, G. and {Andreon}, S. and {Bardelli}, S. and {Baumont}, L. and {Bolzonella}, M. and {Cabanac}, R. and {Cappi}, A. and {Castignani}, G. and {Combet}, C. and {Comparat}, J. and {Farrens}, S. and {Ghaffari}, Z. and {Giles}, P.~A. and {Giocoli}, C. and {Girardi}, M. and {Hatch}, N.~A. and {Ingoglia}, L. and {Iovino}, A. and {Mamon}, G.~A. and {Mei}, S. and {Moscardini}, L. and {Mourre}, S. and {Odier}, J. and {Pratt}, G.~W. and {Tarr{\'\i}o}, P. and {Toni}, G. and {Weller}, J. and {Zucca}, E. and {Aghanim}, N. and {Altieri}, B. and {Amara}, A. and {Auricchio}, N. and {Baccigalupi}, C. and {Baldi}, M. and {Balestra}, A. and {Battaglia}, P. and {Bellagamba}, F. and {Bonchi}, A. and {Branchini}, E. and {Brescia}, M. and {Brinchmann}, J. and {Camera}, S. and {Capobianco}, V. and {Carbone}, C. and {Carretero}, J. and {Casas}, S. and {Castellano}, M. and {Cavuoti}, S. and {Chambers}, K.~C. and {Cimatti}, A. and {Colodro-Conde}, C. and {Congedo}, G. and {Conselice}, C.~J. and {Conversi}, L. and {Copin}, Y. and {Courbin}, F. and {Courtois}, H.~M. and {Cropper}, M. and {Da Silva}, A. and {Degaudenzi}, H. and {De Lucia}, G. and {Di Giorgio}, A.~M. and {Dolding}, C. and {Dole}, H. and {Dubath}, F. and {Duncan}, C.~A.~J. and {Dupac}, X. and {Dusini}, S. and {Ealet}, A. and {Escoffier}, S. and {Farina}, M. and {Farinelli}, R. and {Faustini}, F. and {Ferriol}, S. and {Finelli}, F. and {Fosalba}, P. and {Fotopoulou}, S. and {Frailis}, M. and {Franceschi}, E. and {Fumana}, M. and {Galeotta}, S. and {George}, K. and {Gillis}, B. and {G{\'o}mez-Alvarez}, P. and {Gracia-Carpio}, J. and {Granett}, B.~R. and {Grazian}, A. and {Grupp}, F. and {Guzzo}, L. and {Gwyn}, S. and {Haugan}, S.~V.~H. and {Hoar}, J. and {Hoekstra}, H. and {Holmes}, W. and {Hormuth}, F. and {Hornstrup}, A. and {Hudelot}, P. and {Ili{\'c}}, S. and {Jahnke}, K. and {Jhabvala}, M. and {Joachimi}, B. and {Keih{\"a}nen}, E. and {Kermiche}, S. and {Kiessling}, A. and {Kilbinger}, M. and {Kubik}, B. and {Kuijken}, K. and {K{\"u}mmel}, M. and {Kunz}, M. and {Kurki-Suonio}, H. and {Le Boulc'h}, Q. and {Le Brun}, A.~M.~C. and {Le Mignant}, D. and {Liebing}, P. and {Ligori}, S. and {Lilje}, P.~B. and {Lindholm}, V. and {Lloro}, I. and {Mainetti}, G. and {Maino}, D. and {Maiorano}, E. and {Mansutti}, O. and {Marcin}, S. and {Marggraf}, O. and {Martinelli}, M. and {Martinet}, N. and {Marulli}, F. and {Massey}, R. and {Medinaceli}, E. and {Melchior}, M. and {Mellier}, Y. and {Merlin}, E. and {Meylan}, G. and {Mora}, A. and {Moresco}, M. and {Nakajima}, R. and {Neissner}, C. and {Nichol}, R.~C. and {Niemi}, S.-M. and {Nightingale}, J.~W. and {Padilla}, C. and {Paltani}, S. and {Pasian}, F. and {Pedersen}, K. and {Percival}, W.~J. and {Pettorino}, V. and {Pires}, S. and {Polenta}, G. and {Poncet}, M. and {Popa}, L.~A. and {Pozzetti}, L. and {Raison}, F. and {Rebolo}, R. and {Renzi}, A. and {Rhodes}, J. and {Riccio}, G. and {Romelli}, E. and {Roncarelli}, M. and {Saglia}, R. and {Sakr}, Z. and {S{\'a}nchez}, A.~G. and {Sapone}, D. and {Schewtschenko}, J.~A. and {Schneider}, P. and {Schrabback}, T. and {Secroun}, A. and {Sefusatti}, E. and {Seidel}, G. and {Serrano}, S. and {Simon}, P. and {Sirignano}, C. and {Sirri}, G. and {Skottfelt}, J. and {Stanco}, L. and {Steinwagner}, J. and {Tallada-Cresp{\'\i}}, P. and {Taylor}, A.~N. and {Tereno}, I. and {Toft}, S. and {Toledo-Moreo}, R. and {Torradeflot}, F.},
        title = "{Euclid Quick Data Release (Q1): XXXIV. First detections from the Euclid galaxy cluster workflow}",
      journal = {\aap},
         year = 2026,
        month = jun,
       volume = {711},
          eid = {A34},
        pages = {A34},
          doi = {10.1051/0004-6361/202554937},
archivePrefix = {arXiv},
       eprint = {2503.19196},
 primaryClass = {astro-ph.CO},
       adsurl = {https://ui.adsabs.harvard.edu/abs/2026A&A...711A..34E}
}

@ARTICLE{2018MNRAS.474.1065S,
       author = {{Sanders}, J.~S. and {Fabian}, A.~C. and {Russell}, H.~R. and {Walker}, S.~A.},
        title = "{Hydrostatic Chandra X-ray analysis of SPT-selected galaxy clusters - I. Evolution of profiles and core properties}",
      journal = {\mnras},
         year = 2018,
        month = feb,
       volume = {474},
       number = {1},
        pages = {1065-1098},
          doi = {10.1093/mnras/stx2796},
archivePrefix = {arXiv},
       eprint = {1705.09299},
 primaryClass = {astro-ph.CO},
       adsurl = {https://ui.adsabs.harvard.edu/abs/2018MNRAS.474.1065S}
}

@ARTICLE{2021ApJS..256...21N,
       author = {{Ni}, Qingling and {Brandt}, W.~N. and {Chen}, Chien-Ting and {Luo}, Bin and {Nyland}, Kristina and {Yang}, Guang and {Zou}, Fan and {Aird}, James and {Alexander}, David M. and {Bauer}, Franz Erik and {Lacy}, Mark and {Lehmer}, Bret D. and {Mallick}, Labani and {Salvato}, Mara and {Schneider}, Donald P. and {Tozzi}, Paolo and {Traulsen}, Iris and {Vaccari}, Mattia and {Vignali}, Cristian and {Vito}, Fabio and {Xue}, Yongquan and {Banerji}, Manda and {Chow}, Kate and {Comastri}, Andrea and {Del Moro}, Agnese and {Gilli}, Roberto and {Mullaney}, James and {Paolillo}, Maurizio and {Schwope}, Axel and {Shemmer}, Ohad and {Sun}, Mouyuan and {Timlin}, III, John D. and {Trump}, Jonathan R.},
        title = "{The XMM-SERVS Survey: XMM-Newton Point-source Catalogs for the W-CDF-S and ELAIS-S1 Fields}",
      journal = {\apjs},
         year = 2021,
        month = sep,
       volume = {256},
       number = {1},
          eid = {21},
        pages = {21},
          doi = {10.3847/1538-4365/ac0dc6},
archivePrefix = {arXiv},
       eprint = {2106.10572},
 primaryClass = {astro-ph.GA},
       adsurl = {https://ui.adsabs.harvard.edu/abs/2021ApJS..256...21N}
}

@ARTICLE{2020MNRAS.499.4768I,
       author = {{Ider Chitham}, J. and {Comparat}, J. and {Finoguenov}, A. and {Clerc}, N. and {Kirkpatrick}, C. and {Damsted}, S. and {Kukkola}, A. and {Capasso}, R. and {Nandra}, K. and {Merloni}, A. and {Bulbul}, E. and {Rykoff}, E.~S. and {Schneider}, D.~P. and {Brownstein}, J.~R.},
        title = "{Cosmological constraints from CODEX galaxy clusters spectroscopically confirmed by SDSS-IV/SPIDERS DR16}",
      journal = {\mnras},
         year = 2020,
        month = dec,
       volume = {499},
       number = {4},
        pages = {4768-4784},
          doi = {10.1093/mnras/staa3044},
       adsurl = {https://ui.adsabs.harvard.edu/abs/2020MNRAS.499.4768I}
}

@ARTICLE{2024A&A...688A.210K,
       author = {{Kluge}, M. and {Comparat}, J. and {Liu}, A. and {Balzer}, F. and {Bulbul}, E. and {Ider Chitham}, J. and {Ghirardini}, V. and {Garrel}, C. and {Bahar}, Y.~E. and {Artis}, E. and {Bender}, R. and {Clerc}, N. and {Dwelly}, T. and {Fabricius}, M.~H. and {Grandis}, S. and {Hern{\'a}ndez-Lang}, D. and {Hill}, G.~J. and {Joshi}, J. and {Lamer}, G. and {Merloni}, A. and {Nandra}, K. and {Pacaud}, F. and {Predehl}, P. and {Ramos-Ceja}, M.~E. and {Reiprich}, T.~H. and {Salvato}, M. and {Sanders}, J.~S. and {Schrabback}, T. and {Seppi}, R. and {Zelmer}, S. and {Zenteno}, A. and {Zhang}, X.},
        title = "{The SRG/eROSITA All-Sky Survey. Optical identification and properties of galaxy clusters and groups in the western galactic hemisphere}",
      journal = {\aap},
         year = 2024,
        month = aug,
       volume = {688},
          eid = {A210},
        pages = {A210},
          doi = {10.1051/0004-6361/202349031},
archivePrefix = {arXiv},
       eprint = {2402.08453},
 primaryClass = {astro-ph.CO},
       adsurl = {https://ui.adsabs.harvard.edu/abs/2024A&A...688A.210K}
}

@ARTICLE{2025arXiv251114356R,
       author = {{Ramos-Ceja}, M.~E. and {Fiorino}, L. and {Bulbul}, E. and {Ghirardini}, V. and {Clerc}, N. and {Liu}, A. and {Sanders}, J.~S. and {Bahar}, Y.~E. and {Dietl}, J. and {Kluge}, M. and {Pacaud}, F. and {Artis}, E. and {Balzer}, F. and {Comparat}, J. and {Ding}, Z. and {Malavasi}, N. and {Merloni}, A. and {Mistele}, T. and {Nandra}, K. and {Seppi}, R. and {Zelmer}, S. and {Zhang}, X.},
        title = "{The SRG/eROSITA all-sky survey: X-ray scaling relations of galaxy groups and clusters in the western Galactic hemisphere}",
      journal = {arXiv e-prints},
         year = 2025,
        month = nov,
          eid = {arXiv:2511.14356},
        pages = {arXiv:2511.14356},
          doi = {10.48550/arXiv.2511.14356},
archivePrefix = {arXiv},
       eprint = {2511.14356},
 primaryClass = {astro-ph.CO},
       adsurl = {https://ui.adsabs.harvard.edu/abs/2025arXiv251114356R}
}

@ARTICLE{2022A&A...661A...1B,
       author = {{Brunner}, H. and {Liu}, T. and {Lamer}, G. and {Georgakakis}, A. and {Merloni}, A. and {Brusa}, M. and {Bulbul}, E. and {Dennerl}, K. and {Friedrich}, S. and {Liu}, A. and {Maitra}, C. and {Nandra}, K. and {Ramos-Ceja}, M.~E. and {Sanders}, J.~S. and {Stewart}, I.~M. and {Boller}, T. and {Buchner}, J. and {Clerc}, N. and {Comparat}, J. and {Dwelly}, T. and {Eckert}, D. and {Finoguenov}, A. and {Freyberg}, M. and {Ghirardini}, V. and {Gueguen}, A. and {Haberl}, F. and {Kreykenbohm}, I. and {Krumpe}, M. and {Osterhage}, S. and {Pacaud}, F. and {Predehl}, P. and {Reiprich}, T.~H. and {Robrade}, J. and {Salvato}, M. and {Santangelo}, A. and {Schrabback}, T. and {Schwope}, A. and {Wilms}, J.},
        title = "{The eROSITA Final Equatorial Depth Survey (eFEDS). X-ray catalogue}",
      journal = {\aap},
         year = 2022,
        month = may,
       volume = {661},
          eid = {A1},
        pages = {A1},
          doi = {10.1051/0004-6361/202141266},
archivePrefix = {arXiv},
       eprint = {2106.14517},
 primaryClass = {astro-ph.HE},
       adsurl = {https://ui.adsabs.harvard.edu/abs/2022A&A...661A...1B}
}

@ARTICLE{2025MNRAS.544.2080M,
       author = {{Myles}, J. and {Gruen}, D. and {Jeltema}, T. and {Mantz}, A. and {Allen}, S. and {Fu}, S. and {Kremin}, A. and {Aguilar}, J. and {Ahlen}, S. and {Bianchi}, D. and {Brooks}, D. and {Castander}, F.~J. and {Claybaugh}, T. and {de la Macorra}, A. and {Dey}, Arjun and {Doel}, P. and {Ferraro}, S. and {Forero-Romero}, J.~E. and {Gazta{\~n}aga}, E. and {Gontcho A Gontcho}, S. and {Gutierrez}, G. and {Honscheid}, K. and {Ishak}, M. and {Kehoe}, R. and {Kirkby}, D. and {Kisner}, T. and {Lahav}, O. and {Landriau}, M. and {Le Guillou}, L. and {Manera}, M. and {Meisner}, A. and {Miquel}, R. and {Moustakas}, J. and {Nadathur}, S. and {Newman}, J.~A. and {Palanque-Delabrouille}, N. and {Prada}, F. and {P{\'e}rez-R{\`a}fols}, I. and {Rossi}, G. and {Sanchez}, E. and {Schlegel}, D. and {Schubnell}, M. and {Silber}, J. and {Sprayberry}, D. and {Tarl{\'e}}, G. and {Weaver}, B.~A. and {Zhou}, R.},
        title = "{Spectroscopic Characterization of redMaPPer Galaxy Clusters with DESI}",
      journal = {\mnras},
         year = 2025,
        month = dec,
       volume = {544},
       number = {2},
        pages = {2080-2097},
          doi = {10.1093/mnras/staf1831},
archivePrefix = {arXiv},
       eprint = {2506.06249},
 primaryClass = {astro-ph.CO},
       adsurl = {https://ui.adsabs.harvard.edu/abs/2025MNRAS.544.2080M}
}

@ARTICLE{2025PhRvD.112d3517C,
       author = {{Cao}, Shulei and {Wu}, Hao-Yi and {Costanzi}, Matteo and {Farahi}, Arya and {Grandis}, Sebastian and {Weinberg}, David H. and {Evrard}, August E. and {Rozo}, Eduardo and {Salcedo}, Andr{\'e}s N. and {To}, Chun-Hao and {Yang}, Lei and {Zhou}, Conghao and {DES Collaboration}},
        title = "{Association between optically identified galaxy clusters and the underlying dark matter halos}",
      journal = {\prd},
         year = 2025,
        month = aug,
       volume = {112},
       number = {4},
          eid = {043517},
        pages = {043517},
          doi = {10.1103/r7tt-bzs7},
archivePrefix = {arXiv},
       eprint = {2506.17526},
 primaryClass = {astro-ph.CO},
       adsurl = {https://ui.adsabs.harvard.edu/abs/2025PhRvD.112d3517C}
}

@ARTICLE{2019arXiv191013548K,
       author = {{Kirby}, Matthew and {Rozo}, Eduardo and {Morris}, R. Glenn and {Allen}, Steven W. and {Costanzi}, Matteo and {Jeltema}, Tesla E. and {Mantz}, Adam B. and {Romer}, A. Kathy and {Rykoff}, E.~S. and {von der Linden}, Anja},
        title = "{Improved Cosmological Constraints from SDSS redMaPPer Clusters via X-ray Follow-up of a Complete Subsample of Systems}",
      journal = {arXiv e-prints},
         year = 2019,
        month = oct,
          eid = {arXiv:1910.13548},
        pages = {arXiv:1910.13548},
          doi = {10.48550/arXiv.1910.13548},
archivePrefix = {arXiv},
       eprint = {1910.13548},
 primaryClass = {astro-ph.CO},
       adsurl = {https://ui.adsabs.harvard.edu/abs/2019arXiv191013548K}
}

@ARTICLE{2026OJAp....955863A,
       author = {{Aguena}, M. and {Aiola}, S. and {Allam}, S. and {Andrade-Oliveira}, F. and {Bacon}, D. and {Bahcall}, N. and {Battaglia}, N. and {Battistelli}, E.~S. and {Bocquet}, S. and {Bolliet}, B. and {Bond}, J.~R. and {Brooks}, D. and {Calabrese}, E. and {Carretero}, J. and {Choi}, S.~K. and {da Costa}, L.~N. and {Costanzi}, M. and {Coulton}, W. and {Davis}, T.~M. and {Desai}, S. and {Devlin}, M.~J. and {Dicker}, S. and {Doel}, P. and {Duivenvoorden}, A.~J. and {Dunkley}, J. and {Ferraro}, S. and {Flaugher}, B. and {Frieman}, J. and {Gallardo}, P.~A. and {Gatti}, M. and {Gaztanaga}, E. and {Gill}, A.~S. and {Golec}, J.~E. and {Gruen}, D. and {Gruendl}, R.~A. and {Halpern}, M. and {Hasselfield}, M. and {Hill}, J.~C. and {Hilton}, M. and {Hincks}, A.~D. and {Hinton}, S.~R. and {Hollowood}, D.~L. and {Honscheid}, K. and {Hubmayr}, J. and {Huffenberger}, K.~M. and {Hughes}, J.~P. and {James}, D.~J. and {Klein}, M. and {Knowles}, K. and {Koopman}, B.~J. and {Kosowsky}, A. and {Lahav}, O. and {Lee}, E. and {Lin}, Y. and {Lokken}, M. and {Madhavacheril}, M.~S. and {Malag{\'o}n}, A.~A. Plazas and {Marrewijk}, J. v. and {Marshall}, J.~L. and {McMahon}, J. and {Mena-Fern{\'a}ndez}, J. and {Miquel}, R. and {Miyatake}, H. and {Mohr}, J.~J. and {Moodley}, K. and {Mroczkowski}, T. and {Naess}, S. and {Nati}, F. and {Nicola}, A. and {Niemack}, M.~D. and {Ogando}, R.~L.~C. and {Oguri}, M. and {Orlowski-Scherer}, J. and {Page}, L.~A. and {Partridge}, B. and {da Silva Pereira}, M.~E. and {Porredon}, A. and {Qu}, F.~J. and {Ragavan}, D.~C. and {Guachalla}, B. Ried and {Romer}, A.~K. and {Rosell}, A. Carnero and {Rykoff}, E.~S. and {Samuroff}, S. and {Sanchez}, E. and {Sevilla-Noarbe}, I. and {Sierra}, C. and {Sif{\'o}n}, C. and {Smith}, M. and {Staggs}, S.~T. and {Suchyta}, E. and {Swanson}, M.~E.~C. and {Tucker}, D.~L. and {Vargas}, C. and {Vavagiakis}, E.~M. and {De Vicente}, J. and {Weaverdyck}, N. and {Weller}, J. and {Wollack}, E.~J. and {Zubeldia}, I.},
        title = "{The Atacama Cosmology Telescope: DR6 Sunyaev-Zel'dovich Selected Galaxy Clusters Catalog}",
      journal = {The Open Journal of Astrophysics},
         year = 2026,
        month = jan,
       volume = {9},
        pages = {55863},
          doi = {10.33232/001c.155863},
archivePrefix = {arXiv},
       eprint = {2507.21459},
 primaryClass = {astro-ph.CO},
       adsurl = {https://ui.adsabs.harvard.edu/abs/2026OJAp....955863A}
}


\appendix

\section{Images of the DESY3RM-eRASS-unmatched clusters}
\label{sec:unmatched_images}

Here, we provide images of the 12 DESY3RM-eRASS-unmatched clusters. The images provided are the same as those outlined in Fig.~\ref{fig:cutouts}, namely, a DES optical image (left figures), a soft-band eRASS1 cut-out (middle image), and a DES optical image with cluster members highlighted (right image).

\begin{figure*}
\begin{center}
\begin{tabular}{ccc}
\includegraphics[width=5.4cm]{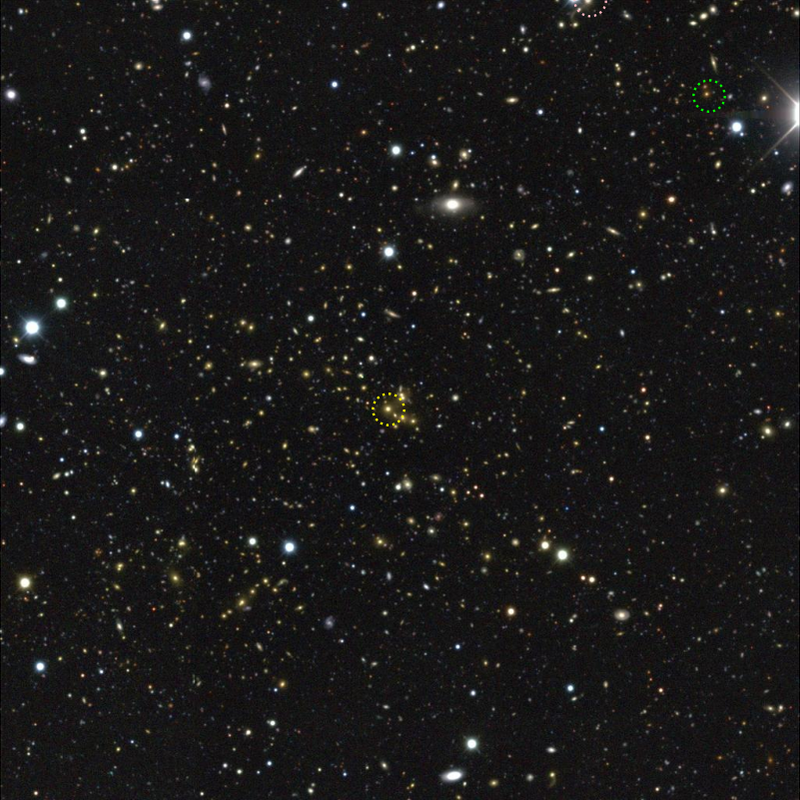} &
\includegraphics[width=5.4cm]{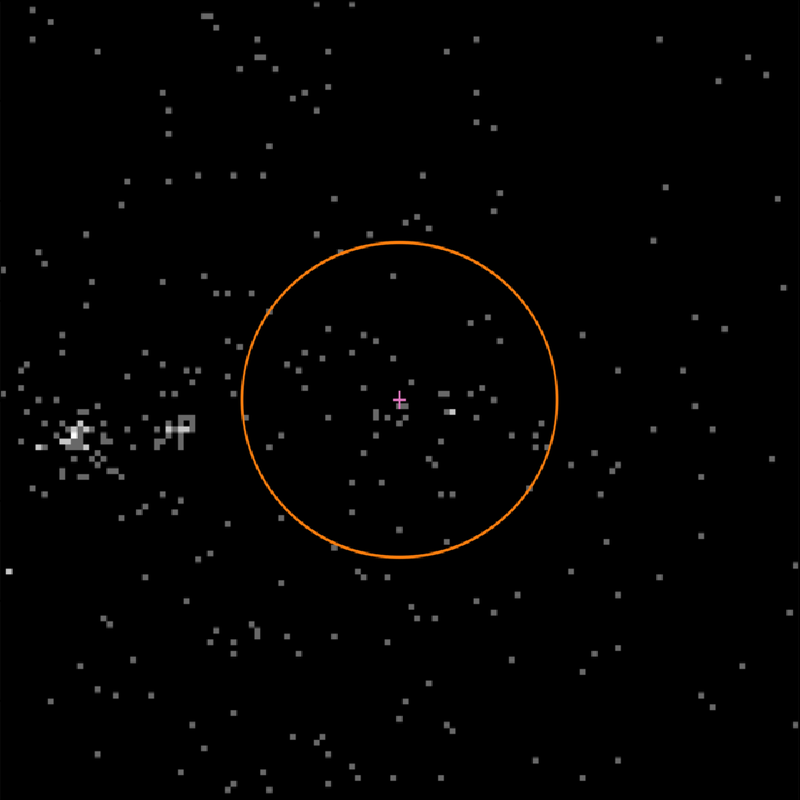} &
\includegraphics[width=5.4cm]{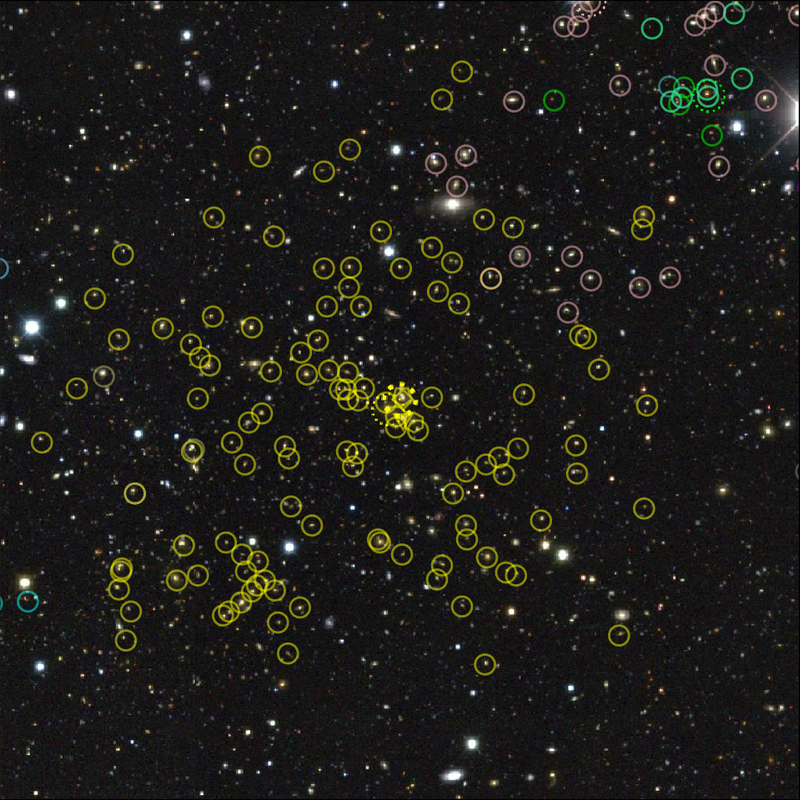} \\
\multicolumn{3}{c}{(a) RM ID 655}\\
\includegraphics[width=5.4cm]{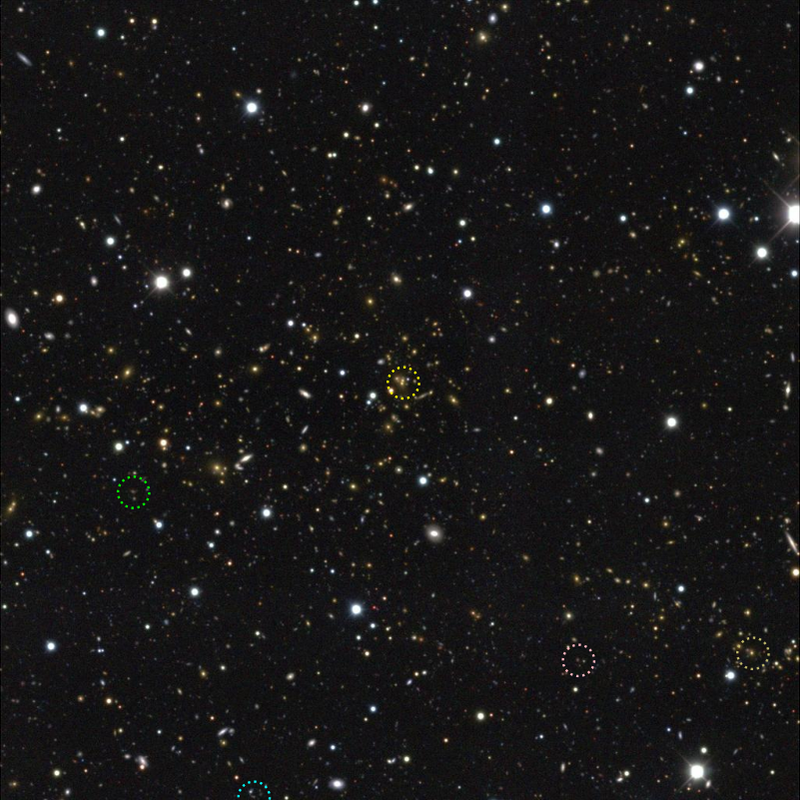} &
\includegraphics[width=5.4cm]{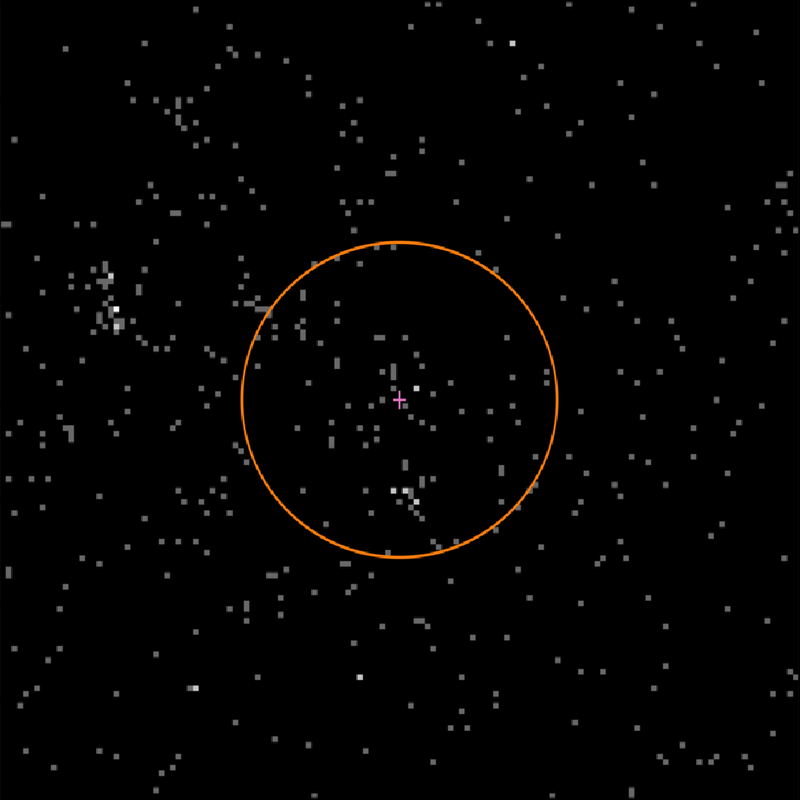} &
\includegraphics[width=5.4cm]{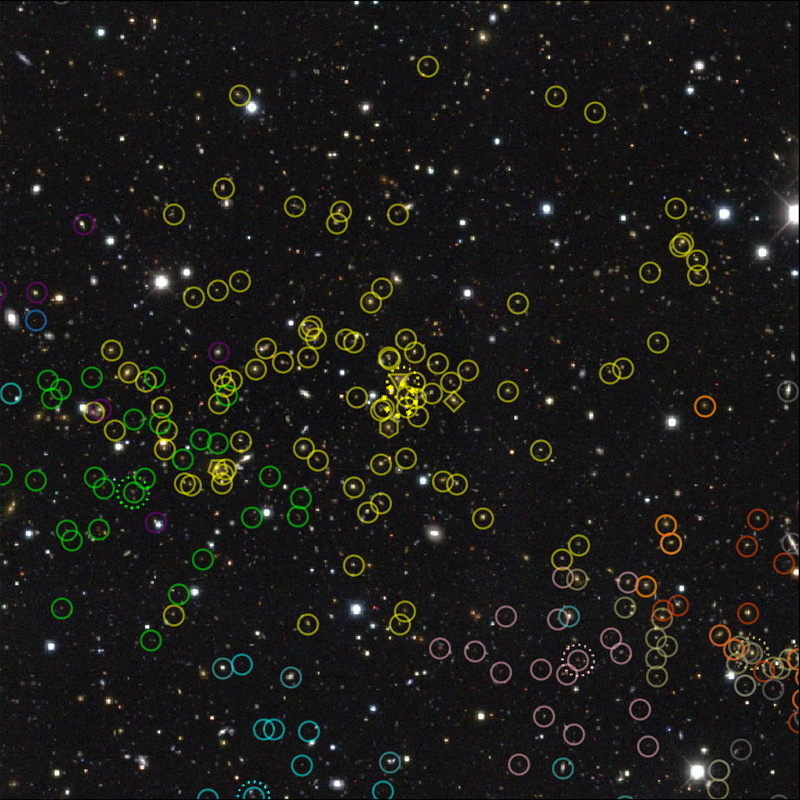} \\
\multicolumn{3}{c}{(b) RM ID 1812}\\
\includegraphics[width=5.4cm]{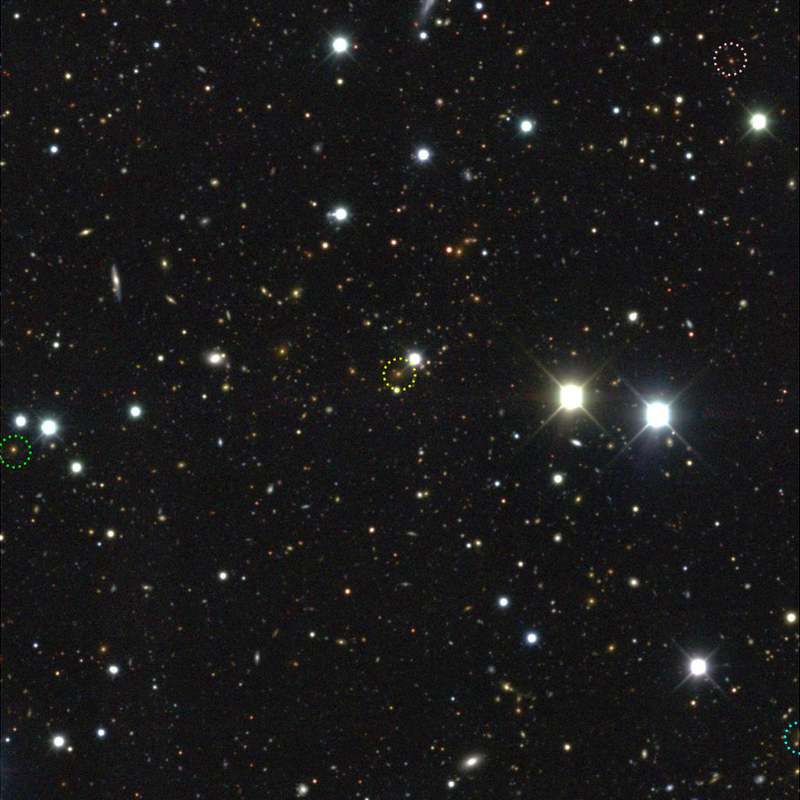} &
\includegraphics[width=5.4cm]{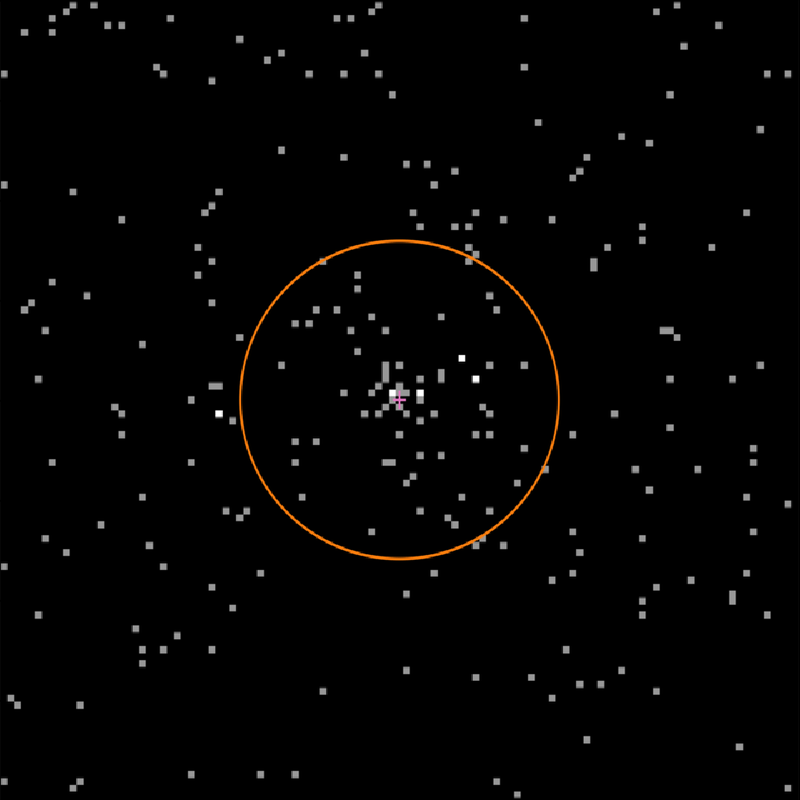} &
\includegraphics[width=5.4cm]{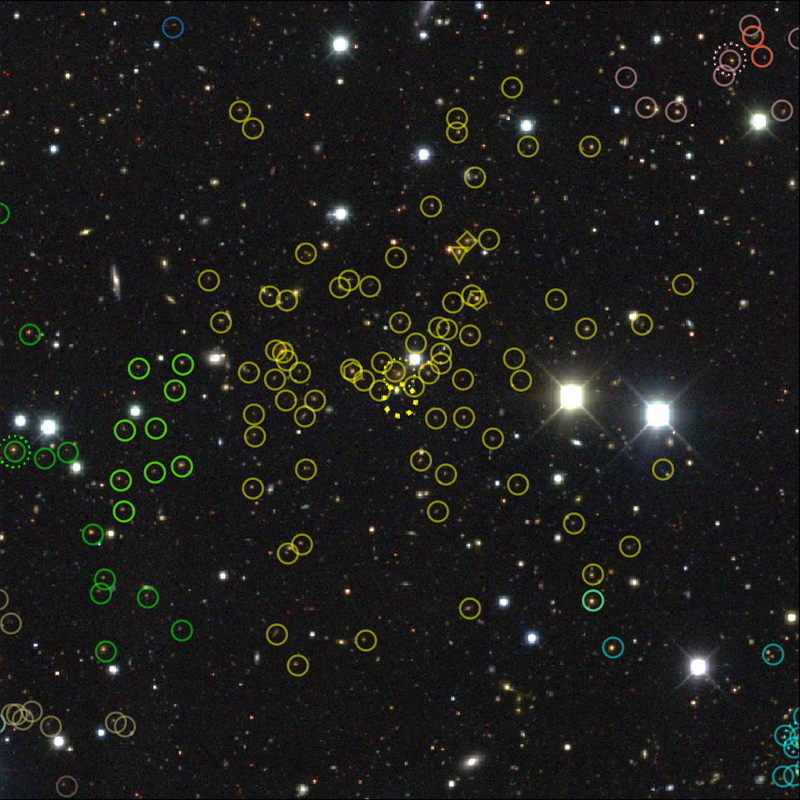} \\
\multicolumn{3}{c}{(c) RM ID 2096}\\
\includegraphics[width=5.4cm]{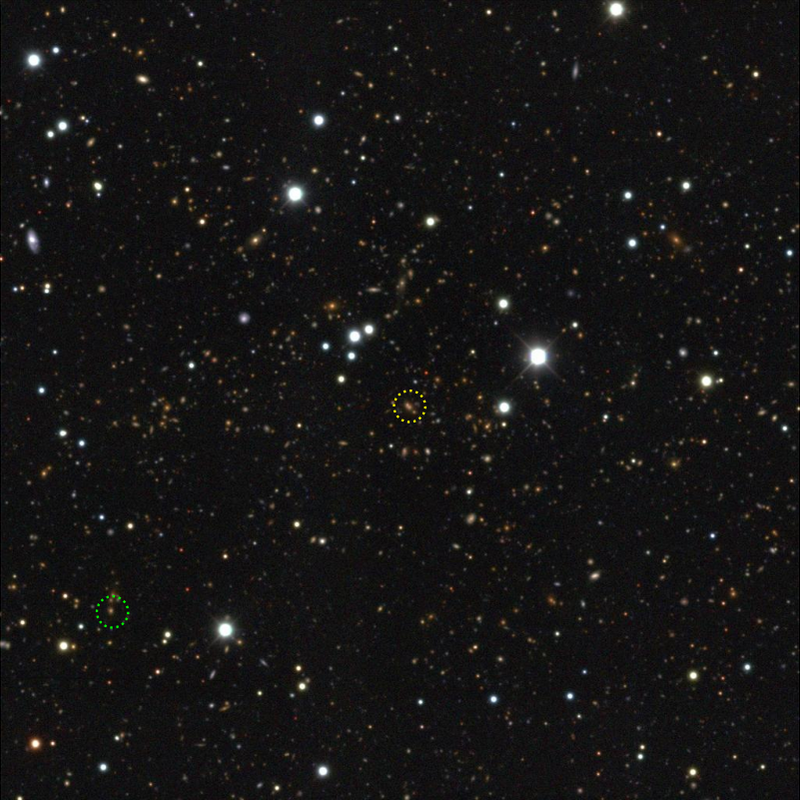} &
\includegraphics[width=5.4cm]{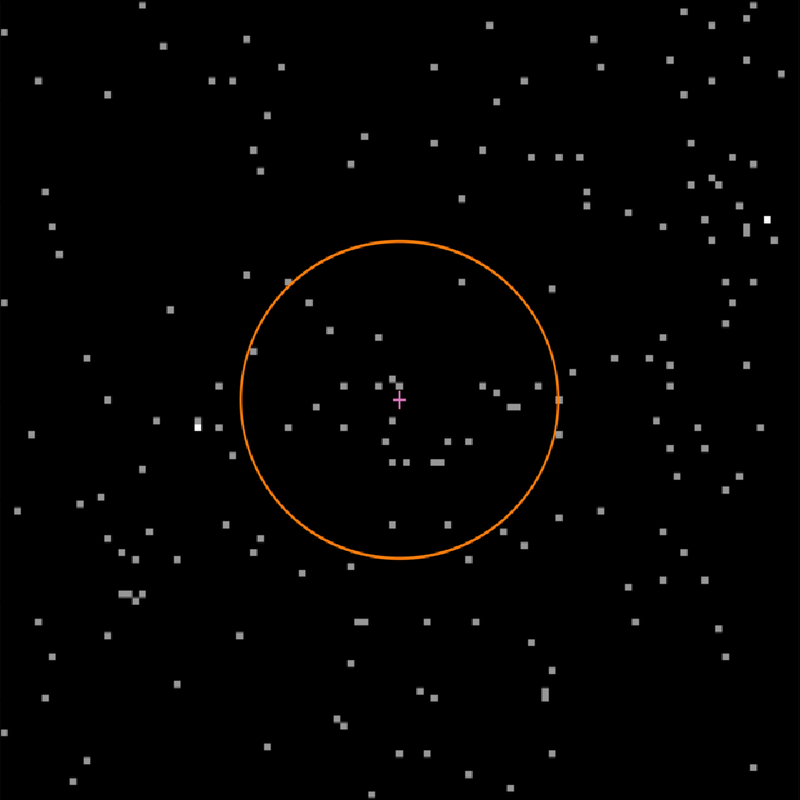} &
\includegraphics[width=5.4cm]{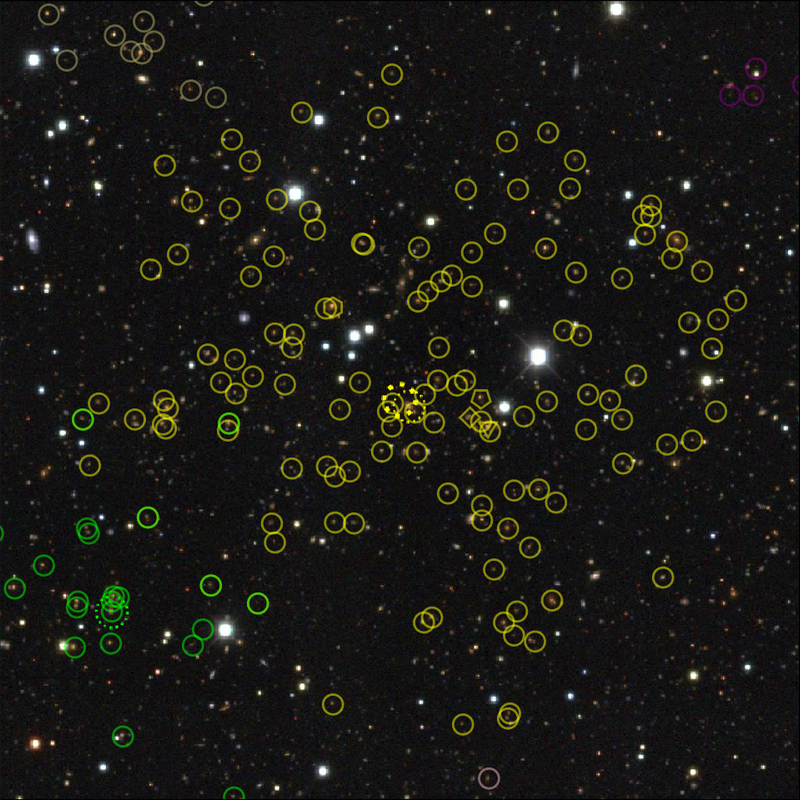} \\
\multicolumn{3}{c}{(d) RM ID 2326}\\
\end{tabular}
\end{center}
\caption[]{Images of the DESY3RM-eRASS-unmatched clusters.  The image descriptions are the same as those in Figure~\ref{fig:cutouts}.}
\label{fig:undet_images}
\end{figure*}

\begin{figure*}
\begin{center}
\begin{tabular}{ccc}
\includegraphics[width=5.4cm]{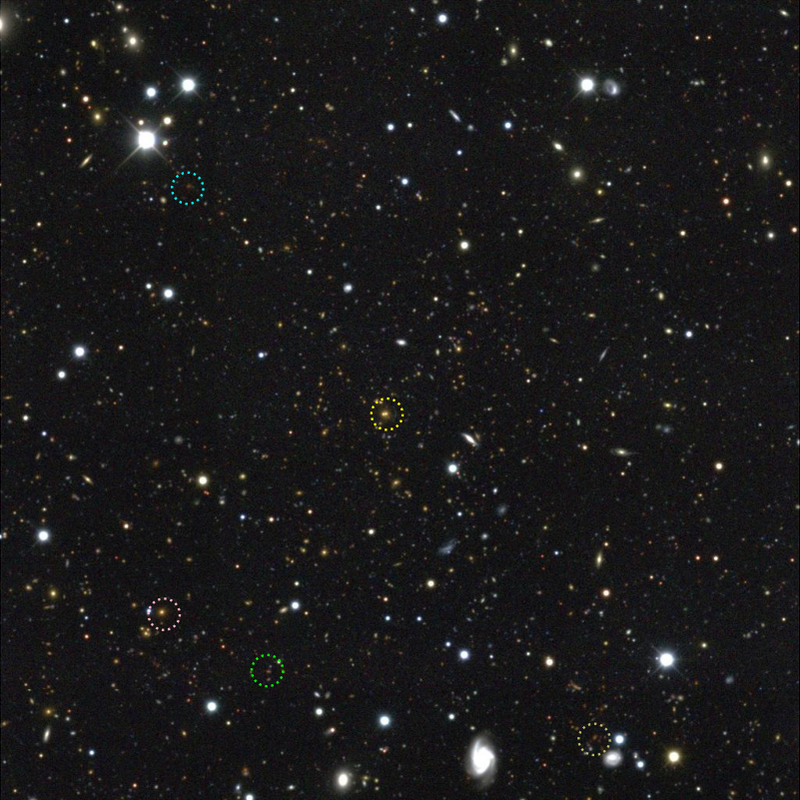} &
\includegraphics[width=5.4cm]{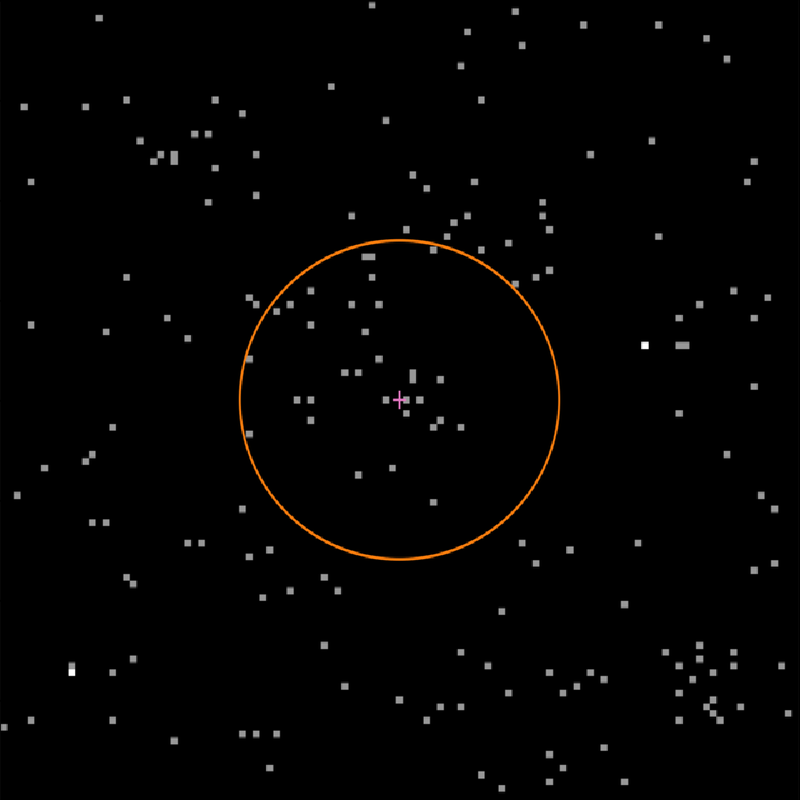} &
\includegraphics[width=5.4cm]{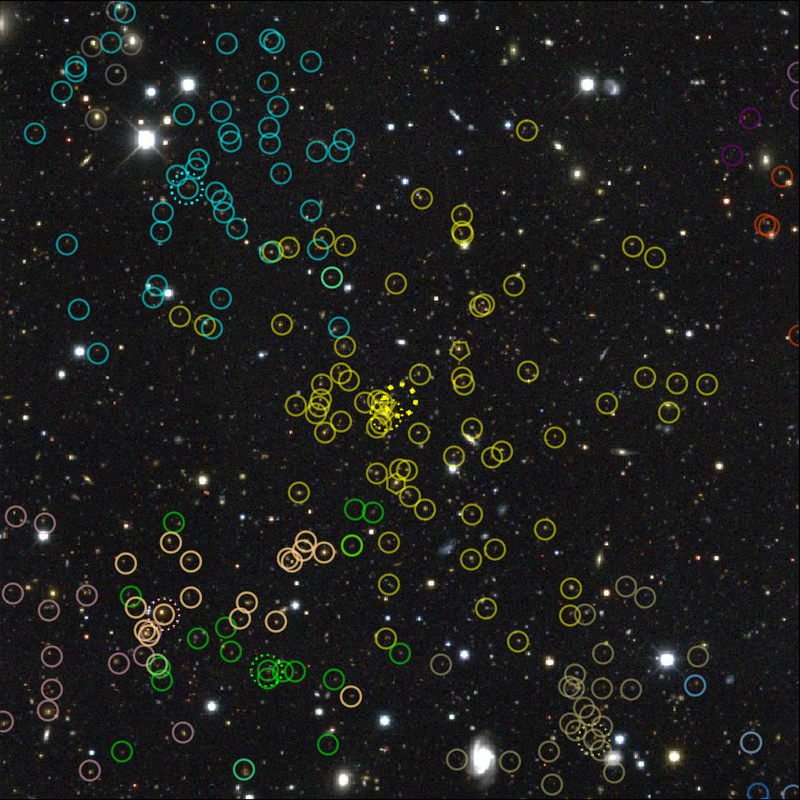} \\
\multicolumn{3}{c}{(e) RM ID 2442}\\
\includegraphics[width=5.4cm]{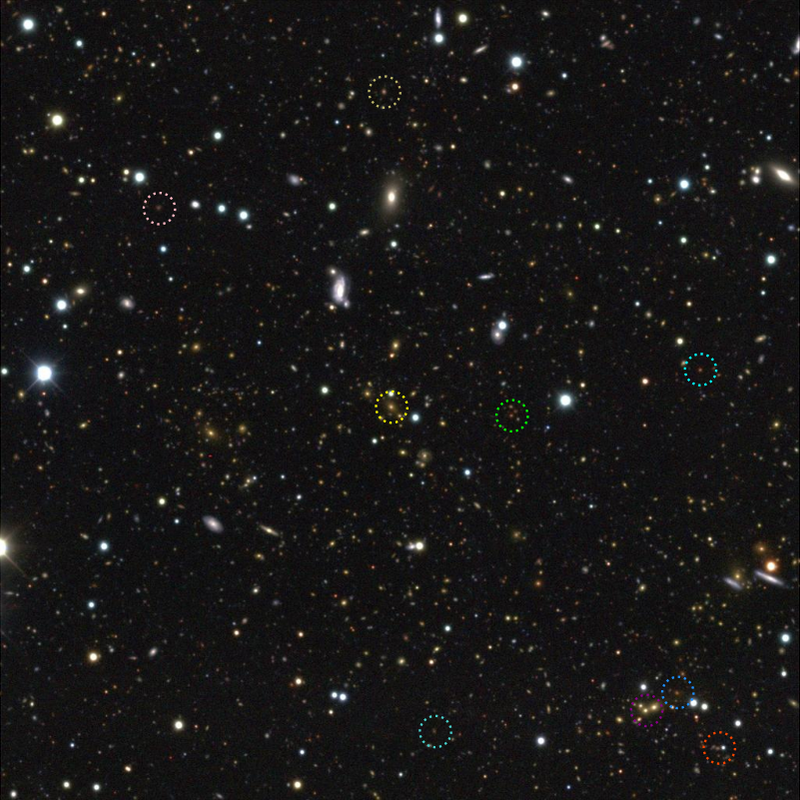} &
\includegraphics[width=5.4cm]{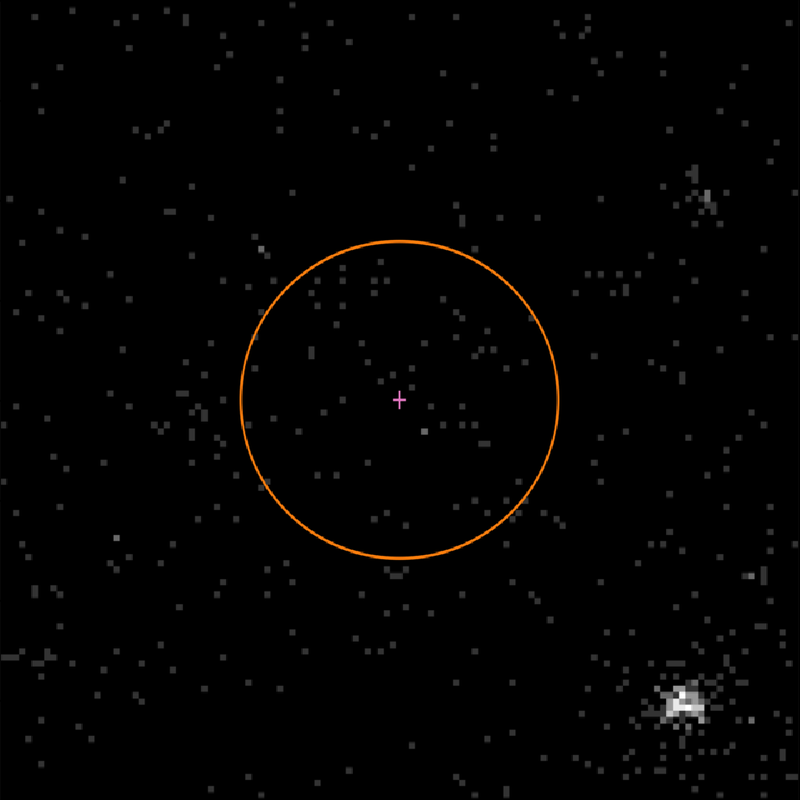} &
\includegraphics[width=5.4cm]{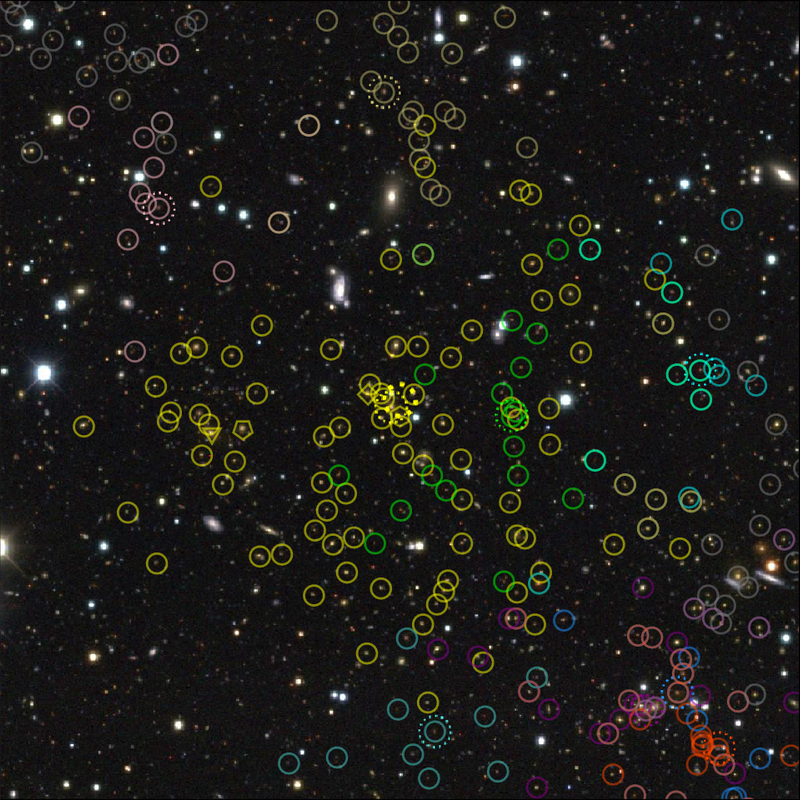} \\
\multicolumn{3}{c}{(f) RM ID 2451}\\
\includegraphics[width=5.4cm]{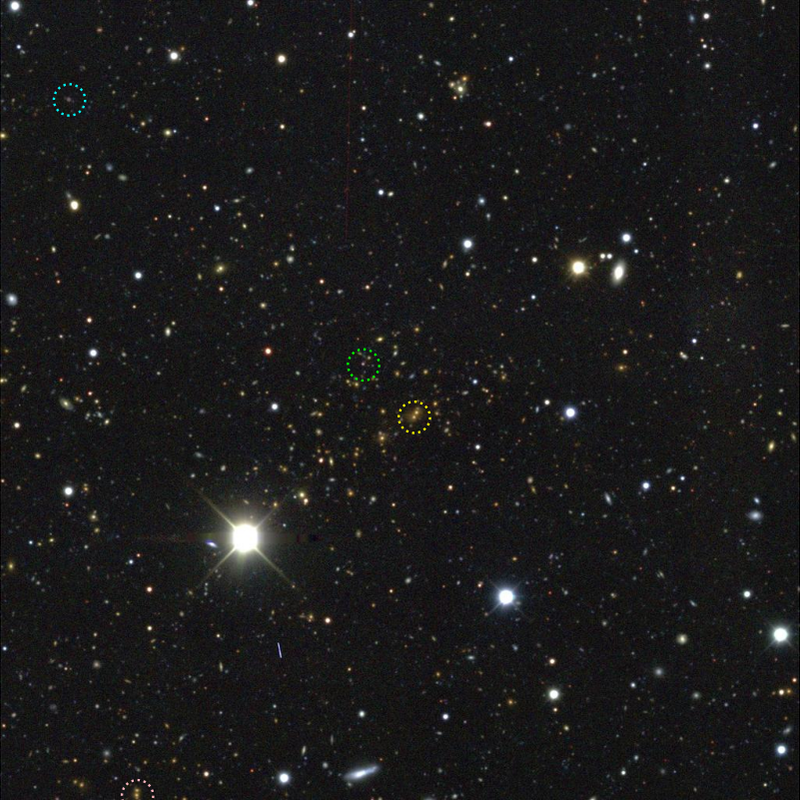} &
\includegraphics[width=5.4cm]{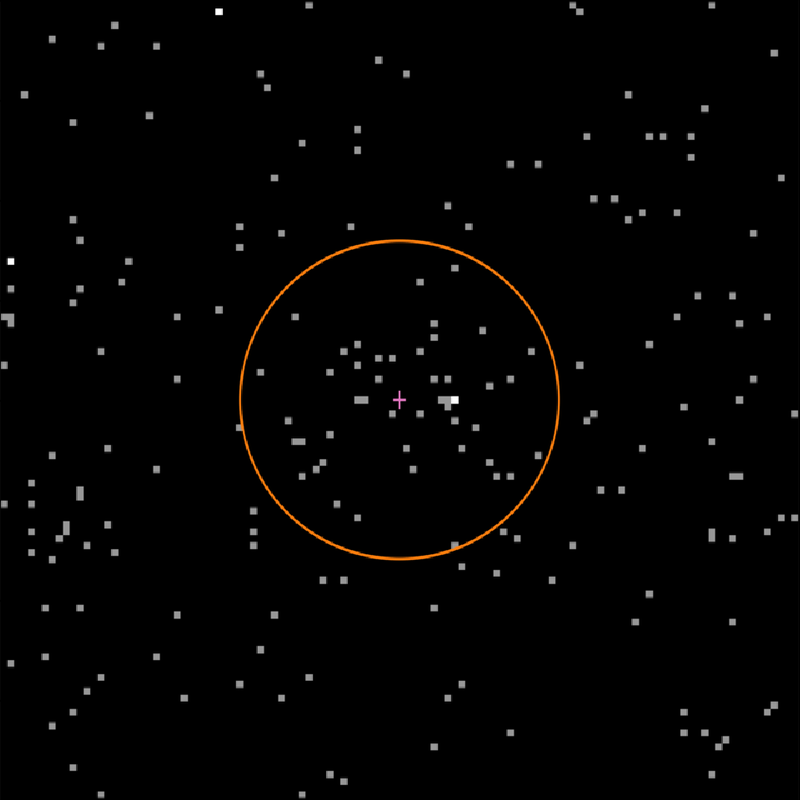} &
\includegraphics[width=5.4cm]{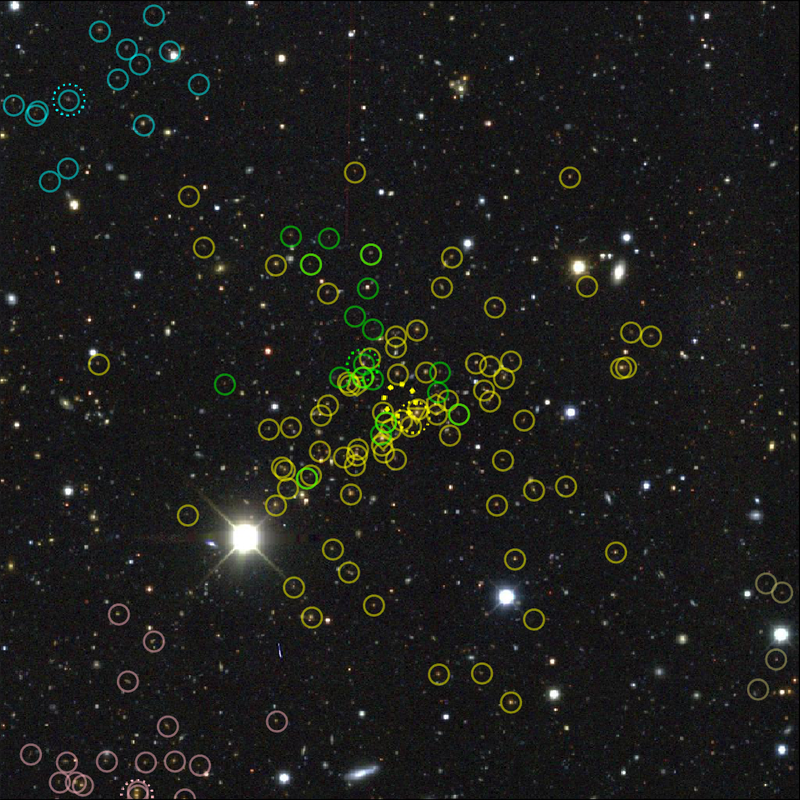} \\
\multicolumn{3}{c}{(g) RM ID 2609}\\
\includegraphics[width=5.4cm]{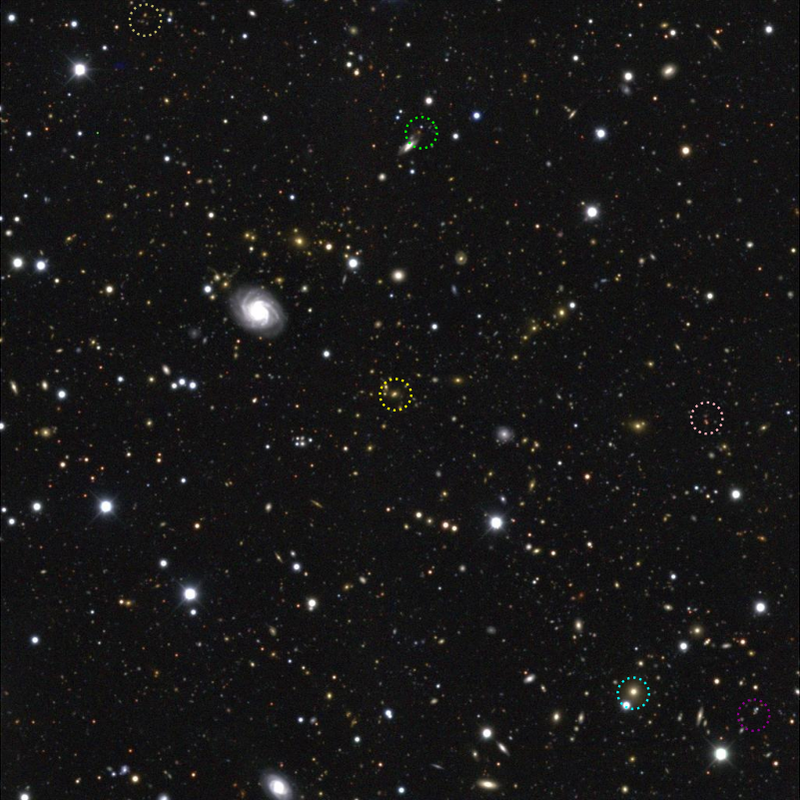} &
\includegraphics[width=5.4cm]{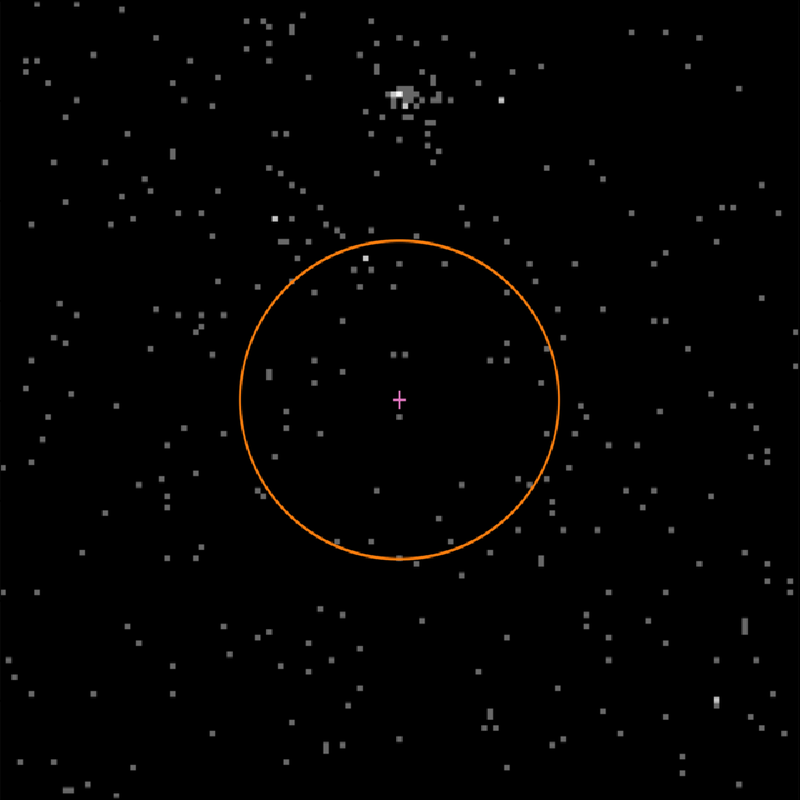} &
\includegraphics[width=5.4cm]{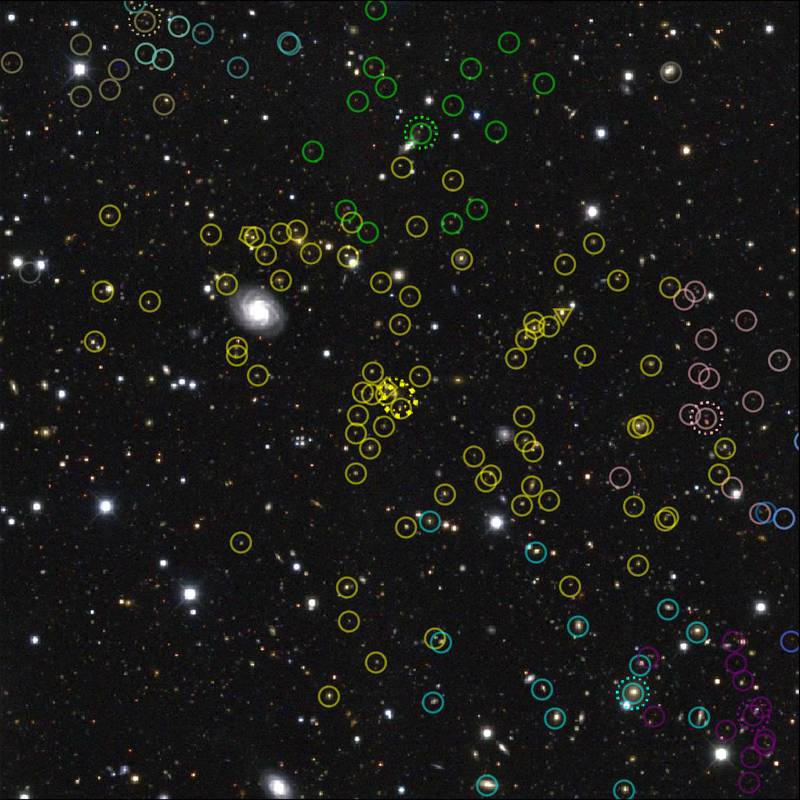} \\
\multicolumn{3}{c}{(h) RM ID 3363}\\
\end{tabular}
\end{center}
\contcaption{}
\end{figure*}

\begin{figure*}
\begin{center}
\begin{tabular}{ccc}
\includegraphics[width=5.4cm]{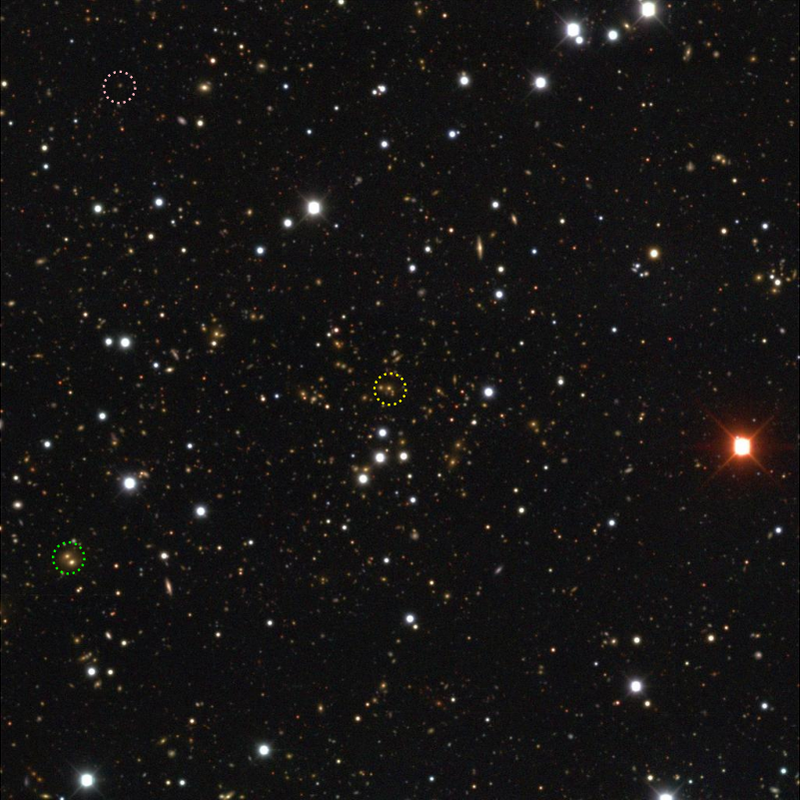} &
\includegraphics[width=5.4cm]{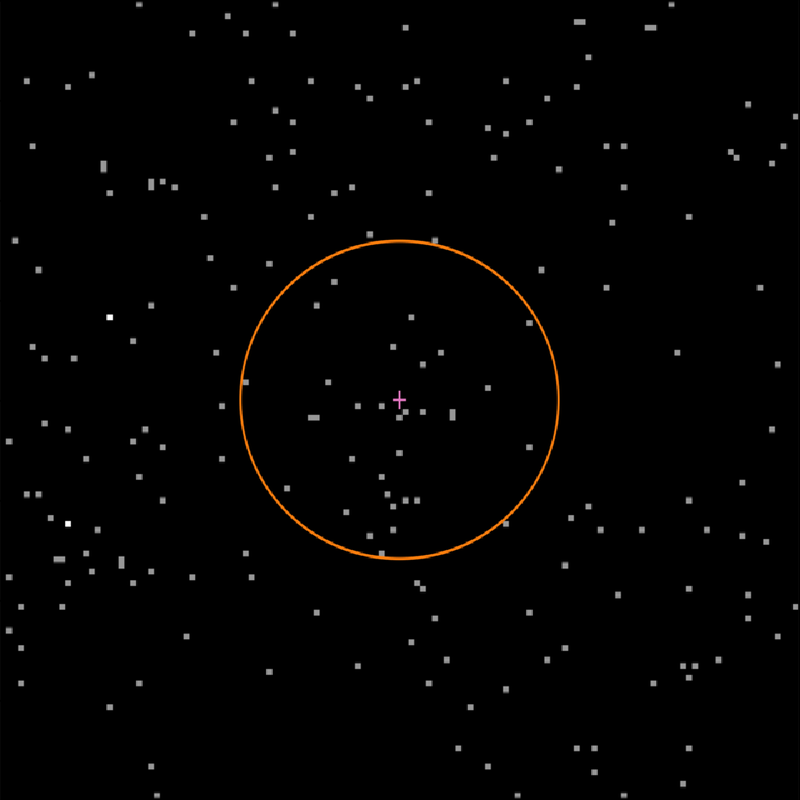} &
\includegraphics[width=5.4cm]{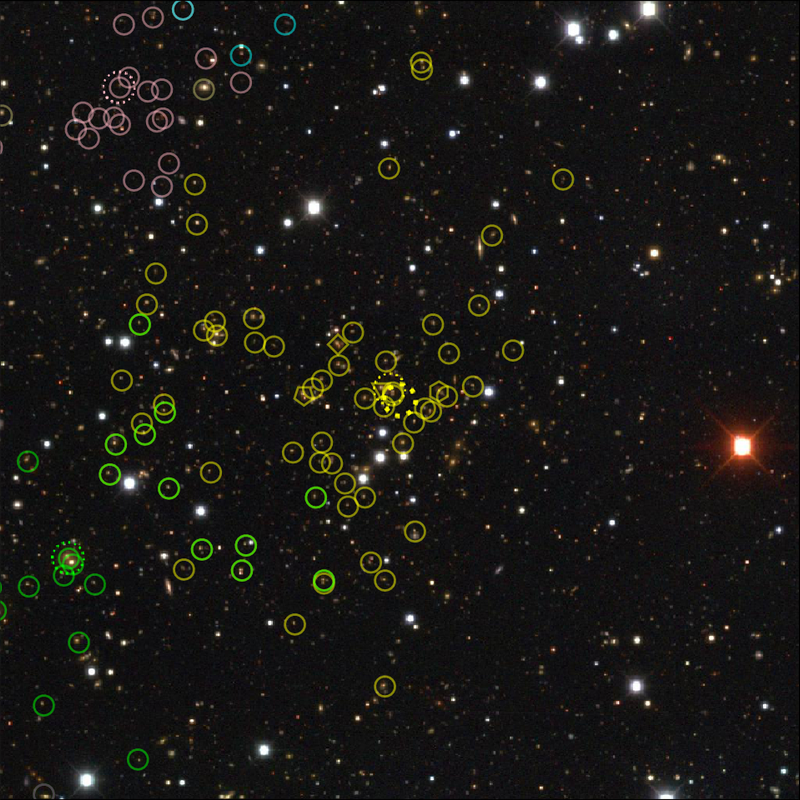} \\
\multicolumn{3}{c}{(i) RM ID 3470}\\
\includegraphics[width=5.4cm]{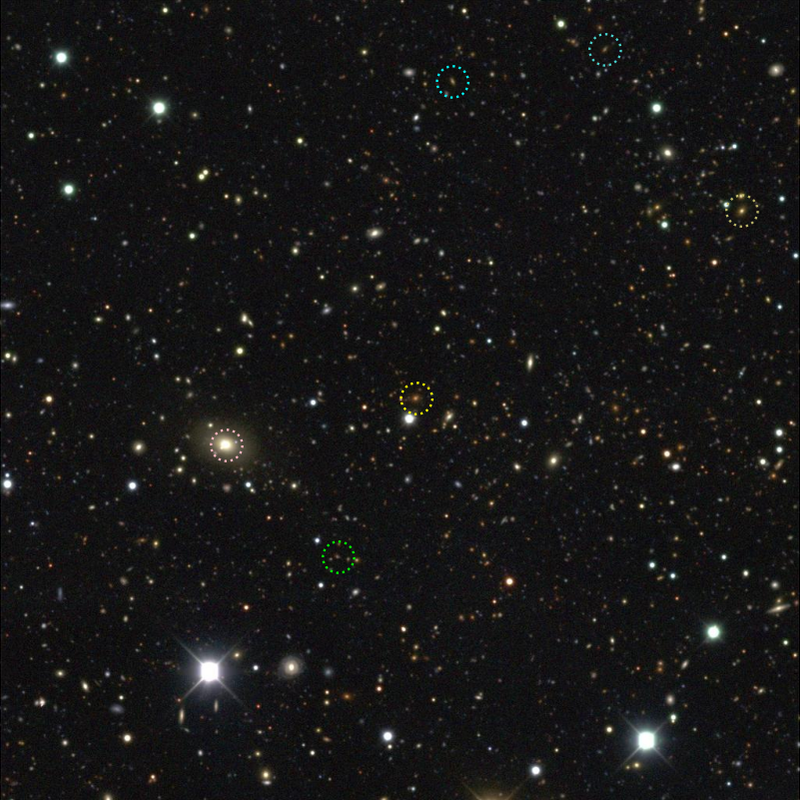} &
\includegraphics[width=5.4cm]{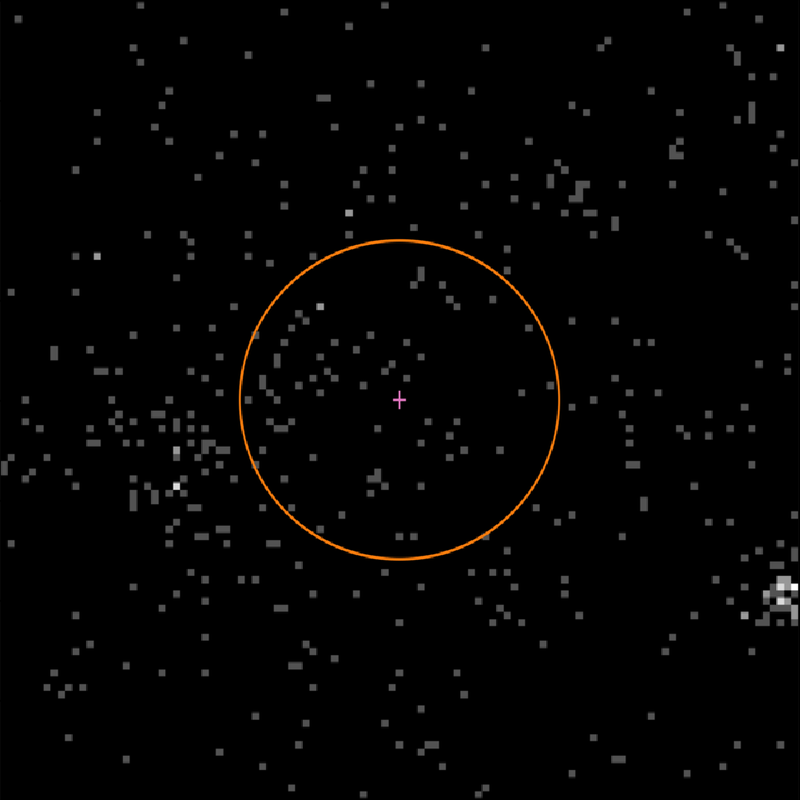} &
\includegraphics[width=5.4cm]{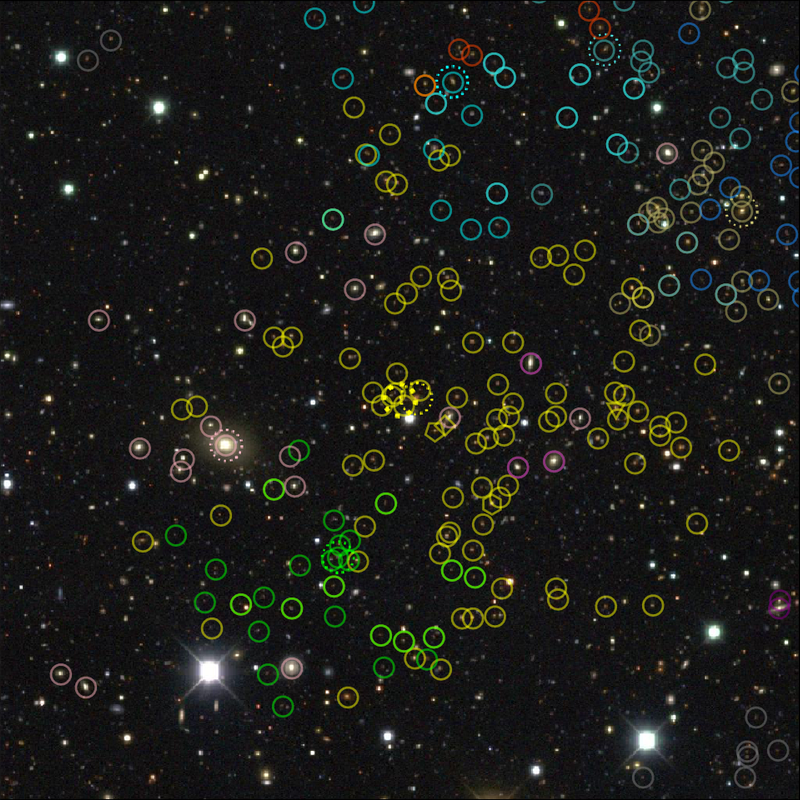} \\
\multicolumn{3}{c}{(j) RM ID 3834}\\
\includegraphics[width=5.4cm]{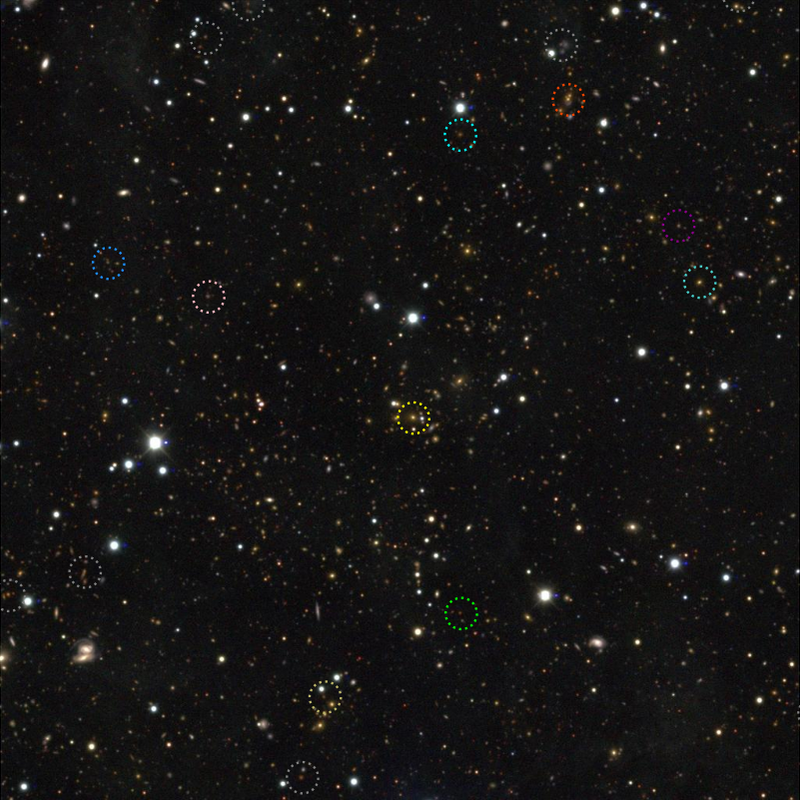} &
\includegraphics[width=5.4cm]{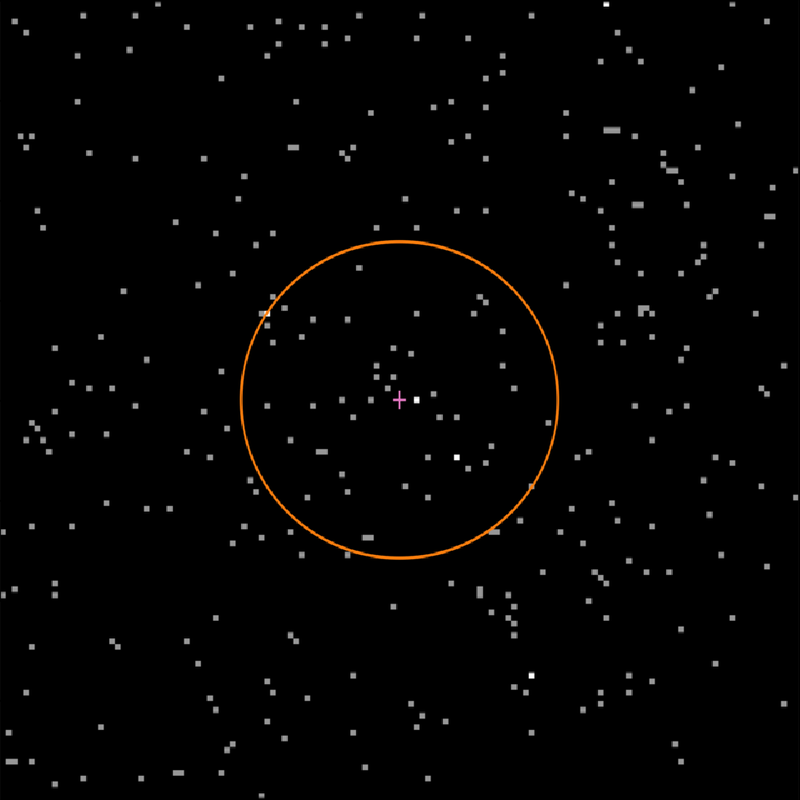} &
\includegraphics[width=5.4cm]{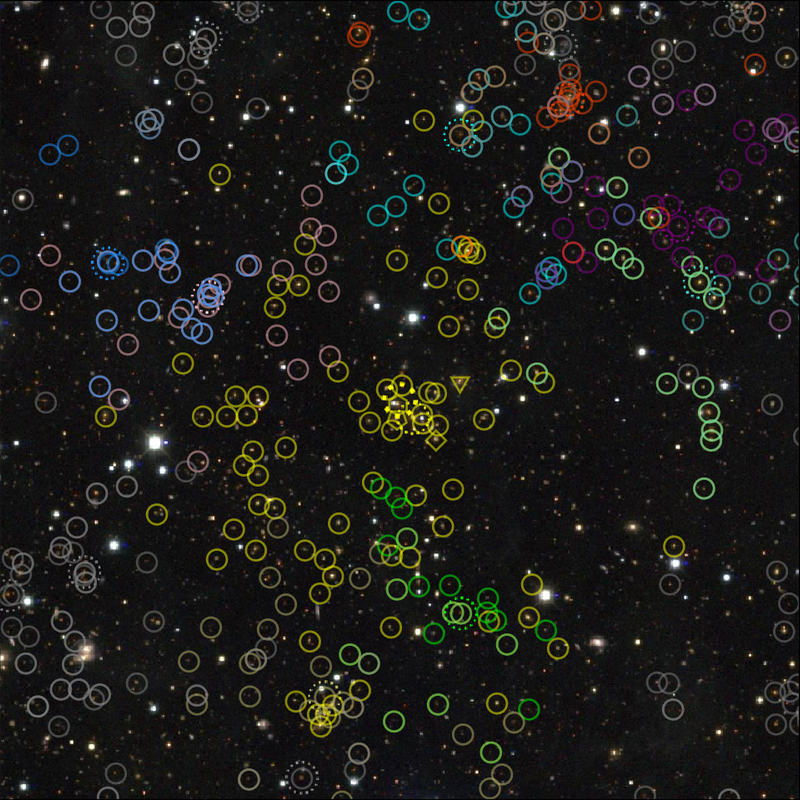} \\
\multicolumn{3}{c}{(k) RM ID 4156}\\
\includegraphics[width=5.4cm]{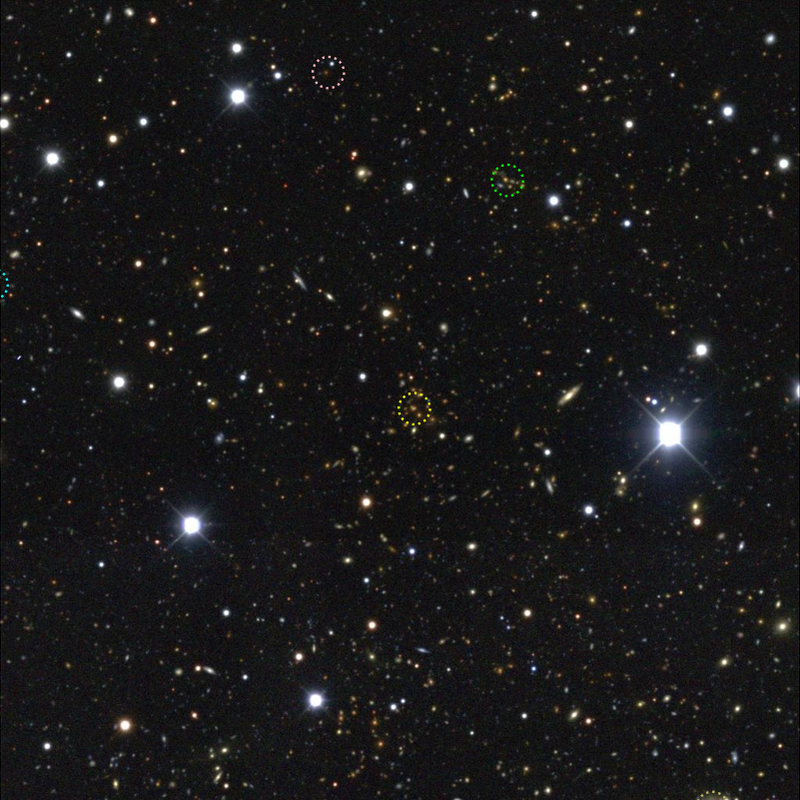} &
\includegraphics[width=5.4cm]{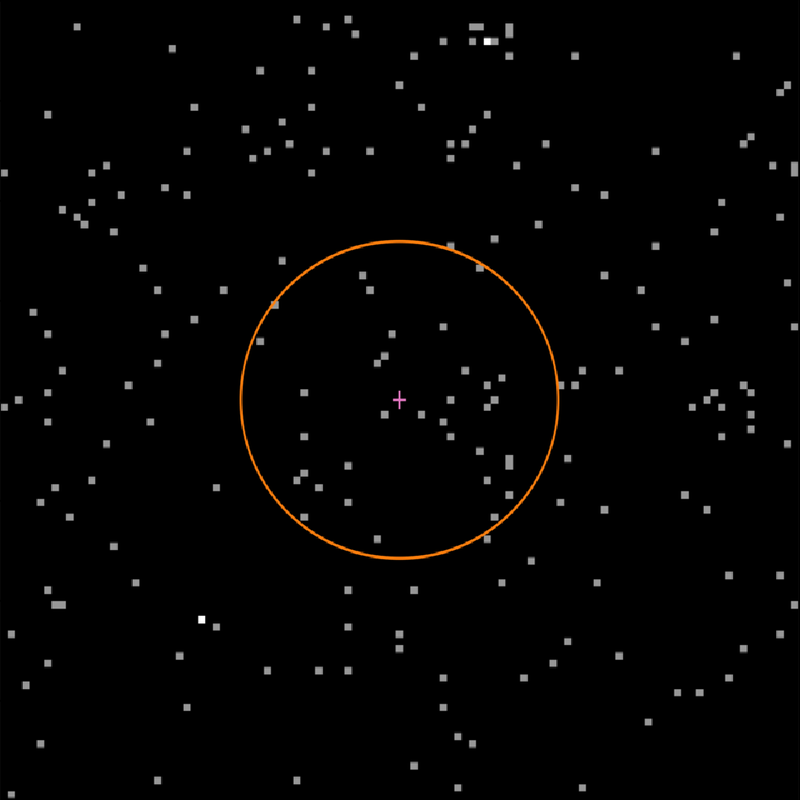} &
\includegraphics[width=5.4cm]{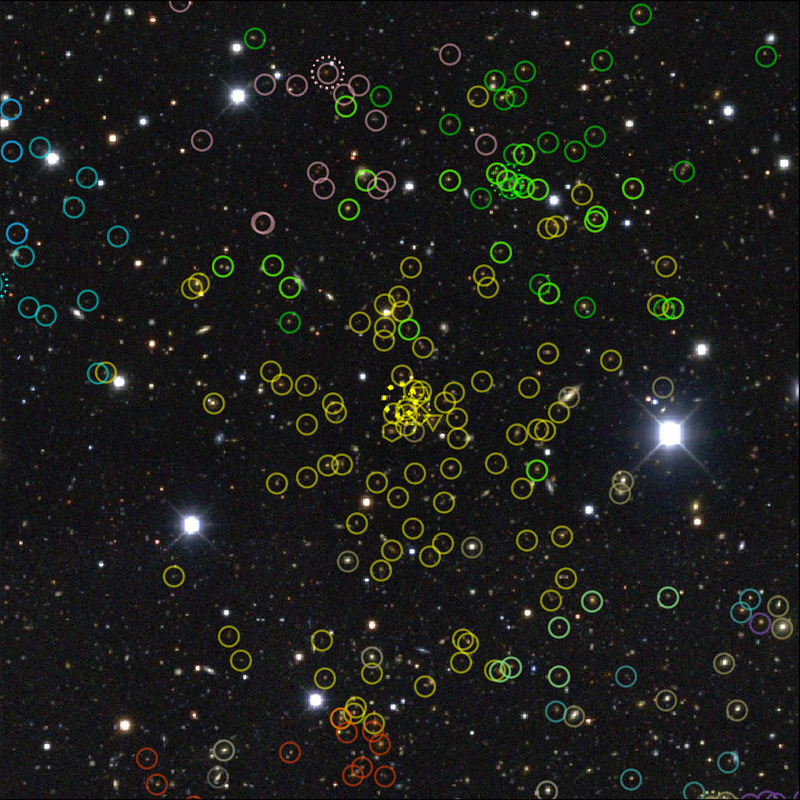} \\
\multicolumn{3}{c}{(l) RM ID 4917}\\
\end{tabular}
\end{center}
\contcaption{}
\end{figure*}

\section{DESY3RM clusters with additional available X-ray data}

Here we provide the properties of the DESY3RM clusters that have been detected by either {\em XMM} or {\em Chandra}.  In Section~\ref{sec:externaldata}, we found 7 DESY3RM clusters were detected and analysed via the {\em XMM} Cluster Survey, and 8 DESY3RM clusters were detected and analysed using {\em Chandra} observations as taken from \cite{kelly24}.  The {\em XMM} and {\em Chandra} properties of these clusters are given in Table~\ref{tab:xmmprops} and Table~\ref{tab:chandradata}.  Additionally, an {\em XMM} image of the DESY3RM cluster RM ID = 2096 (which had no match in the eRASS catalogue) is provided in Figure~\ref{fig:2096XMM}.

\begin{table*}
    \centering
    \begin{tabular}{ccccc}
    \hline
        RM ID & XCS ID & XCS R.A. & XCS Dec. & XCS $L_{X,0520}^{r500}$ \\
         & & (deg) & (deg) & $\times$10$^{42}$~(erg~s$^{-1}$) \\
        \hline\hline
        4 & XMMXCS J051636.6-543120.8 & 79.16 & -54.5 & 416.5$^{+2.1}_{-2.2}$ \\
        6 & XMMXCS J041114.1-481910.9 & 62.80 & -48.33 & 753.6$^{+6.1}_{-6.3}$ \\
        48 & XMMXCS J043818.3-541916.5 & 69.57 & -54.32 & 341.4$^{+10.2}_{-10.9}$ \\
        61 & XMMXCS J055943.5-524937.5 & 89.93 & -52.82 & 160.9$^{+2.9}_{-2.8}$ \\
        629 & XMMXCS J052215.8-481817.2 & 80.57 & -48.30 & -- \\
        1265 & XMMXCS J040352.4-571939.7 & 60.97 & -57.33 & -- \\
        2096$^{\dagger}$ & XMMXCS J042226.4-514025.8 & 65.61 & -51.67 & 99.6$^{+2.8}_{-3.4}$ \\
        \hline
    \end{tabular}
    \caption[]{Properties of DESY3RM clusters with available {\em XMM} data.  The X-ray luminosities were taken from \cite{kelly24} where available. $\dagger$The cluster RM ID 2096 is a part of the DESY3-eRASS-unmatched sample.}
    \label{tab:xmmprops}
\end{table*}

\begin{table*}
    \centering
    \begin{tabular}{ccccc}
    \hline
        RM ID & MATCHa ID & MATCHa R.A. & MATCHa Dec. & MATCHa $L_{X,0520}^{r500}$ \\
         & & (deg) & (deg) & $\times$10$^{42}$~(erg~s$^{-1}$) \\
        \hline\hline
        4 & RMJ051637.36-543001.65 & 79.15 & -54.51 & 466.0$_{-10.2}^{+6.0}$ \\
        6 & RMJ041110.97-481939.64 & 62.82 & -48.31 & 443.0$_{-7.9}^{+8.0}$ \\
        48 & RMJ043816.32-541902.97 & 69.57 & -54.32 & 745.0$_{-17.4}^{+17.3}$ \\
        61 & RMJ055943.22-524927.09 & 89.92 & -52.83 & 352.0$_{-10.1}^{+11.2}$ \\
        629 & RMJ052215.58-481817.53 & 80.57 & -48.30 & 280.0$_{-11.4}^{+12.0}$ \\
        674 & RMJ050921.39-534212.64 & 77.33 & -53.70 & 292.0$_{-11.2}^{+12.1}$ \\
        1437 & RMJ045628.12-511635.09 & 74.11 & -51.28 & 267.0$_{-5.9}^{+12.0}$ \\
        1638 & RMJ042604.12-545531.10 & 66.52 & -54.92 & 208.0$_{-24.2}^{+10.7}$ \\
        \hline
    \end{tabular}
    \caption[]{{\em Chandra} properties of DESY3RM clusters as taken from \cite{kelly24}.  Note the $r_{500}$ luminosities are provided via priv. communication.}
    \label{tab:chandradata}
\end{table*}

\begin{figure}
    \centering
    \includegraphics[width=5.4cm]{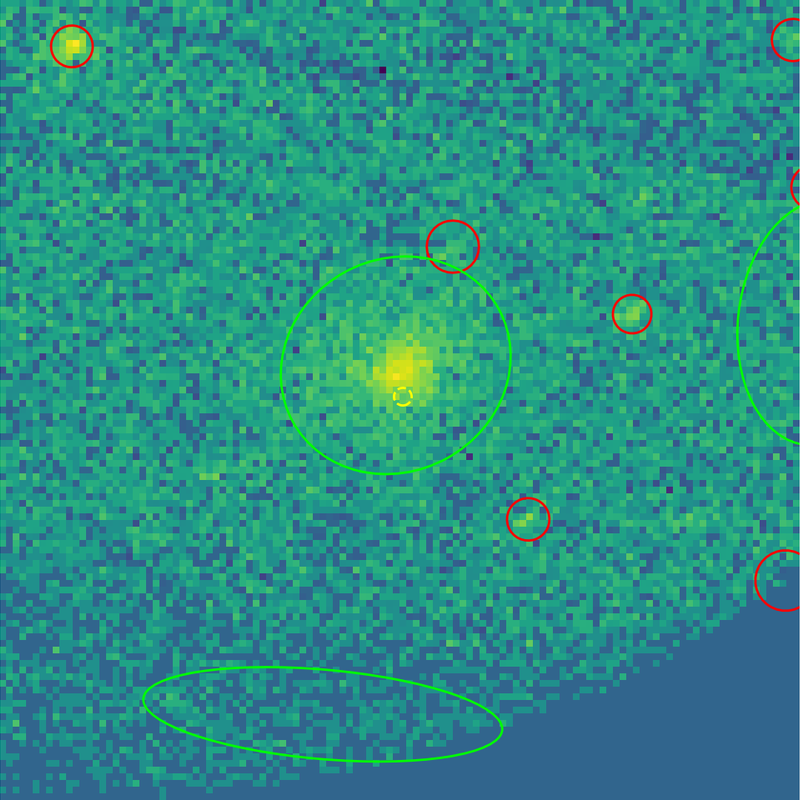}
    \caption{{\em XMM} soft-band (0.5--2.0 keV) image of the DESY3RM cluster RM ID 2096, centered on the redMaPPer coordinates of the cluster.  Extended sources are highlighted by the green regions.  Point sources in the field are highlighted by the red circles.}
    \label{fig:2096XMM}
\end{figure}

\section{AFFLILIATIONS}
{\em $^{1}$Department of Physics and Astronomy, University of Sussex, Brighton, BN1 9QH, UK \\
$^{2}$Center for Space Science and Technology, University of Maryland, Baltimore County (UMBC), Baltimore, MD 21250, USA\\
$^{3}$High Energy Astrophysics Science Archive Research Center (HEASARC), NASA/GSFC, Greenbelt, MD 20771, USA\\
$^{4}$Center for Research and Exploration in Space Science and Technology (CRESST II), NASA/GSFC, Greenbelt, MD 20771, USA\\
$^{5}$Santa Cruz Institute for Particle Physics, University of California, Santa Cruz, 1156 High St, Santa Cruz, CA 95064, USA \\
$^{6}$Departments of Statistics and Data Sciences, University of Texas at Austin, Austin, TX 78712, USA \\
$^{7}$The NSF-Simons AI Institute for Cosmic Origins, University of Texas at Austin, Austin, TX 78712, USA \\
$^{8}$Max Planck Institute for extraterrestrial Physics, Giessenbachstrasse 1, 85748 Garching, Germany \\
$^{9}$Universit\"at Innsbruck, Institut f\"ur Astro- und Teilchen- physik, Technikerstrasse 25, 6020 Innsbruck, Austria \\
$^{10}$Astrophysics Research Centre, University of KwaZulu-Natal, Westville Campus, Durban 4041, SA \\
$^{11}$School of Mathematics, Statistics, and Computer Science, University of KwaZulu-Natal, Westville Campus, Durban 4041, SA \\
$^{12}$Université Côte d’Azur, Observatoire de la Côte d’Azur, CNRS, Laboratoire Lagrange, Bd de l’Observatoire, CS 34229, 06304 Nice cedex 4, France\\
$^{13}$Department of Astronomy, University of Michigan, Ann Arbor, MI 48109, USA \\
$^{14}$Instituto de Astrof\'isica e Ci\^{e}ncias do Espa\c co, Universidade do Porto, CAUP, Rua das Estrelas, 4150-762 Porto, Portugal \\
$^{15}$Departamento de F\'isica e Astronomia, Faculdade de Ci\^{e}ncias, Universidade do Porto, Rua do Campo Alegre, 687, 4169-007 Porto, Portugal \\
$^{16}$INAF-Osservatorio Astronomico di Trieste, via G. B. Tiepolo 11, I-34143 Trieste, Italy \\
$^{17}$Laborat\'orio Interinstitucional de e-Astronomia - LIneA, Av. Pastor Martin Luther King Jr, 126 Del Castilho, Nova Am\'erica Offices, Torre 3000/sala 817 CEP: 20765-000, Brazil \\
$^{18}$Fermi National Accelerator Laboratory, P. O. Box 500, Batavia, IL 60510, USA \\
$^{19}$Department of Physics, University of Michigan, Ann Arbor, MI 48109, USA \\
$^{20}$Physik-Institut, University of Zurich, Winterthurerstrasse 190, CH-8057 Zurich, Switzerland \\
$^{21}$Institute of Cosmology and Gravitation, University of Portsmouth, Portsmouth, PO1 3FX, UK \\
$^{22}$Department of Physics \& Astronomy, University College London, Gower Street, London, WC1E 6BT, UK \\
$^{23}$School of Mathematics and Physics, University of Queensland,  Brisbane, QLD 4072, Australia \\
$^{24}$Instituto de Astrofisica de Canarias, E-38205 La Laguna, Tenerife, Spain \\
$^{25}$Universidad de La Laguna, Dpto. Astrof$\bar{A}$sica, E-38206 La Laguna, Tenerife, Spain \\
$^{26}$Institut de F\'{\i}sica d'Altes Energies (IFAE), The Barcelona Institute of Science and Technology, Campus UAB, 08193 Bellaterra (Barcelona) Spain \\
$^{27}$Astronomy Unit, Department of Physics, University of Trieste, via Tiepolo 11, I-34131 Trieste, Italy \\
$^{28}$Institute for Fundamental Physics of the Universe, Via Beirut 2, 34014 Trieste, Italy \\
$^{29}$Centro de Investigaciones Energ\'eticas, Medioambientales y Tecnol\'ogicas (CIEMAT), Madrid, Spain \\
$^{30}$Department of Physics, IIT Hyderabad, Kandi, Telangana 502285, India \\
$^{31}$Department of Physics and Astronomy, University of Pennsylvania, Philadelphia, PA 19104, USA \\
$^{32}$Universit\'e Grenoble Alpes, CNRS, LPSC-IN2P3, 38000 Grenoble, France\\
$^{33}$Department of Astronomy and Astrophysics, University of Chicago, Chicago, IL 60637, USA\\
$^{34}$Kavli Institute for Cosmological Physics, University of Chicago, Chicago, IL 60637, USA\\
$^{35}$ia-Bellido,"Instituto de Fisica Teorica UAM/CSIC, Universidad Autonoma de Madrid, 28049 Madrid, Spain \\
$^{36}$Institute of Space Sciences (ICE, CSIC),  Campus UAB, Carrer de Can Magrans, s/n,  08193 Barcelona, Spain\\
$^{37}$University Observatory, LMU Faculty of Physics, Scheinerstr. 1, 81679 Munich, Germany \\
$^{38}$Center for Cosmology and Astro-Particle Physics, The Ohio State University, Columbus, OH 43210, USA \\
$^{39}$Department of Physics, The Ohio State University, Columbus, OH 43210, USA\\
$^{40}$Australian Astronomical Optics, Macquarie University, North Ryde, NSW 2113, Australia \\
$^{41}$Lowell Observatory, 1400 Mars Hill Rd, Flagstaff, AZ 86001, USA \\
$^{42}$Jet Propulsion Laboratory, California Institute of Technology, 4800 Oak Grove Dr., Pasadena, CA 91109, USA \\
$^{43}$George P. and Cynthia Woods Mitchell Institute for Fundamental Physics and Astronomy, and Department of Physics and Astronomy, Texas A\&M University, College Station, TX 77843,  USA\\
$^{44}$Aix Marseille Univ, CNRS/IN2P3, CPPM, Marseille, France \\
$^{45}$Universit\'e Grenoble Alpes, CNRS, LPSC-IN2P3, 38000 Grenoble, France \\
$^{46}$Center for Astrophysical Surveys, National Center for Supercomputing Applications, 1205 West Clark St., Urbana, IL 61801, USA \\
$^{47}$Department of Astronomy, University of Illinois at Urbana-Champaign, 1002 W. Green Street, Urbana, IL 61801, USA \\
$^{48}$Instituci\'o Catalana de Recerca i Estudis Avan\c{c}ats, E-08010 Barcelona, Spain \\
$^{49}$School of Mathematics and Physics, University of Surrey, Guildford, Surrey, GU2 7XH, UK\\
$^{50}$Observat\'orio Nacional, Rua Gal. Jos\'e Cristino 77, Rio de Janeiro, RJ - 20921-400, Brazil\\
$^{51}$Kavli Institute for Particle Astrophysics \& Cosmology, P. O. Box 2450, Stanford University, Stanford, CA 94305, USA \\
$^{52}$SLAC National Accelerator Laboratory, Menlo Park, CA 94025, USA\\
$^{53}$Department of Physics, Northeastern University, Boston, MA 02115, USA\\
$^{54}$Physics Department, Lancaster University, Lancaster, LA1 4YB, UK\\
$^{55}$Hamburger Sternwarte, Universit\"{a}t Hamburg, Gojenbergsweg 112, 21029 Hamburg, Germany\\
$^{56}$Computer Science and Mathematics Division, Oak Ridge National Laboratory, Oak Ridge, TN 37831, USA\\
$^{57}$Cerro Tololo Inter-American Observatory, NSF's National Optical-Infrared Astronomy Research Laboratory, Casilla 603, La Serena, Chile\\
$^{58}$Berkeley Center for Cosmological Physics, Department of Physics, University of California, Berkeley, CA 94720, US \\
$^{59}$Lawrence Berkeley National Laboratory, 1 Cyclotron Road, Berkeley, CA 94720, USA \\
$^{60}$Universit\"ats-Sternwarte, Fakult\"at f\"ur Physik, Ludwig-Maximilians Universit\"at M\"unchen, Scheinerstr. 1, 81679 M\"unchen, Germany}

\bsp	
\label{lastpage}
\end{document}